\documentclass[twocolumn,traditabstract]{aa}
\usepackage{graphicx,amsmath}
\usepackage{epsf}
\usepackage{txfonts}
\usepackage[hyphens]{url}
\usepackage{longtable}
\usepackage{lscape}
\usepackage[hyperindex,breaklinks=true, colorlinks, citecolor=blue]{hyperref}  
\usepackage{breakurl}
\usepackage{comment}
\usepackage{natbib}
\usepackage{upgreek}
\usepackage{float}
\usepackage[switch,pagewise]{lineno}
\usepackage{multirow}
\usepackage{caption}

\newcommand{\hralf}[1]{}
\newcommand{\juan}[1]{{#1}}

\newcommand{\commentproof}[1]{{\bf \color{green}#1}}
\renewcommand{\commentproof}[1]{} 

\providecommand{\sorthelp}[1]{}

\def\deg{\ifmmode^\circ\else$^\circ$\fi}
\def\pdeg{\ifmmode $\setbox0=\hbox{$^{\circ}$}\rlap{\hskip.11\wd0 .}$^{\circ} \else \setbox0=\hbox{$^{\circ}$}\rlap{\hskip.11\wd0 .}$^{\circ}$\fi}
\def\arcs{\ifmmode {^{\scriptstyle\prime\prime}} \else $^{\scriptstyle\prime\prime}$\fi}
\def\arcm{\ifmmode {^{\scriptstyle\prime}} \else $^{\scriptstyle\prime}$\fi}
\newdimen\saa  \newdimen\sbb
\def\parcs{\saa=.07em \sbb=.03em
     \ifmmode \hbox{\rlap{.}}^{\scriptstyle\prime\kern -\sbb\prime}\hbox{\kern -\sa}
     \else \rlap{.}$^{\scriptstyle\prime\kern -\sbb\prime}$\kern -\sa\fi}
\def\parcm{\saa=.08em \sbb=.03em
     \ifmmode \hbox{\rlap{.}\kern\saa}^{\scriptstyle\prime}\hbox{\kern-\sbb}
     \else \rlap{.}\kern\saa$^{\scriptstyle\prime}$\kern-\sbb\fi}

\begin{document}

\title{The Galactic dynamics revealed by the filamentary structure in atomic hydrogen emission}
\titlerunning{The Galactic dynamics revealed by the filaments in the H{\sc i} emission}
    \author{
        J.~D.~Soler$^{1}$\thanks{Corresponding author, \email{juandiegosolerp@gmail.com}},
        M.-A.~Miville-Desch\^{e}nes$^{2}$,
        S.~Molinari$^{1}$,
        R.\,S.~Klessen$^{3,4}$,
        P.~Hennebelle$^{2}$,
        L.~Testi$^{5}$,
        N.\,M.~McClure-Griffiths$^{6}$,
        H.~Beuther$^{7}$,
        D.~Elia$^{1}$,
        E.~Schisano$^{1}$,
        A.~Traficante$^{1}$,
        P.~Girichidis$^{3}$,
        S.\,C.\,O.~Glover$^{3}$,
        R.\,J.~Smith$^{8}$,
        M.~Sormani$^{3}$,
        R.~Tre\ss$^{3}$
} 
\institute{
1. Istituto di Astrofisica e Planetologia Spaziali (IAPS). INAF. Via Fosso del Cavaliere 100, 00133 Roma, Italy\\
2. Laboratoire AIM, Paris-Saclay, CEA/IRFU/SAp - CNRS - Universit\'{e} Paris Diderot. 91191, Gif-sur-Yvette Cedex, France.\\
3. Universit\"{a}t Heidelberg, Zentrum f\"{u}r Astronomie, Institut f\"{u}r Theoretische Astrophysik, Albert-Ueberle-Str. 2, 69120, Heidelberg, Germany.\\ 
4. Universit\"{a}t Heidelberg, Interdiszipli\"{a}res Zentrum f\"{u}r Wissenschaftliches Rechnen, 69120 Heidelberg, Germany.\\
5. European Southern Observatory, Karl-Schwarzschild-Strasse 2, 85748, Garching bei M\"{u}nchen, Germany.\\
6. Research School of Astronomy and Astrophysics, The Australian National University, Canberra, ACT, Australia.\\
7. Max-Planck Institute for Astronomy, K\"{o}nigstuhl 17, 69117, Heidelberg, Germany.\\
8. Jodrell Bank Centre for Astrophysics, Department of Physics and Astronomy, University of Manchester, Oxford Road, Manchester M13 9PL, UK.
}
\authorrunning{Soler,\,J.D. et al.}

\date{Received 16 02 2022 / Accepted 06 05 2022}

\abstract{
We present a study of the filamentary structure in the neutral atomic hydrogen (H{\sc i}) emission at the 21\,cm \juan{wavelength} toward the Galactic plane using the 16\parcm2-resolution observations in the H{\sc i} 4$\pi$ (HI4PI) survey.
Using the Hessian matrix method across radial velocity channels, we identified the filamentary structures and quantified their orientations using circular statistics.
We found that the regions of the Milky Way's disk beyond 10\,kpc and up to roughly 18\,kpc from the Galactic center display H{\sc i} filamentary structures predominantly parallel to the Galactic plane.
For regions at lower Galactocentric radii, we found that the H{\sc i} filaments are mostly perpendicular or do not have a preferred orientation with respect to the Galactic plane.
We interpret these results as the imprint of supernova feedback in the inner Galaxy and Galactic rotation and shear in the outer Milky Way.
We found that the H{\sc i} filamentary structures follow the Galactic warp and flaring and that they highlight some of the variations interpreted as the effect of the gravitational interaction with satellite galaxies. 
In addition, the mean scale height of the filamentary structures is lower than that sampled by the bulk of the H{\sc i} emission, thus indicating that the cold and warm atomic hydrogen phases have different scale heights in the outer galaxy.
Finally, we found that the fraction of the column density in H{\sc i} filaments is almost constant up to approximately 18\,kpc from the Galactic center.
This is possibly a result of the roughly constant ratio between the cold and warm atomic hydrogen phases inferred from the H{\sc i} absorption studies.
Our results indicate that the H{\sc i} filamentary structures provide insight into the dynamical processes shaping the Galactic disk.
Their orientations record how and where the stellar energy input, the Galactic fountain process, the cosmic ray diffusion, and the gas accretion have molded the diffuse interstellar medium in the Galactic plane.
}

\maketitle

\section{Introduction}

The diffuse neutral atomic hydrogen (H{\sc i}) is the matrix within which star-forming clouds form and reside and the medium that takes a large portion of the energy injected by the massive stars during their lifecycles \citep[see,][for a review]{dickeyANDlockman1990,kalberla2009,Ballesteros-Paredes2020}.
Therefore, the investigation of the H{\sc i} distribution and dynamics is crucial to understanding the cycle of energy and matter in the interstellar medium \citep[ISM,][]{ferriere2001,klessen2016}.
In this paper, we present a study of the H{\sc i} emission toward the Galactic plane, mainly focusing on the orientation of its filamentary structure and the Galactic dynamics that it reveals.

The H{\sc i} emission at the 21\,cm wavelength has been instrumental for the study of the Milky Way. 
This radio line is not limited by the dust absorption that renders optical methods ineffective at distances greater than a few kiloparsecs \citep{vandeHulst1953}.
Face-on maps of the outer galaxy have been constructed by using the Doppler shift of the emission line in combination with the Galactic rotation curve  \citep{vandeHulst1954,westerhout1957,oort1958}.
The concentration of the H{\sc i} gas in spiral arm features has been evident since the first face-on projections, although the detailed reconstruction of the density structures is complicated by the departures from circular motion, the complexity of the velocity field, and the overlapping of multiple components \citep{kerr1969}.

Much of H{\sc i} is observed to be either a warm neutral medium (WNM) with $T$\,$\approx$\,$10^{4}$\,K or a cold neutral medium (CNM) with $T$\,$\approx$\,$10^{2}$\,K \citep{heilesANDtroland2003}.
In the classical two-phase picture, both phases exist in pressure equilibrium with stable phases for temperatures of $T$\,$\approx$\,300\,K (CNM) and $T$\,$\approx$\,5000\,K (WNM) embedded in a hot ionized phase \citep{mckeeANDostriker1977,wolfire2003}. 
A portion of the CNM is sampled by the absorption of background H{\sc i} emission by cold foreground H{\sc i}, which is generically known as H{\sc i} self-absorption \citep[HISA;][]{heeschen1955,gibson2000,wang2020hisa}.
However, most of our understanding of the multiphase structure of H{\sc i} comes from the study of absorption toward continuum sources \citep[see, for example,][]{strasser2007,stanimirovic2014,murray2018}.
Comparisons between 21-cm H{\sc i} emission and absorption measurements indicate that, in the vicinity of the Sun, the WNM has roughly the same column density as the CNM \citep{falgaronANDlequeux1973,liszt1983}.
Observations of the absorption against strong continuum sources indicate that the CNM to WNM mixture may be constant throughout the Galaxy \citep{dickey2009,dickey2021}.

The first all-sky H{\sc i} observations, provided by the 36\arcm-resolution Leiden/Argentine/Bonn (LAB) survey \citep{kalberla2005}, displayed the Milky Way as a nonaxisymmetric spiral system of four arms that are traced out to at least 25\,kpc from the Galactic center \citep{levine2006,nakanishi2016,koo2017}.
These four arms are designated Sagittarius-Carina (Sgr-Car), Scutum-Centaurus (Sct-Cen), Perseus, and Outer (or Norma-Cygnus).
Their existence, however, is in tension with the star counts toward the tangent points, which favor the existence of only two spiral arms \citep{drimmel2000,benjamin2009,churchwell2009}.

Further studies of the H{\sc i} emission have also identified additional spiral arm structures.
Using the LAB survey observations, \cite{dameANDthaddeus2011} identified a presumed spiral arm structure lying beyond the Outer Arm in the first Galactic quadrant (QI) at approximately 15\,kpc from the Galactic center.
This arm appears to be the continuation of the Scutum-Centaurus Arm in the outer Galaxy, as a symmetric counterpart of the nearby Perseus Arm, for which it is designated as the Outer Scutum-Centaurus (OSC) arm.
Using the 2\arcm-resolution H{\sc i} observations in the interferometric Southern Galactic Plane Survey \citep[SGPS,][]{mcclure-griffiths2005}, \citet{mcclure-griffiths2004} identified a possible distant spiral arm in the fourth quadrant (QIV) of the Milky Way. 
This distinct emission feature can be traced for over 70\deg\ as the most extreme positive velocity feature in the longitude-velocity diagram.
Assuming circular motions around the Galactic center, this structure is located at Galactocentric radii ($R_{\rm gal}$) between 18 and 24\,kpc.

The H{\sc i} emission also reveals the bending of the Galactic plane noticeable at $R_{\rm gal}$\,$>$\,9\,kpc and up to 40\,kpc \citep{henderson1982,levine2006warp,kalberla2007}.
This ``Galactic warp'' is also observed in the dust emission, in the distribution of H{\sc ii} regions, in the CO emission, and the kinematics of stars beyond the solar neighborhood \citep[see, for example,][]{drimmel2001,sun2020,cheng2020,paladini2004}.
Furthermore, the H{\sc i} emission also reveals that the average disk thickness shows a pronounced flaring that can be approximated by an exponential function in the range $5$\,$<$\,$R_{\rm gal}$\,$<$\,35\,kpc \citep{kalberla2008}.

In addition to the spiral-arm structure, the warp, and the flare, the distribution of the H{\sc i} emission also carries the imprint of the energy input by main sequence stars, supernovae (SNe), and other energetic process in the ISM.
The first photographic representations of the H{\sc i} emission toward the Galactic plane revealed a series of shells, arcs, and filaments \citep{weaverANDwilliams1973,heiles1979}.
The H{\sc i} shells are thought to be formed by the combined effects of stellar winds and SNe sweeping up gas and dust 
\citep{cox1974,castor1975,mckeeANDostriker1977}.
The vertical H{\sc i} filaments, or ``worms'' seemingly ``crawling away from the Galactic plane'', have been interpreted as the walls of bubbles which have broken through the thin gaseous disk \citep{heiles1984,koo1992}.
 
Dedicated studies of the filamentary structures in the H{\sc i} emission, based on single-dish observations at 4\arcmin\ and 30\arcmin\ resolutions, have shown that they have a preferential orientation parallel to the interstellar magnetic field in the local ISM \citep{clark2014,kalberla2016}.
This alignment may be the product of magnetically-induced velocity anisotropies, the collapse of material along field lines, shocks, or anisotropic density distributions \citep[see, for example,][]{lazarianANDpogosyan2000,heitsch2001a,chenANDostriker2015,inoueANDinutsuka2016,solerANDhennebelle2017,moczANDburkhart2018}.
More recently, \cite{soler2020} presented the analysis of the 40\arcsec-resolution H{\sc i} interferometric observations in The H{\sc i}/OH/recombination-line (THOR) survey \citep[][]{beuther2016,wang2020hi}, revealing that the majority of the filamentary structures in the H{\sc i} emission are aligned with the Galactic plane, but that there are portions of the first Galactic quadrant dominated by vertical H{\sc i} structures that are most likely resulting from the combined effect of SN feedback and the interstellar magnetic fields.

This paper presents the extension of the \cite{soler2020} analysis to the whole Galactic disk, hence extending in Galactic longitude and Galactic latitude.
The filament orientation is a characteristic of the H{\sc i} emission that is less explicitly dependent on the angular resolution than other properties, such as the filament width and the length.
If the filaments are understood as features in the intensity field and not as physical objects, this analysis characterizes the anisotropy in the emission distribution.
This characteristic is particularly relevant given the natural axis of symmetry in the Galactic plane. The gravitational potential of the Galaxy and the stretching due to the circular motions and the spiral arms could naturally produce horizontal structures.
Departures from this natural anisotropy may indicate energetic processes that lift the gas in the disk or infall.
Thus, rather than defining the filaments as objects, this study aims at studying the general anisotropy of the atomic gas distribution across the Galactic disk.

To cover the entire Galactic plane, we used the state-of-the-art H{\sc i} emission observations that we introduce in Sec.~\ref{section:observations}.
We describe the methods for identifying the filamentary structures and quantifying their preferential orientation in Sec.~\ref{section:methods}.
We present the main results of our analysis in Sec.~\ref{section:results}.
We provide a discussion of our results in Sec.~\ref{section:discussion} and present our general conclusions in Sec.~\ref{section:conclusions}.
We reserve additional analyses for a set of appendices as follows.
Appendix~\ref{appendix:method} presents details on the Hessian matrix filament identification method, detailing the use of circular statistics and the selection of the analysis parameters.
We present the method for de-projecting the radial velocities into Galactocentric radii and the calculation of the disk's warp and flaring in App.~\ref{app:deprojection}.
Appendix~\ref{appendix:HI4PIandGALFAHI} presents the comparison between the filamentary structures found in the HI4PI and GALFA-HI surveys.
Finally, App.~\ref{appendix:HVCs} presents the results of the filament analysis toward two high-velocity cloud complexes.

\section{Data}\label{section:observations}

We used the publicly-available observations of the emission by atomic hydrogen at 21\,cm wavelength in the H{\sc i} 4$\pi$ (HI4PI) survey \citep[][]{hi4pi2016}.
The HI4PI survey is based on data from the Effelsberg-Bonn H{\sc i} Survey \citep[EBHIS,][]{kerp2011} and the Galactic All-Sky Survey \cite[GASS,][]{McClure-Griffiths2009,kalberla2010}.
This data comprises 21-cm neutral atomic hydrogen data over the whole sky in a radial velocity range $-600$\,$<$\,$v_{\rm LSR}$\,$<$\,$600$\,km\,s$^{-1}$ for declination $\delta$\,$>$\,0\deg\ and $-470$\,$<$\,$v_{\rm LSR}$\,$<$\,470\,km\,s$^{-1}$ for $\delta$\,$<$\,0\,\deg, as observed with the 100-m radio telescope at Effelsberg and the 64-m radio telescope at Parkes.
These observations have full spatial sampling, thus overcoming the major issue of the LAB survey, and provide a final data product with an angular resolution of 16\parcm2 and sensitivity of 43\,mK per 1.29-km\,s$^{-1}$ velocity channel.

We used the data distributed in FITS-format binary tables containing lists of spectra sampled on a Hierarchical Equal Area isoLatitude Pixelization (HEALPix\footnote{\url{http://healpix.sf.net}}) grid with $N_{\rm side}$\,=\,$1024$, which corresponds to a pixel angular size of 3\parcm44 \citep{gorski2005,zonca2019}.
We arranged these spectra in all-sky HEALPix maps corresponding to each of the 1.29-km\,s$^{-1}$ velocity channels.
Using the tools in the Python {\tt healpy} package, we produced a Cartesian projection for each one of these velocity channels to obtain maps covering the range $-180$\deg\,$<$\,$l$\,$<$\,180\deg\ and $-10$\deg\,$<$\,$b$\,$<$\,10\deg\ with a pixel size $\delta l$\,$=$\,$\delta b$\,$=$\,3\parcm44.
We combined these velocity-channel maps into a single position-position-velocity (PPV) cube, which we segmented for the analysis of particular $l$ and $b$ ranges using the {\tt spectral-cube\footnote{\url{http://spectral-cube.readthedocs.io}}} package.

We also used the publicly-available observations from the Galactic Arecibo L-Band Feed Array H{\sc i} survey (GALFA-H{\sc i}) data release 2 \citep[DR2,][]{peek2018}.
This survey covers the H{\sc i} emission from $-650$ to $650$\,km\,s$^{-1}$, with 0.184\,km\,s$^{-1}$ channel spacing, 4\arcmin\ angular resolution, and 150\,mK  root mean square (rms) noise per 1\,km\,s$^{-1}$ velocity channel.
The GALFA-H{\sc i} DR2 observations cover the entirety of the sky available from the William E. Gordon 305\,m antenna at Arecibo, from $\delta$\,$=$\,$-1$\deg17\arcmin\ to $+37$\deg57\arcmin\ across the whole right ascension ($\alpha$) range.
We chose these data to extend our analysis to two portions of the Galactic plane at higher angular and spectral resolution.
In the range $|b|$\,$\leq$10\deg, GALFA-H{\sc i} DR2 provides observations in the ranges 36\pdeg5\,$<$\,$l$\,$<$\,70\pdeg5 and 175\pdeg3\,$<$\,$l$\,$<$\,209\pdeg4.

We obtained the GALFA-H{\sc i} DR2 distributed in FITS-format cubes in the ``narrow'' velocity range ($v$\,$\leq$188\,km\,s$^{-1}$). through the GALFA collaboration website\footnote{\url{https://purcell.ssl.berkeley.edu/}}.
We used the Python {\tt spectral-cube} and {\tt reproject\footnote{\url{https://reproject.readthedocs.io}}} packages to arrange these cubes into two mosaics covering the aforementioned $l$ and $b$ ranges with a pixel size $\delta l$\,$=$\,$\delta b$\,$=$\,1\parcm0 and spectral resolution of 0.184\,km\,s$^{-1}$.

\section{Methods}\label{section:methods}

\subsection{Analysis of filamentary structure}\label{ssec:HIfils}

We selected the filamentary structures using the Hessian matrix technique as implemented in \cite{soler2021} and coded in the {\tt hessiananalysis} routine of the {\tt magnetar} software package\footnote{\url{https://github.com/solerjuan/magnetar}} \citep{magnetar2020}.
This method is less computationally demanding than other filament identification methods, such as the rolling Hough transform \citep[RHT,][]{clark2014} or FILFINDER \citep{koch2015}, and yields similar results for the filament orientation, as detailed in Appendix C of \cite{soler2020}.
The main parameters used in the analysis are summarized in Table~\ref{table:HessianParameters} and were applied as follows.

The HI4PI data was divided in 3\deg\,$\times$\,20\deg\,$\times$\,1.29-km\,s$^{-1}$ velocity channel maps centered on Galactic latitude $b$\,$=$\,$0$\deg\ and covering the Galactic longitude in the range 4\deg\,$<$\,$l$\,$<$\,356\deg\ and the radial velocity range $-200$\,$<$\,$v_{\rm LSR}$\,$<$\,200\,km\,s$^{-1}$.
Throughout this paper we refer to these velocity channels as ``tiles''.
The rectangular shape of the tiles aims to maximize the $b$ coverage and maintain the fine sampling in $l$, covering both the Galactic warp and flaring.
This selection does not imply any loss of generality in our results, as we demonstrate for different tile sizes in App.~\ref{appendix:method}.

For each tile centered on the position $l_{0}$ and $b_{0}$ and $v_{\rm LSR}$\,$=$\,$v_{0}$, $I(l_{0},b_{0},v_{0})$, we estimated the derivatives with respect to the local coordinates $(x,y)$ and built the Hessian matrix,
\begin{equation}
\mathbf{H}(x,y) \equiv \begin{bmatrix} 
H_{xx} & H_{xy} \\
H_{yx} & H_{yy} 
\end{bmatrix},
\end{equation}
where $H_{xx}$\,$\equiv$\,$\partial^{2} I/\partial x^{2}$, $H_{xy}$\,$\equiv$\,$\partial^{2} I/\partial x \partial y$, $H_{yx}$\,$\equiv$\,$\partial^{2} I/\partial y \partial x$, $H_{yy}$\,$\equiv$\,$\partial^{2} I/\partial y^{2}$, and $x$ and $y$ are related to the Galactic coordinates $(l, b)$ as $x$\,$\equiv$\,$l\cos b$ and $y$\,$\equiv$\,$b$, so that the $x$-axis is parallel to the Galactic plane.
We obtained the partial spatial derivatives by convolving $I(l_{0},b_{0},v_{0})$ with the second derivatives of a two-dimensional Gaussian function.
In practice, we used the {\tt gaussian\_filter} function in the open-source software package {\tt SciPy} \citep{2020SciPy-NMeth}.
We report the results obtained with a 18\arcm\ FWHM derivative kernel, but present the results of different kernel sizes in App.~\ref{appendix:method}.

The two eigenvalues ($\lambda_{\pm}$) of the Hessian matrix were found by solving the characteristic equation,
\begin{equation}\label{eq:lambda}
\lambda_{\pm} = \frac{(H_{xx}+H_{yy}) \pm \sqrt{(H_{xx}-H_{yy})^{2}+4H_{xy}H_{yx}}}{2}.
\end{equation}
Both eigenvalues define the local curvature of $I(l_{0},b_{0},v_{0})$.
In particular, the minimum eigenvalue ($\lambda_{-}$) highlights filamentary structures or ridges, as detailed in \cite{planck2014-XXXII}.
The eigenvector corresponding to $\lambda_{-}$ defines the orientation of intensity ridges with respect to the Galactic plane, which is characterized by the angle
\begin{equation}\label{eq:theta}
\theta = \frac{1}{2}\arctan\left[\frac{H_{xy}+H_{yx}}{H_{xx}-H_{yy}}\right].
\end{equation}
We estimated $\theta$ for each of the pixels in $I(l_{0},b_{0},v_{0})$.
This angle, however, is only meaningful in regions of the map that are rated as filamentary according to selection criteria based on the values of $\lambda_{-}$ and on the noise properties of the data.

\begin{table}
\caption{Hessian analysis parameters.}              
\label{table:HessianParameters}      
\centering                                      
\begin{tabular}{l l l}          
\hline\hline                        
Parameter & & Value \\    
\hline                                   
Tile size & & 3\deg\,$\times$\,20\deg\,$\times$\,1.29\,km\,s$^{-1}$ \\      
Kernel size & & 18\arcmin\ FWHM \\
Intensity threshold & & 0.3\,K \\
Curvature threshold ($\lambda^{C}_{-}$) & & $-1.8$\,K/deg$^{2}$ \\
\hline                                             
\end{tabular}
\end{table}

\subsection{Data selection for statistical analysis}\label{sssec:selection}

\juan{In each tile, we selected the filamentary structures based on the criterion $\lambda_{-}$\,$<$\,$\lambda^{C}_{-}$\,$<$\,$0$, where $\lambda^{C}_{-}$ is a curvature threshold that we determined from the data.}
Additionally, we chose regions where $I(l_{0},b_{0},v_{0})$\,$>$\,7$\sigma_{I}$.
\juan{We picked $\sigma_{I}$\,$=$\,0.43\,mK to match the brightness temperature noise level for a 1.29 km\,s$^{-1}$ velocity channel reported in \cite{hi4pi2016}, which was estimated using emission line-free spectral channels to compute the standard deviation and iteratively clip outliers following the procedure described in \cite{winkel2016a}.}
\juan{We empirically tested different signal-to-noise ratio (S/N) thresholds for a fixed $\lambda^{C}_{-}$ and found no significant differences in our results for $I(l_{0},b_{0},v_{0})$\,$>$\,3$\sigma_{I}$, 5$\sigma_{I}$, and up to 9$\sigma_{I}$, as shown in App.~\ref{appendix:hessian}.}
\juan{Most of the differences introduced by the intensity threshold were found in a few tiles beyond the terminal velocities, which are dominated by fainter emission and are dismissed with the highest S/N cutoffs.}
\juan{We found that the distribution of $V$ across tiles is more sensitive to $\lambda^{C}_{-}$ than to the intensity S/N, thus, we chose 5$\sigma_{I}$ as a reference value to charaterize the curvature threshold.}

Following the method introduced in \cite{planck2014-XXXII}, we estimated $\lambda^{C}_{-}$ in noise-dominated velocity channels.
\juan{For each tile group centered on a particular Galactic longitude, we estimated $\lambda_{-}$ in five velocity channels with low S/N and determined the minimum value of $\lambda_{-}$ in each of them.}
We used the median of these five $\lambda_{-}$ values as the threshold value, $\lambda^{C}_{-}$.
We employed the median to reduce the effect of outliers, but in general the values of $\lambda_{-}$ in the noise-dominated channels are similar and this selection does not imply any loss of generality.
\juan{In the filament orientation analysis, we exclusively considered regions of each velocity channel map where $\lambda_{-}$\,$<$\,$\lambda^{C}_{-}$, which corresponds to the selection of filamentary structures with curvatures in $I(l_{0},b_{0},v_{0})$ larger than those present in the noise-dominated channels.}
\juan{We report the results of applying a common $\lambda^{C}_{-}$ estimated from the values found in all of the tile groups, but also evaluated the selection of a different $\lambda^{C}_{-}$ for each tile group, which we present in App.~\ref{appendix:hessian}.}
\juan{The outcome of the two approaches is very similar and the selection of one instead of the other does not have a significant effect on the results of our analysis.}

\subsection{Circular statistics}\label{sssec:circStats}

Once the filamentary structures were selected, we used the angles derived from Eq.~\ref{eq:theta} to study their orientation with respect to the Galactic plane.
For a systematic evaluation of the preferential orientation, we applied the projected Rayleigh statistic ($V$) \citep[see, for example,][]{batschelet1981}, which is a test to determine whether the distribution of angles is not uniform and peaked at a particular angle.
This test is equivalent to the modified Rayleigh test for uniformity proposed by \cite{durandANDgreenwood1958} for the specific directions of interest $\theta$\,$=$\,$0\deg$ and $90\deg$ \citep{jow2018}, such that $V$\,$>$\,0 or $V$\,$<$\,0 correspond to preferential orientations parallel or perpendicular to the Galactic plane, respectively.
It is defined as
\begin{equation}\label{eq:myprs}
V = \frac{\sum_{ij}w_{ij}\cos(2\theta_{ij})}{\sqrt{\sum_{ij}w_{ij}/2}},
\end{equation}
where the indices $i$ and $j$ run over the pixel locations in the two spatial dimensions $(l,b)$ for a given velocity channel and $w_{ij}$ is the statistical weight of each angle $\theta_{ij}$.
In our application, we accounted for the spatial correlations introduced by the telescope beam by choosing $w_{ij}$\,$=$\,$(\delta l/\Delta)^{2}$, where $\delta l$ is the pixel size and $\Delta$ is the diameter of the derivative kernel that we selected to calculate the gradients.
\juan{We present the results of the filament orientation trends in terms of the mean orientation angle, $\left<\theta\right>$, and the Rayleigh test for uniformity, $Z$, in App.~\ref{appendix:circstats}.}

\juan{In this paper, we report the results obtained with an 18\parcm0 FWHM two-dimensional Gaussian derivative kernel.
We chose this kernel size to be close to the 16\parcm2 FWHM resolution of the HI4PI survey, which sets the angular scale of the independent measurements with the Hessian matrix.
The selection of a larger kernel reduces the number of independent gradients and thus, the significance of the orientation trends quantified by $V$, as illustrated in App.~\ref{appendix:hessian}.
We chose a kernel slightly larger than the angular resolution of the observations motivated by an initial choice that mitigated the effect of noise in the analysis of the interferometric observations in the THOR survey \citep{soler2020}.
This early selection was carried forward in this work although its results are essentially identical to those found using 16\parcm2 and 17\parcm0 FWHM derivative kernels.}

The values of $V$ \juan{indicate a significant preferred orientation} if there is sufficient clustering around the orientations $\theta$\,$=$\,$0\deg$ and $ 90\deg$.
The null hypothesis implied in $V$ is that the angle distribution is uniform.
In the particular case of independent and uniformly distributed angles, and for a large number of samples, values of $V$\,$\approx$\,$1.64$ and $2.57$ correspond to the rejection of the null hypothesis with a probability of 5\% and 0.5\%, respectively \citep{batschelet1972}.
Thus, a value of $V$\,$\approx$\,2.87 is roughly equivalent to a 3$\sigma$ confidence interval.

\juan{The statistical significance represented by $V$ is determined numerically by taking random samples from a von Mises probability distribution, which is a circular analog of a Gaussian distribution \citep{mardia2009directional}.
It is defined as
\begin{equation}
\mathcal{F}(\theta | \mu,\kappa) = \frac{e^{\kappa\cos(\theta)-\mu}}{2\pi I_{0}(\kappa)},
\end{equation}
where $I_{0}(\kappa)$ is the modified Bessel function of order 0, $\mu$ is the mean of the distribution, and $1/\kappa^{2}$ characterizes the dispersion.
Following a procedure like the one described in \cite{jow2018}, the significance is determined by drawing $N$ samples from a von Mises distribution and computing $V$ for a given null hypothesis rejection probability, for example,  5\% and 0.5\%.
In a similar way to the chi-square test probabilities, $V$ and its corresponding null hypothesis rejection probability are reported in the classical circular statistics literature as tables of ``critical values'', for example in \cite{batschelet1972}, or computed in the circular statistics packages, such as {\tt circstats} in {\tt astropy} \citep{astropy2018}.}

\section{Results}\label{section:results}

\begin{figure*}[ht]
\centerline{
\includegraphics[width=0.495\textwidth,angle=0,origin=c]{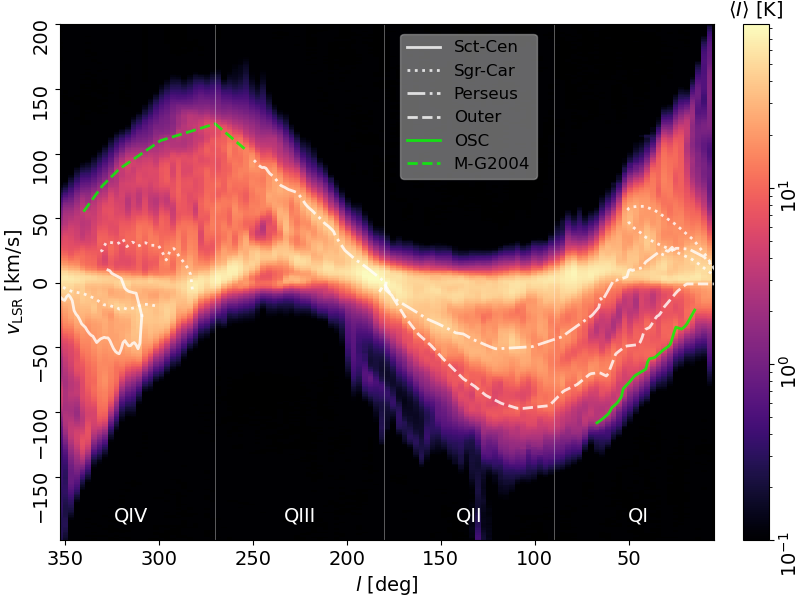}
\includegraphics[width=0.495\textwidth,angle=0,origin=c]{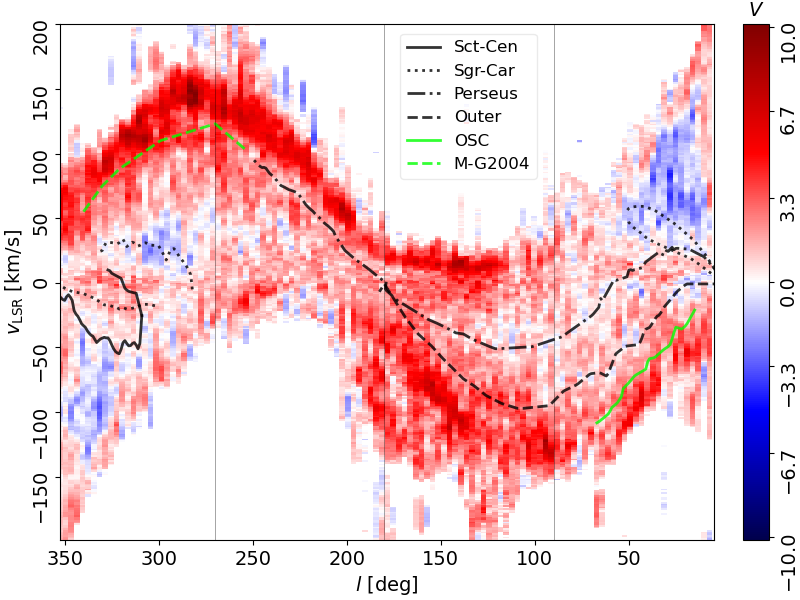}
}
\caption{Longitude-velocity ($lv$) diagrams of the mean intensity ($\left<I\right>$, left) and the filament orientation quantified by the projected Rayleigh statistic ($V$, right).
Each of the pixel elements in the diagrams corresponds to a 3\deg\,$\times$\,20\deg\,$\times$\,1.29\,km\,s$^{-1}$ velocity channel map centered on $b$\,$=$\,0\deg, which we call tile throughout out this paper.
These results correspond to the Hessian analysis performed using an 18\arcmin\ FHWM derivative kernel.
Values of $V$\,$>$\,0 (red) or $V$\,$<$\,$0$ (blue) indicate a preferential orientation of the filaments parallel, $\theta$\,$=$\,$0$\deg, or perpendicular, $\theta$\,$=$\,$90$\deg, to the Galactic plane.
The 3$\sigma$ statistical significance for these two orientations corresponds to $V$\,$>$\,2.87 or $V$\,$<$\,$-2.87$, respectively.
The overlaid curves correspond to the main spiral arms features presented in \cite{reid2016}, the outer Scutum-Centaurus arm \citep[OSC,][]{dameANDthaddeus2011}, and the extended outer arm \citep[M-G2004,][]{mcclure-griffiths2004}.
}
\label{fig:lvdiagrams}
\end{figure*}

The general results of the H{\sc i} filament orientation analysis are presented in Fig.~\ref{fig:lvdiagrams}.
The longitude-velocity ($lv$) diagrams show the values of the mean intensity, $\left<I\right>$, and the filament orientation quantified by the projected Rayleigh statistic, $V$, for each of the 3\deg\,$\times$\,20\deg\,$\times$\,1.29\,km\,s$^{-1}$ tiles around the Galactic plane.
For the sake of reference, we have included lines that indicate the position of the main features that have been traditionally assigned to the Sgr-Car, Sct-Cen, Perseus, and Outer spiral arms, as derived from the trigonometric parallaxes and proper motions for masers associated with young high-mass stars measured in the Bar and Spiral Structure Legacy (BeSSeL) survey \citep{reid2016}.
We also included the OSC arm and the distant arm reported in \cite{mcclure-griffiths2004}.

The $\left<I\right>$ $lv$-diagram, shown on the left-hand side panel of Fig.~\ref{fig:lvdiagrams}, corresponds to the average of the intensity over the area of each tile.
It is presented as a reference for the interpretation of the $V$ $lv$ diagram and not for the particular study of the spiral arm structure.
However, the $\left<I\right>$ $lv$ diagram clearly shows the tracks assigned to the Perseus and Outer arms in the first and second Galactic quadrants. 
Other features, such as the Sct-Cen and the Sgr-Car arms, are less conspicuous.

The right-hand side panel of Fig.~\ref{fig:lvdiagrams} suggests that there is no apparent correlation between the locations of the spiral arms in the $lv$-diagrams and the preferential filament orientations characterized by $V$.
The most prominent features in the $V$ $lv$-diagram, shown on the right-hand side panel of Fig.~\ref{fig:lvdiagrams}, are two groups of filaments parallel to the Galactic plane at the minimum and maximum radial velocities toward the first two (QI and QII) and the last two (QIII and QIV) Galactic quadrants, respectively.
These bands, characterized by $V$\,$>$\,5, extend over a few tens of kilometers per second and appear on radial velocities beyond the $\left<I\right>$ peaks and the locations of the OSC and the \cite{mcclure-griffiths2004} spiral arms.
The high-$V$ band in QIII and QIV appears to extend into QII in the range $0$\,$<$\,$v_{\rm LSR}$\,$<$\,30\,km\,s$^{-1}$ down to $l$\,$\approx$\,110\deg.
Both high-$V$ bands are distinguishable around $l$\,$\approx$180\deg, where they appear to be separated by a gap of around 20\,km\,s$^{-1}$.
The extension in $l$ of these two bands depends on the $b$ coverage of the analysis; that is, they appear truncated or are less prominent when considering ranges $|b|$\,$\leq$\,$1.5$, $3.0$, or $5.0$, as shown in App.~\ref{appendix:method}.

Around 10\% of the H{\sc i} emission over the whole sky is coming from intermediate- and high-velocity clouds (IVCs and HVCs, respectively), which are parcels of gas at velocities incompatible with a simple model of differential Galactic rotation \cite[see][for reviews]{wakkerANDvanWoerden1997,putman2012}.
Most of these structures are located at $|b|$\,$>$\,10\deg.
The majority of the HVCs in the range $|b|$\,$<$\,10\deg\ are too small to dominate the trends reported in this work, for example, the objects in the GALFA-HI Compact Cloud Catalog \citep{saul2012}.
However, there are two exceptions: (i) the complex H \citep{hulsbosch1975}, located around 110\deg\,$<$\,$l$\,$<$\,150\deg, $-15$\deg\,$<$\,$b$\,$<$\,15\deg\ and $-220$\,$<$\,$v_{\rm LSR}$\,$<$\,$-80$\,km\,s$^{-1}$; (ii) the anticenter (AC) complexes \citep{wakker1991} around 155\deg\,$<$\,$l$\,$<$\,190\deg, $-55$\deg\,$<$\,$b$\,$<$\,$-5$\deg\ and $-340$\,$<$\,$v_{\rm LSR}$\,$<$\,$-135$\,km\,s$^{-1}$. 
Figure~\ref{fig:lvdiagrams} indicates that both of these regions are beyond the high-$V$ bands and are composed of mainly horizontal H{\sc i} filaments.
We present the results of the filament analysis for these two HVC complexes in App.~\ref{appendix:HVCs}.

\begin{figure}[ht!]
\centerline{
\includegraphics[width=0.495\textwidth,angle=0,origin=c]{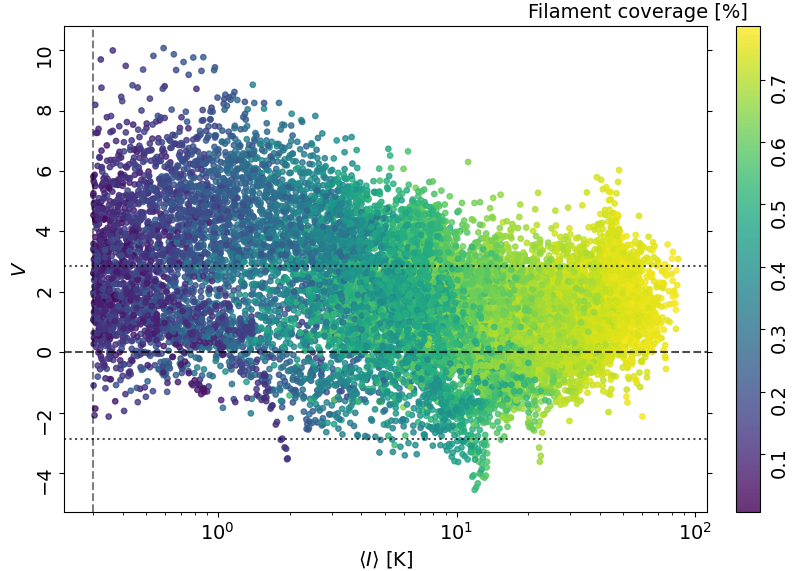}
}
\caption{Mean H{\sc i} emission intensity ($\left<I\right>$) and projected Rayleigh statistic ($V$, Eq.~\ref{eq:myprs}) for the 3\deg\,$\times$\,20\deg\,$\times$\,1.29\,km\,s$^{-1}$ tiles.
The colors indicate the tile percentage covered by the filamentary structures selected following the criteria described in Sec.~\ref{sssec:selection}.
The horizontal \juan{dotted} lines show $V$\,$=$\,$\pm2.87$, which indicate the 3$\sigma$ significance in the preferential orientation parallel, $\theta$\,$=$\,$0$\deg, or perpendicular, $\theta$\,$=$\,$90$\deg, to the Galactic plane, respectively.
\juan{The vertical dashed line indicates the intensity threshold.}
}
\label{fig:PRSvsmeanI}
\end{figure}

Figure~\ref{fig:PRSvsmeanI} indicates that there is no evident correlation between the mean intensity in the tiles, $\left<I\right>$, and the orientation of the filamentary structure, characterized by $V$.
Around 83\% of the tiles show values $V$\,$>$\,0 and roughly 23\% present $V$\,$>$\,2.87.
Only 0.34\% of the tiles display $V$\,$<$\,$-2.87$, which indicates that filamentary structures perpendicular to the Galactic plane are rare.
The tiles with the highest $\left<I\right>$ generally indicate a preference for filaments parallel to the Galactic plane, as shown by $V$\,$>$\,0 in the range $\left<I\right>$\,$\gtrsim$\,60\,K.
However, most of the tiles with a high significance in preferential orientation, $|V|$\,$>$\,2.87, are located in the range $\left<I\right>$\,$\lesssim$\,60\,K, thus suggesting that the overlapping of multiple components in the tiles with the highest $\left<I\right>$ randomizes the orientation of the filamentary structures.

\juan{Figure~\ref{fig:PRSvsmeanI} also shows the filament coverage, which we define as the number of pixels identified as filaments according to the criteria described in Sec.~\ref{sssec:selection} over the total number of pixels in each tile.
We found that the tiles with the} highest $\left<I\right>$ show a \juan{larger} coverage of filamentary structures, which may result from the selection based on the intensity S/N.
Still, both high and low $V$ values appear in tiles with similar filament coverage, which indicates that the preferential orientations are not the product of the number of filaments identified in a tile.
Tiles with low $\left<I\right>$ and filament coverage are more likely to be dominated by a single intensity structure, which explains why they display the most significant values of $V$.

\begin{figure*}[h]%
\centering
\includegraphics[width=0.95\textwidth]{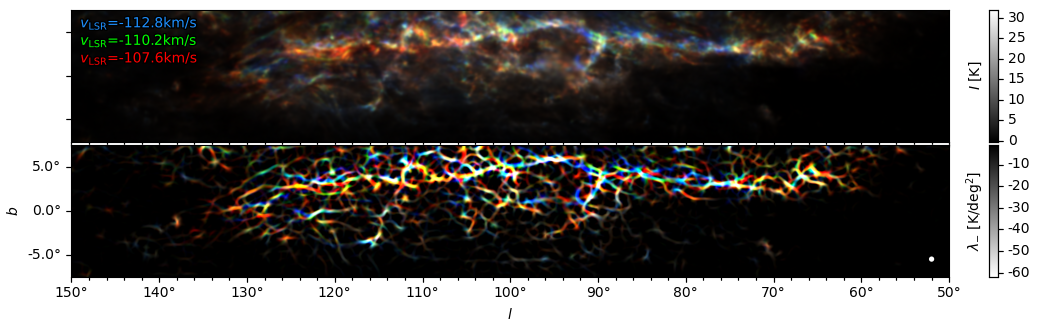}
\includegraphics[width=0.95\textwidth]{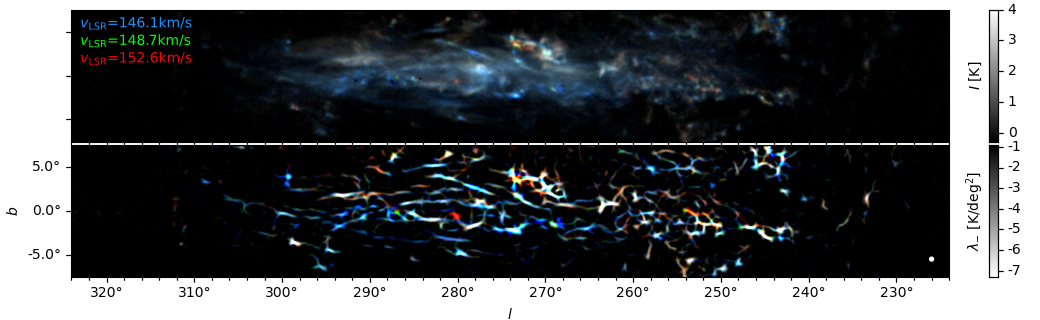}
\caption{Examples of the H{\sc i} emission and filamentary structures toward the $l$ and $v_{\rm LSR}$ ranges with predominantly horizontal filamentary structures.
The top image in each panel corresponds to the H{\sc i} emission intensity in three velocity-channel maps presented in red, green, and blue colors.
The bottom image corresponds to the second eigenvalue of the Hessian matrix, $\lambda_{-}$, which highlights the filamentary structures in each velocity-channel map.
The white disk in the lower right corner corresponds to the size of the derivative kernel used to calculate the Hessian matrix.
}\label{fig:emission}
\end{figure*}

Figure~\ref{fig:emission} presents an example of the filamentary structure in the emission toward the high-$V$ bands reported in Fig.~\ref{fig:lvdiagrams}.
The emission maps show that the filament orientation trends are not dominated by a monolithic object but rather by a collection of filaments where the orientation parallel to the Galactic plane seems to be the most frequent.
The most prominent horizontal structures in the emission maps appear over a background of fainter and seemingly randomly oriented filamentary structures.
In general, the width of the filamentary structures matches the kernel size used to calculate the Hessian matrix.
This fact makes the general definition of the length of the filaments problematic since the apparent overlap of the most extended structures may not be a product of spatial coherence but an artifact of the available angular resolution.

\subsection{Distribution of H{\sc i} filamentary structures in the Galactic plane}\label{subsection:results:filamentorientation}

We analyzed the distribution of H{\sc i} filamentary structures in the Galactic plane by assuming circular motions around the Galactic center and a standard rotation curve, as detailed in App.~\ref{app:deprojection}.
First, we focused on the behavior of the filament orientations with heliocentric radius to study the combined effect of the angular resolution and distance.
Second, we compared the filament orientations at different heliocentric kinematic distances.
Finally, we studied the variations with Galactocentric radius and produced a face-on view of the H{\sc i} filament orientation in the Galactic plane.

\begin{figure}[ht!]
\centerline{\includegraphics[width=0.495\textwidth,angle=0,origin=c]{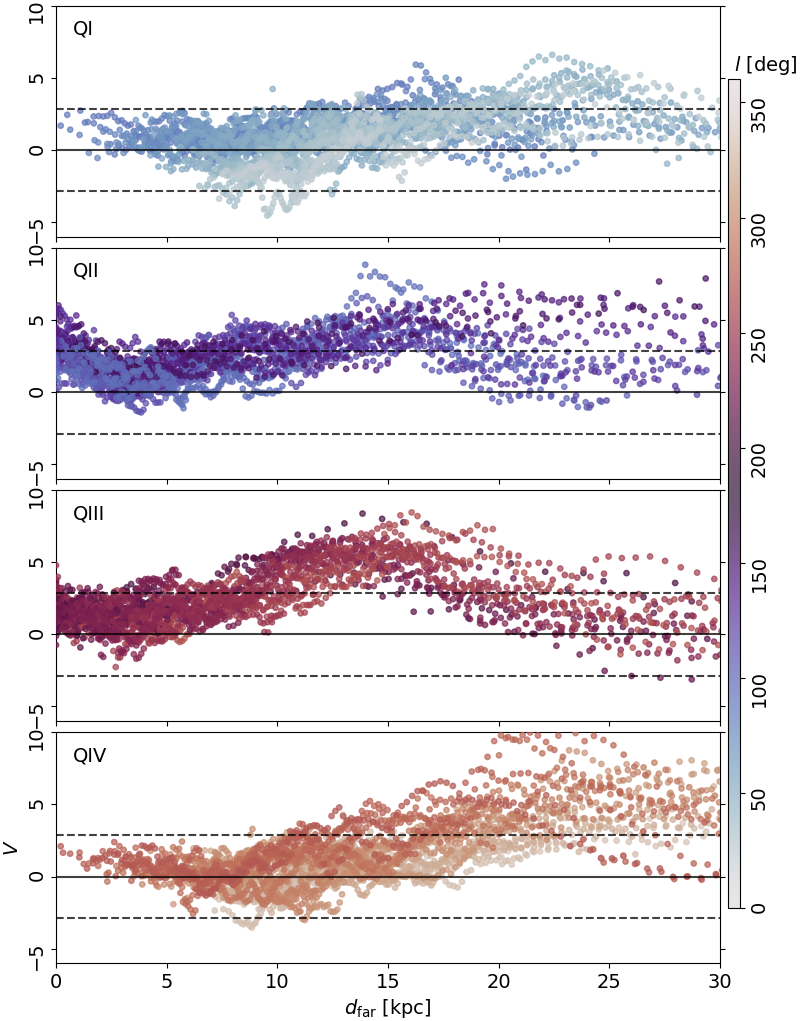}}
\centerline{\includegraphics[width=0.495\textwidth,angle=0,origin=c]{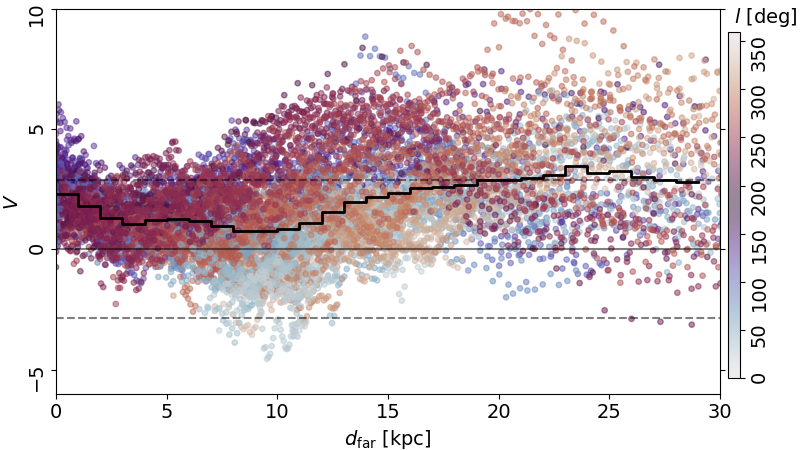}}
\caption{\juan{Kinematic} heliocentric distance ($d$) and projected Rayleigh statistic (V, Eq.~\ref{eq:myprs}) for the 3\deg\,$\times$\,20\deg\,$\times$\,1.29\,km\,s$^{-1}$ tiles.
\juan{We report the largest (far) solution to the heliocentric distance equation, as detailed in App.~\ref{app:deprojection}.}
\juan{From top to bottom, the panels correspond to each Galactic quadrant and to the whole Galactic longitude range.}
\juan{The colors indicate the Galactic longitude of the tiles.}
The horizontal dashed lines indicate $V$\,$=$\,$\pm2.87$, which correspond to the 3$\sigma$ significance in the preferential orientation around $\theta$\,$=$\,$0$\deg\ or $\theta$\,$=$\,$90$\deg, respectively.
\juan{The solid line in the bottom panel corresponds to the mean $V$ values in 1-kpc  $R_{\rm gal}$ bins.}
}
\label{fig:PRSvsDist}
\end{figure}

\subsubsection{Filament orientations and heliocentric distance}


\juan{Figure~\ref{fig:PRSvsDist} shows the distribution of $V$ and the far kinematic heliocentric distance, $d_{\rm far}$, derived for each tile.
We chose to report $d_{\rm far}$ to focus on the effect of distance on $V$, but similar conclusions can be drawn from the near kinematic distances, $d_{\rm near}$, as shown in App.~\ref{app:deprojection}.
We found an increase in $V$ with heliocentric distance that is particularly prominent toward QIII and QIV.
However, we did not find that the tiles corresponding to the largest distances present notably large $V$, thus suggesting that the increase in $V$ is a product of heliocentric distance alone.}

The mean $V$ values in 1-kpc \juan{$R_{\rm gal}$} bins, shown by the black \juan{line in the bottom panel of} Fig.~\ref{fig:PRSvsDist}, reveal a general preference for $V$\,$>$\,$0$.
However, these mean $V$ values are below the $2.87$ limit corresponding to a 3$\sigma$ significance of a preferential orientation parallel to the Galactic plane.
\juan{Most of the scatter in the distribution of $V$ for a particular $R_{\rm gal}$ seems to come from the azimuthal variations, as can be inferred from the comparison between the general trend and that toward each Galactic quadrant.}
\juan{Toward QII and QIII}, we found a systematic increase in $V$ up to $d$\,$\approx$\,15\,kpc followed by a systematic decrease at larger distances.
This may be the product of the decrease in the emission at the edge of the Galactic disk.
\juan{Toward QI and QIV, we found a decrease in $V$ up to $d_{\rm far}$\,$\approx$\,10\,kpc, or $d_{\rm near}$\,$\approx$\,5\,kpc, as illustrated in App.~\ref{app:deprojection}.}

To \juan{further} establish if the prevalence of $V$\,$>$\,0 at large distances is an effect introduced by the distance, we repeated the analysis using larger derivative kernels in App.~\ref{appendix:kernelsize}.
\juan{Using derivative kernels with diameters two and three times larger, which is equivalent to locating the sources two or three times further away, results in a systematic reduction of $V$ rather than a bias toward $V$\,$>$\,0.}
We interpret that result as a strong indication that the relative orientation trends are not exclusively the product of \juan{the heliocentric distance}.
However, we also conducted a test using a derivative kernel fixed to a particular physical scale, which we describe in the following section.

\begin{figure}[ht]
\centerline{
\includegraphics[width=0.375\textwidth,angle=0,origin=c]{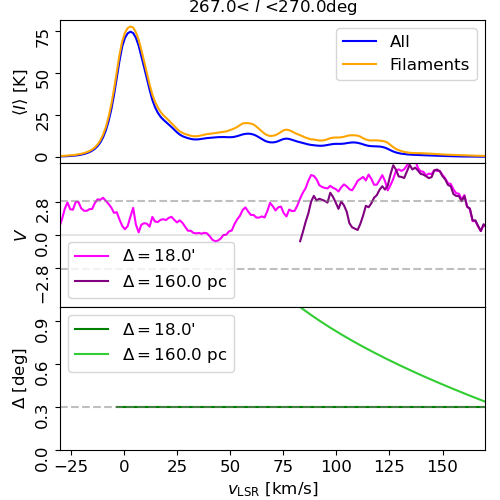}
}
\caption{Mean intensity ($\left<I\right>$, {\it top}), projected Rayleigh statistic ($V$, {\it middle}), and derivative kernel size ($\Delta$, {\it bottom}) for 3\deg\,$\times$\,20\deg\,$\times$\,1.29\,km\,s$^{-1}$ toward the indicated positions.
The two lines in the middle panel indicate the results obtained with a derivative kernel with a fixed angular size and a kernel with a fixed physical size, as indicated on the legend.
The kernel with a fixed physical size is exclusively used in the radial velocity range corresponding to kinematic distances where the angular size of the kernel is below 1\deg.
}\label{fig:MultiKernelTest}
\end{figure}

\subsubsection{Filament orientations near and far}

By fixing the size of the derivative kernel in the Hessian analysis, we are effectively selecting a physical size scale that changes with the distance.
For example, at a distance of approximately 4.5\,kpc the 18\arcmin\ FWHM derivative kernel would select filamentary structures of minimum widths around 25\,pc.
Those structures would be blended into 100-pc-width structures sampled at 18\,kpc with the same beam.

We evaluated the effect of distance by considering a derivative kernel with a fixed physical size.
We ran this test toward the positions $l$\,$\approx$\,90\deg\ and $l$\,$\approx$\,270\deg, where the mapping between $v_{\rm LSR}$ and $d$ has the largest dynamic range.
\juan{We limited} the kernel angular size to $\Delta$\,$<$\,1\deg\ to avoid artifacts \juan{produced by the limited sampling of the 3\deg\,$\times$\,20\deg\,$\times$\,1.29\,km\,s$^{-1}$ tiles}.
\juan{This limit excludes the radial velocity range where most of the emission lies but covers the range in which the tiles with the highest $V$ are found.}
Figure~\ref{fig:MultiKernelTest} shows the comparison of the results obtained with the 18\arcmin\ kernel and a kernel with varying angular size corresponding to $\Delta$\,$=$\,160\,pc at different distances.

We found that the increasing kernel sizes produced lower values of $V$.
This indicates that rather than producing a bias toward higher $V$, as would be the case if multiple structures with a random orientation were blended into a single horizontal filament, the smoothing results in a loss of the preferential orientation parallel to Galactic plane.
This trend toward lower values of $V$ was also found when using derivative kernels with larger angular sizes, as detailed in App.~\ref{appendix:method}.
These results strongly suggest that the high-$V$ bands in the $lv$-diagrams in Fig.~\ref{fig:lvdiagrams} can not be explained by an effect \juan{exclusively} related to \juan{the heliocentric distance}.

Figure~\ref{fig:emission} presents the emission toward the range in $l$ and $v_{\rm lsr}$ corresponding to the high-$V$ bands in Fig.~\ref{fig:lvdiagrams}.
Rather than being concentrated, the emission appears as a series of horizontal filaments distributed in a broad range of Galactic latitudes.
This distribution is the product of the Galactic flaring and indicates that the smoothing by the derivative kernel is not blending the emission into a single monolithic structure at large distances.
Therefore, we conclude that the high values of $V$ are not the result of an observational effect but a property of the H{\sc i} structure.

\subsubsection{Filament orientations and Galactocentric distance}

\begin{figure*}[ht!]
\centerline{
\includegraphics[width=0.49\textwidth,angle=0,origin=c]{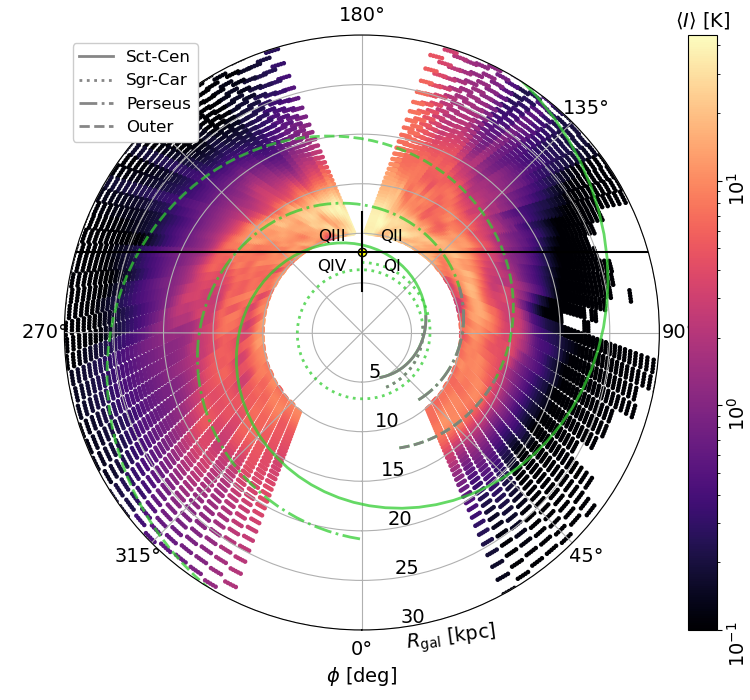}
\includegraphics[width=0.49\textwidth,angle=0,origin=c]{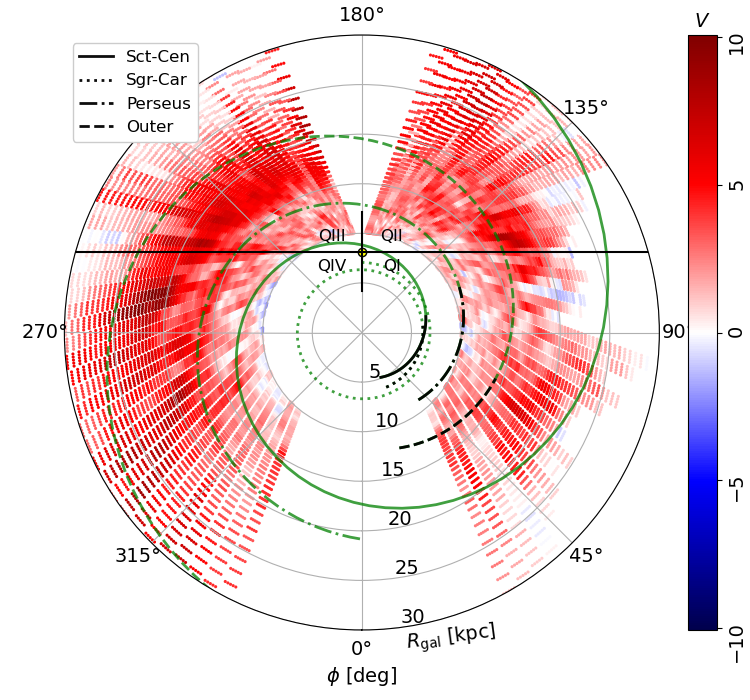}
}
\caption{Face-on view of the mean H{\sc i} emission intensity ($\left<I\right>$; left) and the filament orientation quantified by the projected Rayleigh statistic ($V$; right).
Values of $V$\,$>$\,2.87 or $V$\,$<$\,$-2.87$ indicate a statistical significance above 3$\sigma$ in the preferential orientation around $\theta$\,$=$\,$0$\deg\ or $\theta$\,$=$\,$90$\deg, respectively.
The yellow circular marker indicates the position of the Sun.
The overlaid curves correspond to the spiral arm models identified in \cite{reid2019}, shown in black color, and their extrapolation to larger Galactocentric radii, shown in green.
}
\label{fig:deprojections}
\end{figure*}

We computed a de-projection of the $\left<I\right>$ and $V$ $lv$-diagrams into a face-on view of the Milky Way, shown in Fig.~\ref{fig:deprojections}.
This reconstruction does not aim to produce a complete 3D density distribution but rather identify the Galactocentric radii and azimuth to which the features in the $V$ $lv$-diagrams belong.
Following preceding works, such as \cite{levine2006warp} and \cite{kalberla2007}, we excluded from the analysis the Galactic longitude ranges $l$\,$<$\,15\deg, $l$\,$>$\,345\deg\, and 165\deg\,$<$\,$l$\,$<$\,195\deg, as these regions have radial velocities that are too small with respect to their random velocities to establish reliable distances.
We also excluded from this reconstruction points with $R_{\rm gal}$\,$<$\,10\,kpc, to avoid the complications introduced by the near-far distance ambiguity and the artifacts produced by the emission around $v_{\rm LSR}$\,$\approx$\,0\,km\,s$^{-1}$.

The face-on reconstruction of the mean intensity, shown on the left-hand-side panel of Fig.~\ref{fig:deprojections}, displays some of the spiral arm features identified in the LAB survey around 90\deg\,$\lesssim$\,$\phi$\,$\lesssim$150\deg\ and 200\deg\,$\lesssim$\,$\phi$\,$\lesssim$\,250\deg\ around 10\,$<$\,$R_{\rm gal}$\,$<$\,15\,kpc \citep{levine2006,nakanishi2016,koo2017}.
The identification of these arms is not the main focus of our analysis, but they appear highlighted by the emission concentration. 
The arms in QI and QIV are less conspicuous and not clearly identifiable without additional processing, such as, the median filtering applied in \cite{levine2006} or the spectral peak identification used in \cite{koo2017}.

\begin{figure}[ht!]
\centerline{\includegraphics[width=0.495\textwidth,angle=0,origin=c]{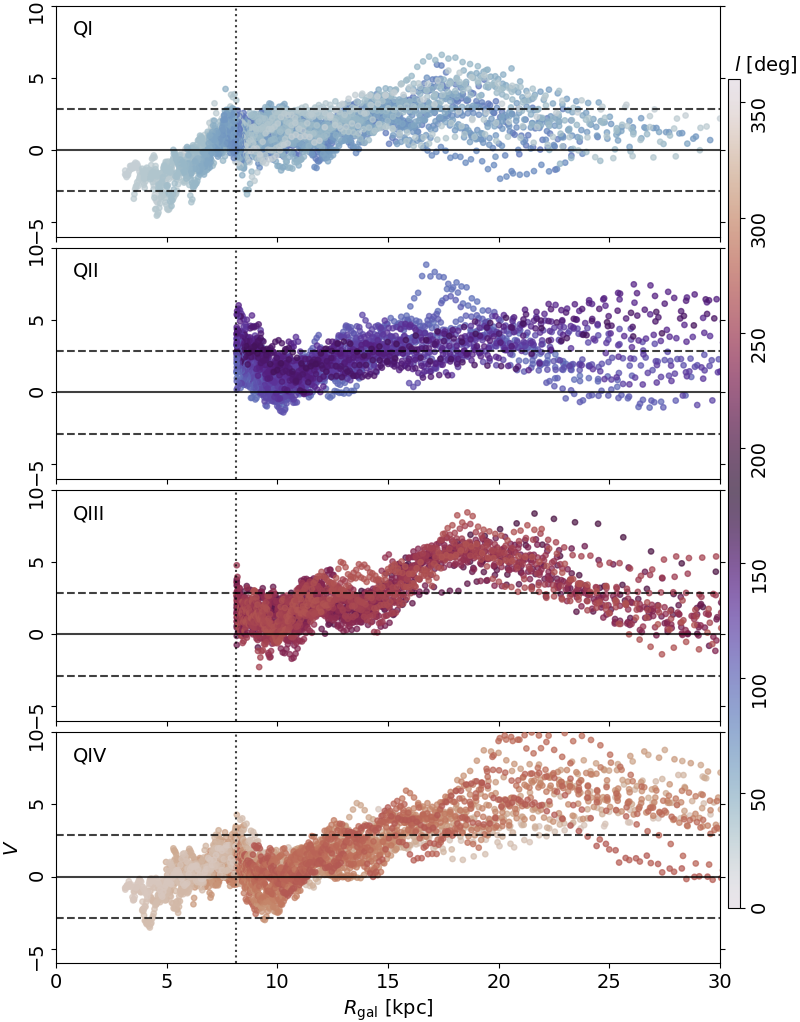}}
\centerline{\includegraphics[width=0.495\textwidth,angle=0,origin=c]{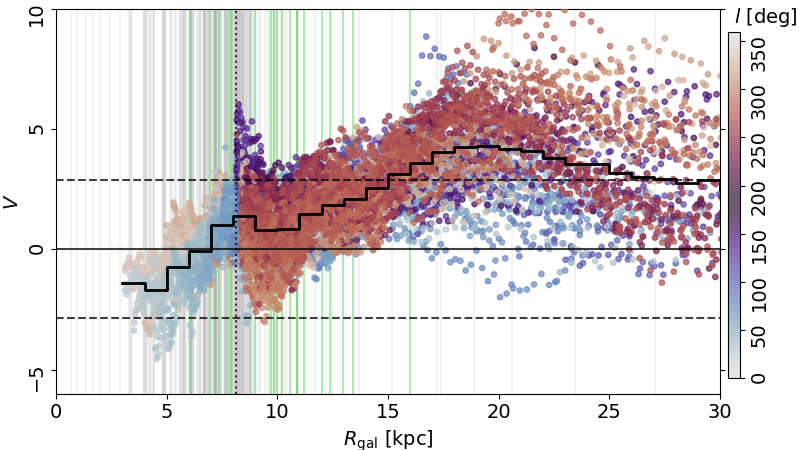}}
\caption{Galactocentric radius ($R_{\rm gal}$) and projected Rayleigh statistic ($V$, Eq.~\ref{eq:myprs}) for the 3\deg\,$\times$\,20\deg\,$\times$\,1.29\,km\,s$^{-1}$ tiles.
\juan{From top to bottom, the panels correspond to each Galactic quadrant and to the whole Galactic longitude range.}
\juan{The colors indicate the Galactic longitude of the tiles.}
The horizontal dashed lines indicate $V$\,$=$\,$\pm2.87$, which indicate the 3$\sigma$ significance in the preferential orientation around $\theta$\,$=$\,$0$\deg\ or $\theta$\,$=$\,$90$\deg, respectively.
The vertical dotted line indicates the radius of the solar orbit.
\juan{The solid line in the bottom panel corresponds to the mean $V$ values in 1-kpc  $R_{\rm gal}$ bins.}
The vertical solid lines mark the central locations of the H{\sc i} shells identified in the LAB survey \citep{ehlerova2013} and the SGPS \citep{mcclure-griffiths2002} in the range $|b|$\,$<$\,10\deg, shown in gray and green colors, respectively.
}
\label{fig:PRSvsRgal}
\end{figure}

The face-on reconstruction of the filament orientation trends, presented on the right-hand-side panel of Fig.~\ref{fig:deprojections}, shows a clustering of high $V$ at $R_{\rm gal}$\,$\gtrsim$\,15\,kpc on both sides of the Galaxy.
This collection of high-$V$ structures, which corresponds to the bands in the $lv$-diagrams in Fig.~\ref{fig:lvdiagrams}, does not appear simply as a ring but seems to have a pitch angle.
This pitch angle impression is produced by the displacement of the peak $V$ values from $R_{\rm gal}$\,$\approx$\,18\,kpc at $\phi$\,$\approx$\,210\deg\ to $R_{\rm gal}$\,$\approx$\,23\,kpc at $\phi$\,$\approx$\,270\deg.
The pitch angle is less noticeable in the other side of the Galaxy, where there is a prominent exception to the general $V$\,$>$\,0 trend around $R_{\rm gal}$\,$\approx$\,17\,kpc and $l$\,$\approx$\,95\deg.
This feature, which corresponds to the negative $V$ tiles at $d$\,$\gtrsim$\,25\,kpc in Fig.~\ref{fig:PRSvsRgal}, was also identified in the Galactic warp and flare studies reported in \cite{levine2006}. 

Figure~\ref{fig:deprojections} also shows the asymmetry in the H{\sc i} between the first two and the last two quadrants.
Toward QI and QII there is a drop in the amount of emission at $R_{\rm gal}$\,$\approx$\,25\,kpc, which is noticeable in the lack of filaments and relative orientation trends beyond that radius in the range 45\,$\lesssim$\,$\phi$\,$\lesssim$\,130\deg.
However, there is enough emission to suggest a drop in $V$ at $R_{\rm gal}$\,$\approx$\,25\,kpc with respect to its peak value, which we explored further by computing radial profiles.

Figure~\ref{fig:PRSvsRgal} shows the distribution of $V$ with respect to the $R_{\rm gal}$ derived from the radial velocity of each tile and the average radial profile computed in 1-kpc-width bins in the range 2\,$\leq$\,$R_{\rm gal}$\,$\leq$\,30\,kpc.
We included the observations at $R_{\rm gal}$\,$<$\,10\,kpc  since the $R_{\rm gal}$ estimates are not affected by the heliocentric distance ambiguity.
The general trend is a progressive increase of $V$ with growing $R_{\rm gal}$ until it reaches its maximum values around $R_{\rm gal}$\,$\gtrsim$\,20\,kpc.
This tendency is equivalent to the continuous transition from H{\sc i} filaments slightly perpendicular or having no preferred orientation to mostly parallel to the Galactic plane with increasing distance from the Galactic center.
The significance of the relative orientation in the tiles dominated by mostly perpendicular filaments increases when considering broader tiles in Galactic longitude, as illustrated in App.~\ref{appendix:tilesize}.
However, \juan{the broader} tile size reduces the sampling in Galactic longitude and the resolution in the face-on reconstructions presented in Fig.~\ref{fig:deprojections}.

Figure~\ref{fig:PRSvsRgal} also shows the central locations of the H{\sc i} shells in the catalogs produced with the LAB survey \citep{ehlerova2013} and the SGPS observations \citep{mcclure-griffiths2002} in the range $|b|$\,$<$\,10\deg.
Most of the H{\sc i} shells are found within the range $R_{\rm gal}$\,$<$\,10\,kpc, and their number tends to decrease with increasing Galactocentric radius.
For $R_{\rm gal}$\,$>$\,15\,kpc, where the values of $V$ reveal a very significant prevalence of H{\sc i} filaments parallel to the Galactic plane, there are very few identified H{\sc i} shells.
The likely relationship between the values of $V$ and the density of H{\sc i} shells with $R_{\sc i}$ suggest a connection between the H{\sc i} filament orientation and SN feedback, which we discuss further in Sec.~\ref{section:discussion}.

In the vicinity of the solar circle, $R_{\rm gal}$\,$=$\,8.15\,kpc \citep[][]{reid2019}, we found both negative and positive values of $V$ that show a significant departure from the general trend.
Some of these fluctuations are due to H{\sc i} emission in the solar neighborhood, whose proximity makes the local atomic gas dynamics more prominent than the large scale trend.
The peak in negative $V$ around $R_{\rm gal}$\,$\approx$\,10\,kpc is most likely associated with the large H{\sc i} shells identified in the range 250\deg\,$\lesssim$\,$l$\,$\lesssim$\,360\deg\ in radial velocities that roughly correspond to that position in the Galactic plane \citep{mcclure-griffiths2002}.
Additional fluctuations in $V$ are also introduced by noncircular motions, which are particularly dominant around the anticenter, thus explaining why the largest $V$ are observed around $l$\,$\approx$\,160\deg.
It is likely that the emission around that Galactic longitude is produced further away and that the distance estimates based on the line kinematics are \juan{inaccurate \citep[see, for example,][]{burton1971,peek2022}}.

At $R_{\rm gal}$\,$\gtrsim$\,16\,kpc the dispersion in $V$ increases and displays relatively large values of $V$ and an excursion into $V$\,$<$\,0 for some of the tiles.
The points corresponding to $V$\,$<$\,0 around $R_{\rm gal}$\,$\approx$\,20\,kpc can be traced into an individual feature around $\phi$\,$\approx$\,100\deg\ that is also found as an outlier in the Galactic warp \citep{levine2006warp}.
Most of the tiles with high values of $V$ at $R_{\rm gal}$\,$\gtrsim$\,20\,kpc are found in QIV, where the high $V$ band in the $lv$-diagram is the broadest, but for most of the lines of sight $V$ decreases beyond $R_{\rm gal}$\,$\gtrsim$\,20\,kpc.

\subsection{Scale height of the H{\sc i} filaments}\label{subsection:results:warpANDflare}

In addition to the analysis of the filament orientation distribution with $R_{\rm gal}$, we considered two other H{\sc i} emission properties that also change with the distance from the Galactic center: the midplane height and the dispersion around the midplane.
We computed the Galactic height $z$ for each of the tiles and obtained its first and second intensity-weighted moments, $\left<z\right>$ and $\sigma_{z}$ following the procedure detailed in App.~\ref{app:deprojection}.
We estimated $\left<z\right>$ and $\sigma_{z}$ both for all of the H{\sc i} emission in each tile and exclusively for the emission in the structures classified as filaments; hence we obtained two separate estimates of the warp and the flare.

\begin{figure*}[h]%
\centering{
\includegraphics[width=0.49\textwidth]{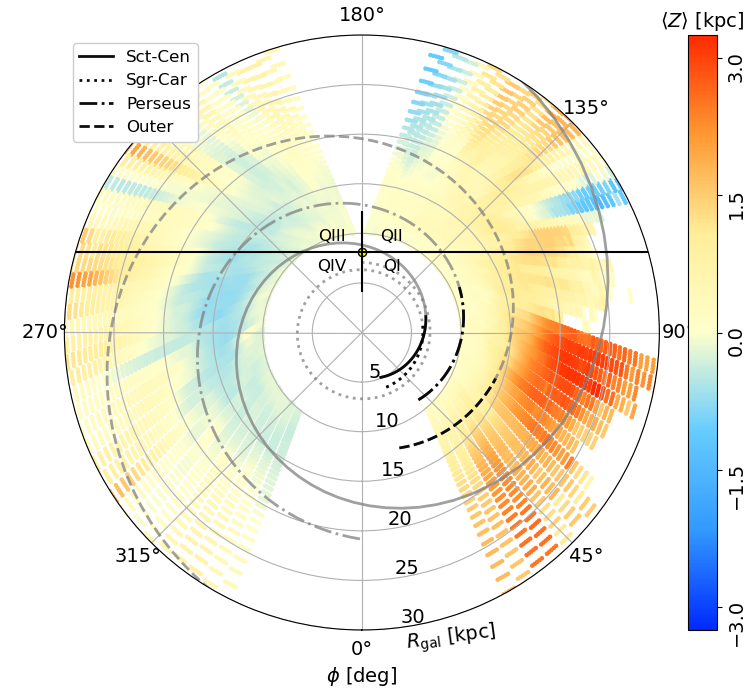}
\includegraphics[width=0.49\textwidth]{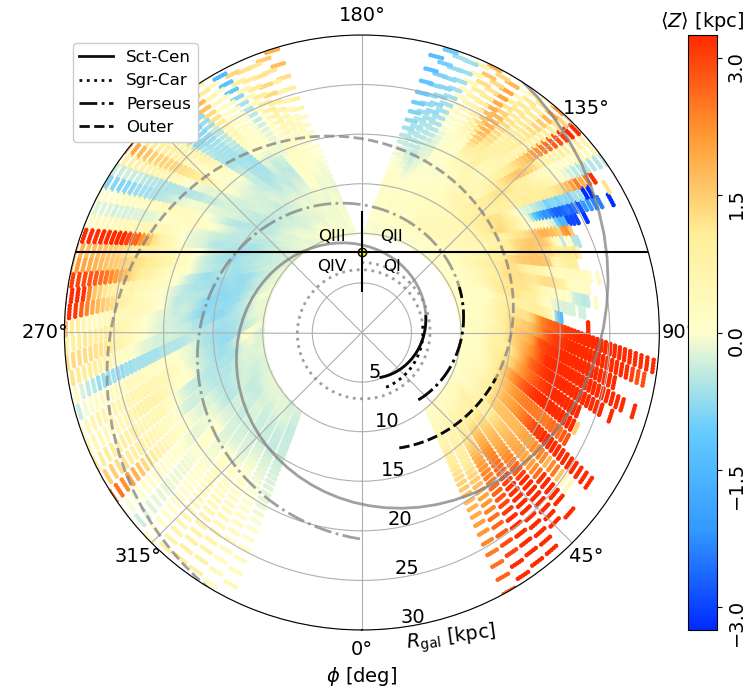}
}
\caption{Face-on view of the H{\sc i} midplane height ($\left<z\right>$) reconstructed from all the emission in the range $|b|$\,$<$\,10\deg\ (left) and the emission in filamentary structures (right).
}\label{fig:scaleheight}
\end{figure*}

Figure~\ref{fig:scaleheight} shows the two face-on reconstructions of $\left<z\right>$.
The midplane height map obtained using all of the H{\sc i} emission reproduces some of the main features obtained in the previous analysis of lower resolution data \citep{levine2006warp,kalberla2007}.
First, the warp of the Milky Way, which corresponds to the variation in midplane height from $\left<z\right>$\,$\approx$\,4\,kpc toward QI and QII to $\left<z\right>$\,$\approx$\,$-1$\,kpc at $R_{\rm gal}$\,$\lesssim$\,20\,kpc toward QIII and QIV.
Second, the finding of $\left<z\right>$\,$>$\,0 and $\left<z\right>$\,$<$\,0 for adjacent tiles toward QIII and QIV at $R_{\rm gal}$\,$\gtrsim$\,20\,kpc, which has been identified as Galactic disk ``scalloping'' \citep{kulkarni1982,levine2006warp}.
Finally, a sharp dip around $\phi$\,$\approx$\,90\deg\ and $R_{\rm gal}$\,$\gtrsim$\,20\,kpc.

The $\left<z\right>$ found with the H{\sc i} filaments shows features that are very similar to the midplane heights obtained using the bulk of the H{\sc i} emission.
However, the contrast in the warp, the scalloping, and the sharp dip is much higher.
This result is most likely a consequence of filtering the diffuse emission implicit in the filament identification.

\begin{figure*}[h]
\centering{
\includegraphics[width=0.49\textwidth]{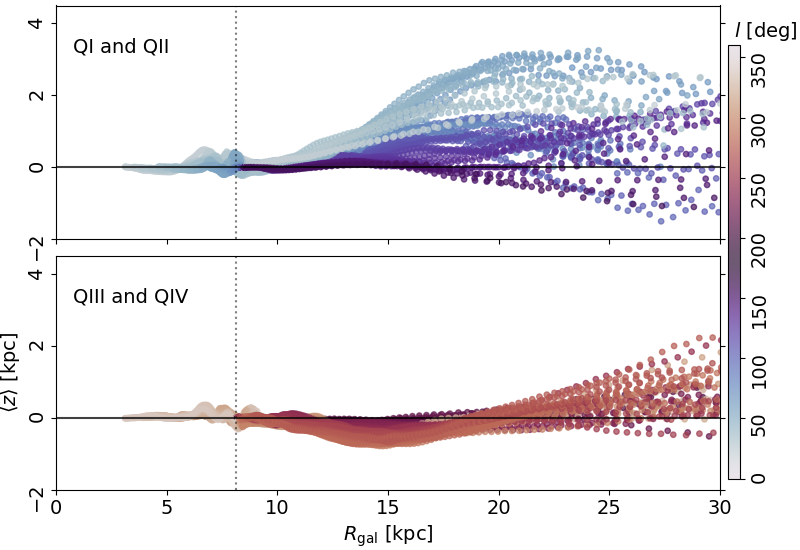}
\includegraphics[width=0.49\textwidth]{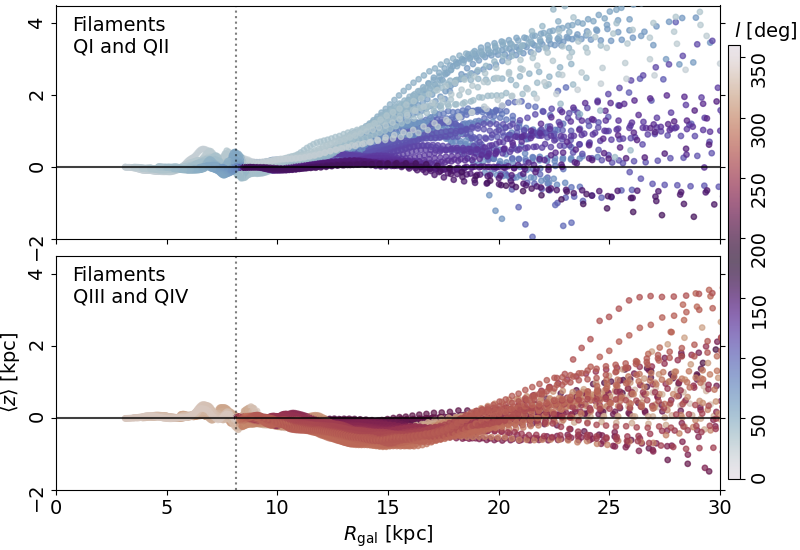}
}
\caption{
Midplane height ($\left< z\right>$) and Galactocentric radius ($R_{\rm gal}$) for the 3\deg\,$\times$\,20\deg\,$\times$\,1.29\,km\,s$^{-1}$ tiles.
The left-hand side panel corresponds to all the H{\sc i} emission in the $|b|$\,$<$\,10\deg\ range.
The right-hand side panel corresponds to the emission in the structures classified as filaments.
The colors indicate the Galactic longitude, $l$.
The vertical dotted line indicates the radius of the solar orbit around the Galactic center.
}\label{fig:profilesMeanZ}
\end{figure*}

The $\left<z\right>$ radial profiles, presented in Fig.~\ref{fig:profilesMeanZ}, are very similar for all the H{\sc i} emission and for the H{\sc i} filaments in the range $R_{\rm gal}$\,$\lesssim$\,20\,kpc.
For larger $R_{\rm gal}$, the H{\sc i} filaments reveal that the Galactic warp extends up to approximately 4\,kpc for tiles in the first Galactic quadrant.
The H{\sc i} filaments also show higher values of $\left<z\right>$ at $R_{\rm gal}$\,$\gtrsim$\,20\,kpc toward QIII and QIV, which strongly suggest that specific filamentary structures produce the apparent disk scalloping.


\begin{figure*}[h]
\centering{
\includegraphics[width=0.49\textwidth]{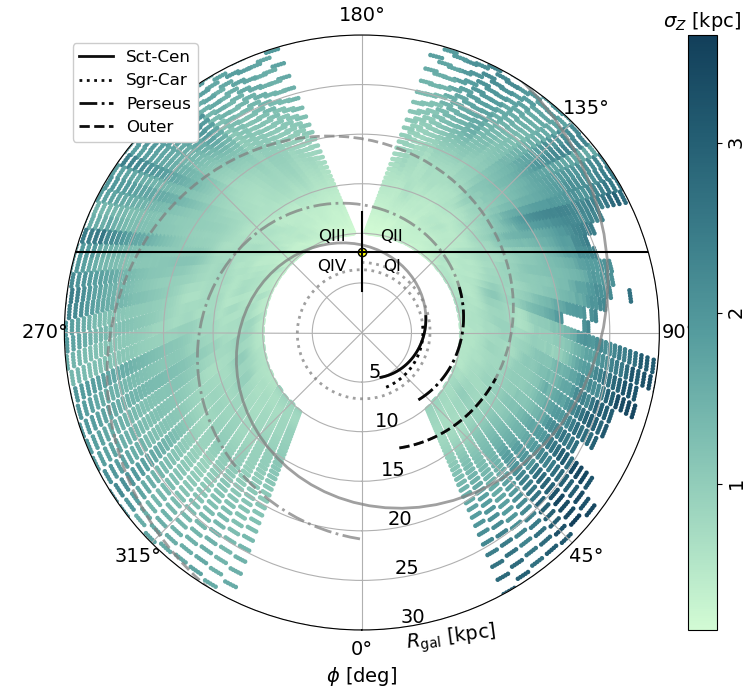}
\includegraphics[width=0.49\textwidth]{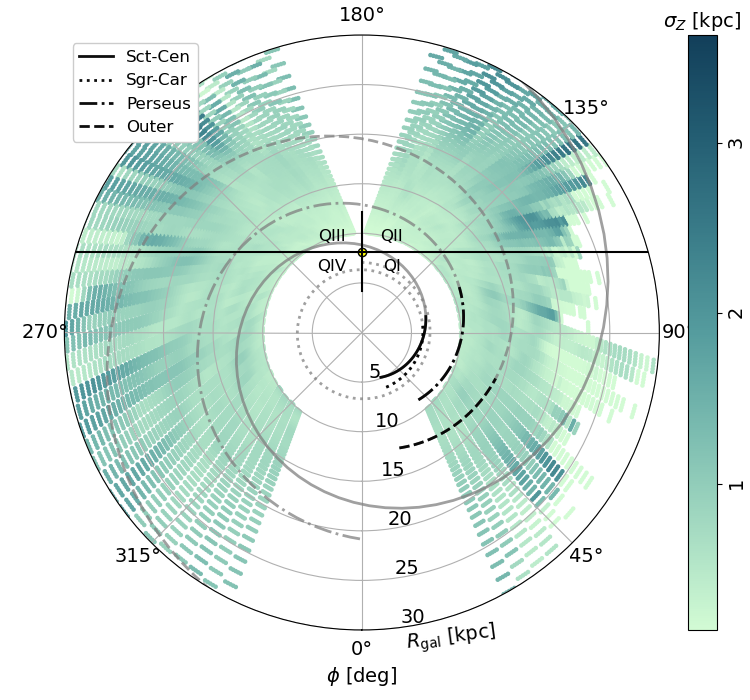}}
\caption{Face-on view of the dispersion around the midplane height ($\sigma_z$) reconstructed from all the emission in the range $|b|$\,$<$\,10\deg\ (left) and the emission in filamentary structures (right).
}\label{fig:flaring}
\end{figure*}

\begin{figure*}[h]%
\centering{
\includegraphics[width=0.49\textwidth]{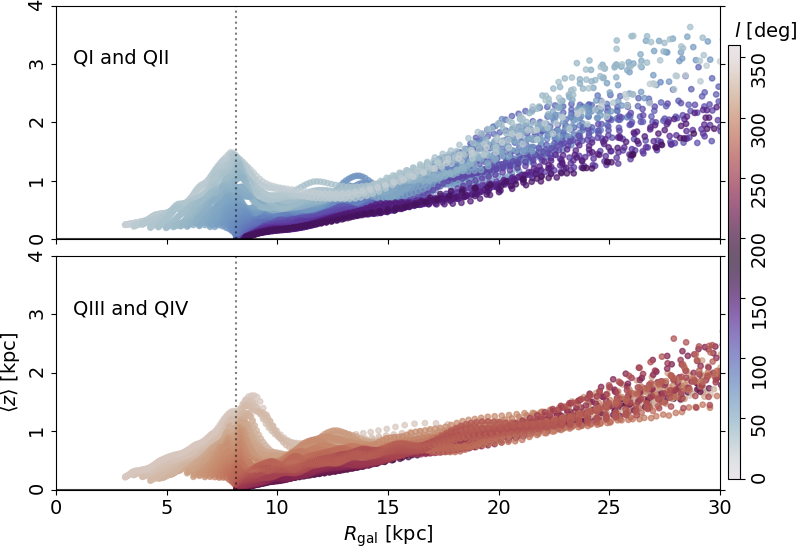}
\includegraphics[width=0.49\textwidth]{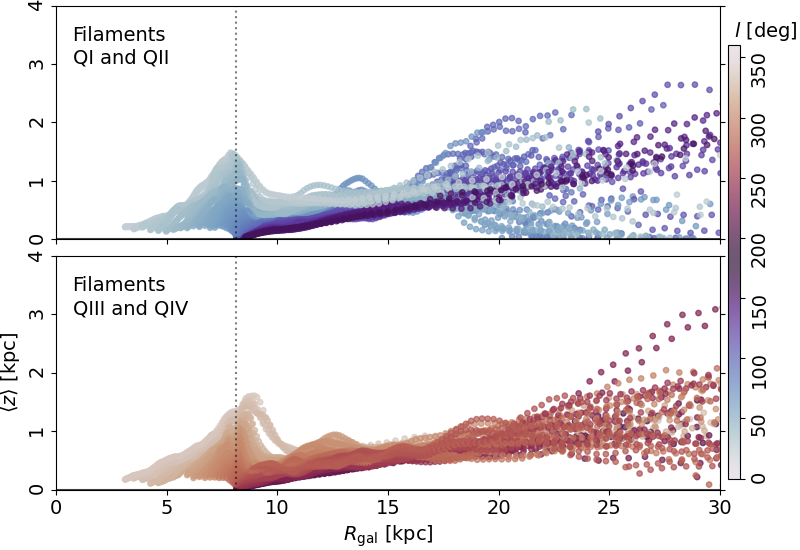}
}
\caption{Same as Fig.~\ref{fig:profilesMeanZ}, but for the dispersion around the midplane height ($\sigma_z$).
}\label{fig:profilesSigmaZ}
\end{figure*}

Figure~\ref{fig:flaring} shows the two face-on reconstructions of $\sigma_z$.
The dispersion of the H{\sc i} emission around the midplane shows the flaring of the Galactic disk at $R_{\rm gal}$\,$>$\,15\,kpc, both in all the emission and just in the filaments.
However, the radial profiles shown in Fig.~\ref{fig:profilesSigmaZ} indicate that the average trend is for the filamentary structures to have a lower \juan{dispersion than} the bulk of the H{\sc i} emission by about a factor of two at $R_{\rm gal}$\,$>$\,15\,kpc.
This general trend is followed in most of the $l$ range, particularly around 250\deg\,$<$\,$l$\,$<$\,350\deg, but there are clear exceptions where the $\sigma_z$ values traced by the filaments drop abruptly.
This can be attributed to the lower emission in radial velocities corresponding to larger distances toward \juan{QI, which makes difficult the filament identification at $R_{\rm gal}$\,$\gtrsim$\,25\,kpc in that azimuth range.}


\subsection{Global properties of the H{\sc i} filaments}\label{subsection:results:properties}

The results of the Hessian analysis provide a description of the H{\sc i} emission morphology across velocity channels, which does not imply that the highlighted structures are filaments or sheets in three-dimensional space.
Our finding is that the high-aspect-ratio features in the H{\sc i} emission have a preferential orientation with respect to the Galactic plane.
Without any loss of generality, we can also estimate what fraction of the H{\sc i} column density is in these high-aspect-ratio features.

We calculated the H{\sc i} column density in each tile using the assumption of optically thin emission \cite[see, for example,][]{kulkarni1987}, such that,
\begin{equation}
N_{\rm H} = 1.83\times10^{18} \left(\frac{T_{b}}{\mbox{K}}\right)\left(\frac{\Delta v}{\mbox{km s}^{-1}}\right)\mbox{cm}^{-2},
\end{equation}
where $T_{b}$ is the \juan{brightness temperature} and $\Delta v$ is the width of the velocity channel.
This assumption reduces our calculation to the ratio between the sum of $T_{b}$ in the filamentary structures and the total $T_{b}$ in each tile.
We note, however, that the optically-thin assumption underestimates the amount of H{\sc i} in the Galactic plane \citep{bihr2015,wang2020hi}.
The optical depth mainly affects the inferred CNM column density, since typically $\tau$\,$\ll$\,1 for the WNM \citep[see, for example,][]{heilesANDtroland2003}.
Insofar as the ratio of CNM to WNM in the filaments may be different than in the total gas, our results for $N^{\rm fil}_{\rm H}/N_{\rm H}$ are only indicative of the global trend.

\begin{figure}[ht]
\centering{
\includegraphics[width=0.49\textwidth]{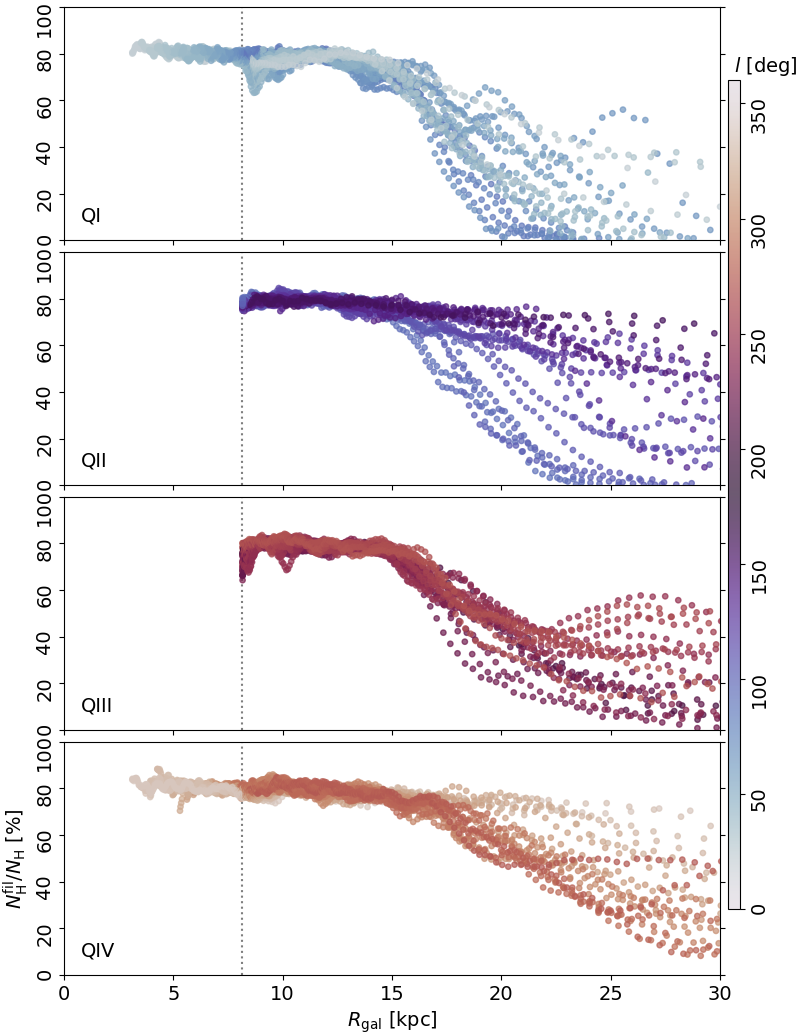}
\includegraphics[width=0.49\textwidth]{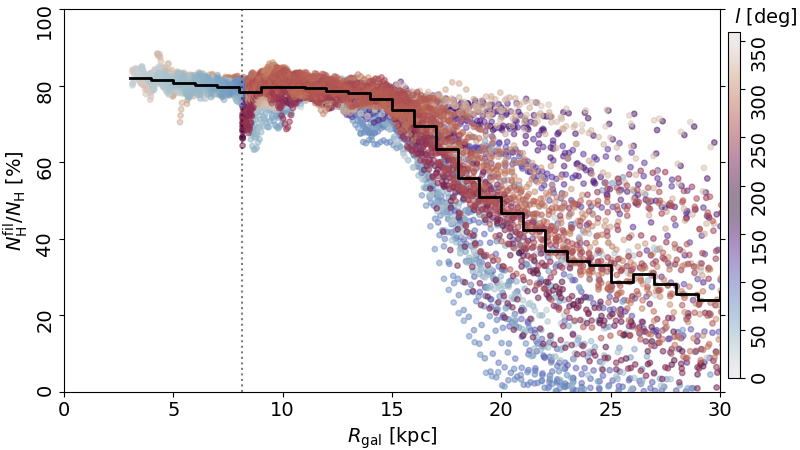}
}
\caption{Radial profiles of the ratio between the column density in the filamentary structures ($N^{\rm fil}_{\rm H}$) and the total column density ($N_{\rm H}$).
\juan{From top to bottom, the panels correspond to each Galactic quadrant and to the whole Galactic longitude range.}
\juan{The colors indicate the Galactic longitude of the tiles.}
The vertical dotted line indicates the radius of the solar orbit.
\juan{The solid line in the bottom panel indicates the mean values of $N^{\rm fil}_{\rm H}/N_{\rm H}$ in 1-kpc $R_{\rm gal}$ bins.}
}
\label{fig:NHfracVsRgal}
\end{figure}

Figure~\ref{fig:NHfracVsRgal} shows the distribution of $N^{\rm fil}_{\rm H}/N_{\rm H}$ as a function of distance from the Galactic center.
The $N^{\rm fil}_{\rm H}/N_{\rm H}$ radial profile is tightly distributed around 80\% for $R_{\rm gal}$\,$\lesssim$\,18\,kpc.
For larger $R_{\rm gal}$, the mean value of $N^{\rm fil}_{\rm H}/N_{\rm H}$ decreases sharply, although we found a large dispersion around this general trend.
The general decrease in $N^{\rm fil}_{\rm H}/N_{\rm H}$ for $R_{\rm gal}$\,$\lesssim$\,18\,kpc is also found when fixing the $I$ S/N threshold using larger derivative kernels, as illustrated in App.~\ref{appendix:kernelsize}.
Thus, it is most likely that this trend represents a transition toward less structured H{\sc i} emission rather than a drop due to the decrease in the emission.

The azimuthal variations in the $N^{\rm fil}_{\rm H}/N_{\rm H}$ radial profile are most likely a product of the asymmetry in the Galactic disk.
In the range $l$\,$\lesssim$\,90\deg\ we found a sharp decrease in $N^{\rm fil}_{\rm H}/N_{\rm H}$ that corresponds to the drop in emission toward that particular direction.
In contrast, we obtained $N^{\rm fil}_{\rm H}/N_{\rm H}$ around 70\% up to $R_{\rm gal}$\,$\approx$\,30\,kpc for some tiles around $l$\,$\approx$\,150\deg.

The sharp drop in the mean values of $N^{\rm fil}_{\rm H}/N_{\rm H}$ with Galactocentric radius roughly coincides with the transition from mostly horizontal to randomly oriented H{\sc i} filaments at $R_{\rm gal}$\,$\approx$\,18\,kpc, illustrated in Fig.~\ref{fig:PRSvsRgal}.
Likely, the large dispersion in $V$ found at $R_{\rm gal}$\,$\gtrsim$\,18\,kpc is the product of a reduction in the number of filamentary structures, as sampled by $N^{\rm fil}_{\rm H}$, and the azimuthal variations in the emission.

\subsection{Comparison with GALFA-HI}\label{subsection:results:observationalEffects}

We have considered the orientation of the H{\sc i} filaments found with the Hessian method fixing a particular angular scale through the selection of the derivative kernel size and in velocity channels set by the spectral resolution of the HI4PI survey.
There is no survey covering the same $b$ range across the whole Galactic plane. 
Still, the 4\arcmin\ and 0.184\,km\,s$^{-1}$ resolution GALFA-HI provides an insight into the effects of the angular and velocity resolutions, which we discuss in what follows.

\begin{table}
\caption{Hessian analysis parameters for the GALFA-HI data.}              
\label{table:HessianParametersGALFAHI}      
\centering                                      
\begin{tabular}{l l l}          
\hline\hline                        
Parameter & & Value \\    
\hline                                   
Tile size & & 3\deg\,$\times$\,20\deg\,$\times$\,0.184\,km\,s$^{-1}$ \\ 
              & & 3\deg\,$\times$\,20\deg\,$\times$\,1.290\,km\,s$^{-1}$ \\      
Kernel size & & 5\arcmin\ FWHM \\
Intensity threshold & & 0.3\,K \\
Curvature threshold ($\lambda^{C}_{-}$) & & $-90.0$\,K/deg$^{2}$ \\
\hline                                             
\end{tabular}
\end{table}

\subsubsection{Effect of the angular resolution}

We performed the Hessian analysis in the two portions of the first and the third Galactic quadrants covered in the 4\arcmin-resolution GALFA-HI data.
To independently study the effects of the difference in angular and spectral resolution, we projected the GALFA-HI observations into the same spectral axis of the HI4PI data. 
We performed the Hessian analysis using an 5\arcmin\ FWHM derivative kernel \juan{and applied the selection criteria presented in Table~\ref{table:HessianParametersGALFAHI}.}
\juan{Further details on the implementation of the Hessian matrix method  in the GALFA-HI data and the results obtained with a larger kernel are presented in App.~\ref{appendix:HI4PIandGALFAHI}.}

\begin{figure*}[ht!]
\centerline{
\includegraphics[width=0.4\textwidth,angle=0,origin=c]{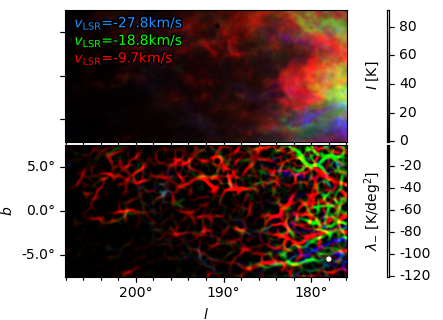}
\includegraphics[width=0.4\textwidth,angle=0,origin=c]{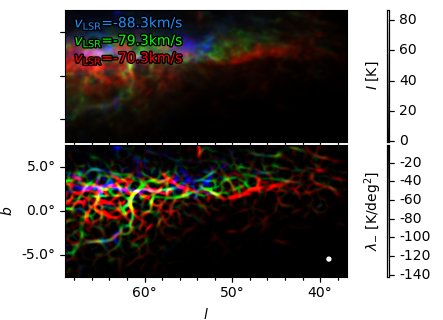}
}
\centerline{
\includegraphics[width=0.4\textwidth,angle=0,origin=c]{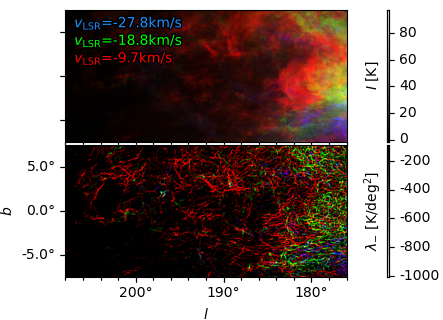}
\includegraphics[width=0.4\textwidth,angle=0,origin=c]{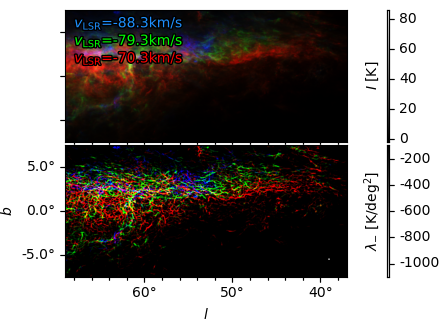}
}
\caption{\juan{Examples of the H{\sc i} emission and filamentary structures identified in the HI4PI and the GALFA-HI data, shown in the top and bottom panels respectively.
The disk in the lower right corner of each panel indicates the size of the derivative kernels used for the calculation of the Hessian matrix, 18\arcmin\ and 5\arcmin\ for the HI4PI and the GALFA-HI observations, respectively.}}
\label{fig:HI4PIandGALFArgb}
\end{figure*}

\begin{figure*}[ht]
\centerline{
\includegraphics[width=0.48\textwidth,angle=0,origin=c]{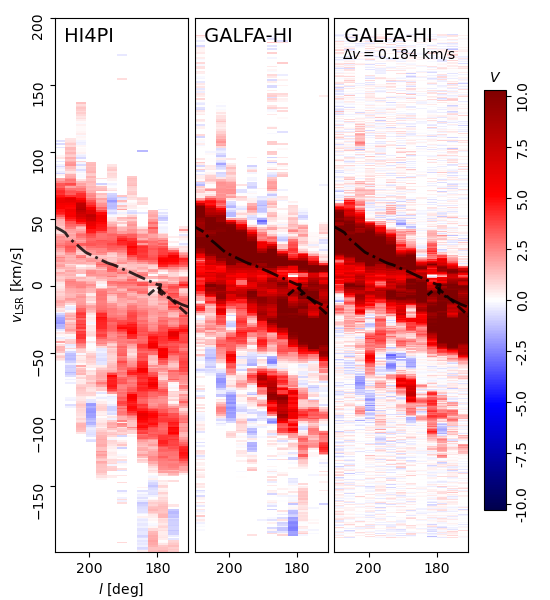}
\includegraphics[width=0.48\textwidth,angle=0,origin=c]{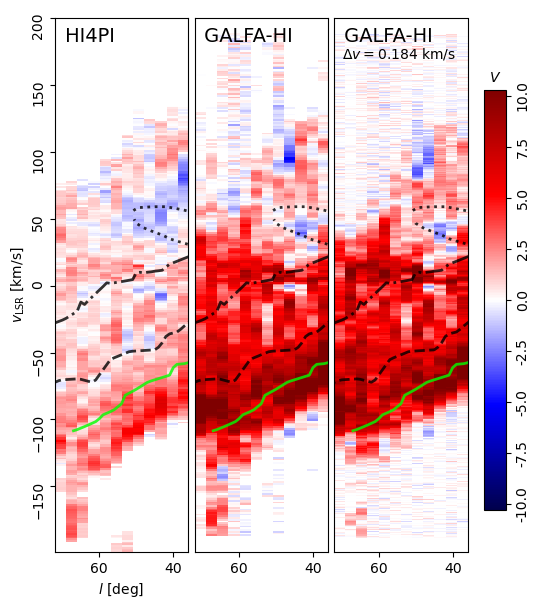}
}
\caption{Longitude-velocity ($lv$) diagrams of the filament orientation quantified by the projected Rayleigh statistic ($V$, right) in the two portions of the Galactic plane covered in the range $|b|$\,$\leq$\,10\deg\ by both HI4PI and GALFA-HI.
The GALFA-HI observations were analyzed at 1.29 and 0.184\,km\,s$^{-1}$ resolution using a 5\arcmin\ FHWM derivative kernel.
The curves correspond to the same spiral arm features introduced in Fig.~\ref{fig:lvdiagrams}.
}
\label{fig:GALFAandHI4PIlvdiagrams}
\end{figure*}

\begin{figure}[ht!]
\centerline{
\includegraphics[width=0.45\textwidth,angle=0,origin=c]{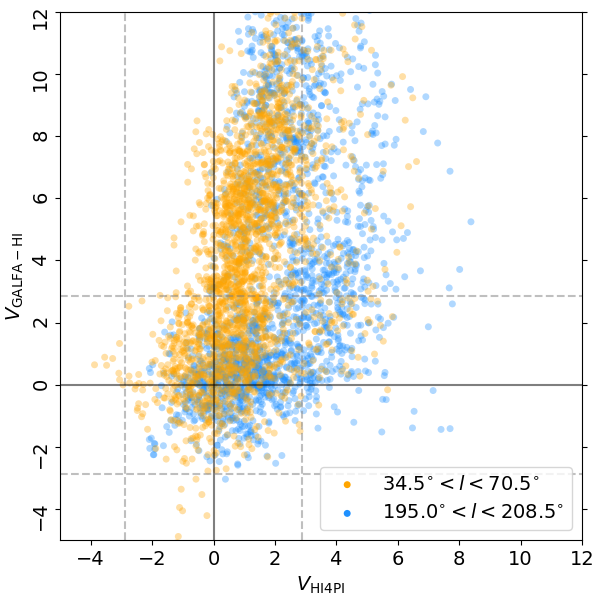}
}
\caption{
Projected Rayleigh statistic ($V$) calculated for the filamentary structure in the HI4PI and GALFA-HI observations.
Each point corresponds to a 3\deg\,$\times$\,20\deg\,$\times$\,1.29\,km\,s$^{-1}$ tile.
The gray dashed lines indicate $V$\,$=$\,2.87 and $V$\,$=$\,$-2.87$, which correspond to a 3$\sigma$ significance in the preferential orientation around $\theta$\,$=$\,$0$\deg\ or $\theta$\,$=$\,$90$\deg, respectively.
}
\label{fig:HI4PIandGALFAprs}
\end{figure}

\juan{The results of the GALFA-HI data analysis}, presented in Fig.~\ref{fig:GALFAandHI4PIlvdiagrams}, indicate at first glance that the global trends in relative orientation are roughly the same in the higher resolution data.
The most substantial difference between the analysis results in the two datasets is the increase in the significance in the GALFA-HI, which is further illustrated in Fig.~\ref{fig:HI4PIandGALFAprs}.
The global trend is for $V$ to increase by approximately a factor of two, which is most likely the combination of two effects.
First, the relative orientation trends observed in HI4PI are generally the same at the scales sampled by GALFA-HI.
If the primarily horizontal filaments responsible for the highest values of $V$ in HI4PI resolved into randomly oriented structures, we would find that the high-$V$ tiles would produce low $V$ in GALFA-HI, which is not what we observed in most of the tiles.
Second, the increase in $V$ is related to the higher angular resolution, \juan{because a smaller beam produces more independent gradient measurements for the same tile area.}
\juan{This does not imply that there is a higher number of filaments in the higher resolution but that the statistical trend is still found when including more independent orientation samples.}
\juan{We present examples of the filaments identified tiles in the two data sets in Fig.~\ref{fig:HI4PIandGALFArgb}.}

The higher values of $V$ in the higher resolution data indicate that the distribution of filament orientations still peaks at 0\deg\ and the prominence of that peak is more elevated.
These results do not exclude cases in which the higher resolution observations reveal no preferential orientation, as in some tiles with $V$\,$>$\,2.87 in HI4PI and $V$\,$\approx$\,0 in GALFA-HI.
Still, the significance in $V$ is generally higher in the GALFA-HI data by roughly a factor of two.

The increase in $V$ with the higher resolution observations produces a more significant portion of the $lv$-diagram, as illustrated in Fig.~\ref{fig:GALFAandHI4PIlvdiagrams}.
This general increase in $V$ does not imply a departure from our observation in the HI4PI data that the filaments tend to be more parallel to the Galactic plane with increasing $R_{\rm gal}$.
It means that the change in filament orientation is smoother and more significant at higher angular resolution.
This trend is confirmed by the analysis of the THOR survey \citep{soler2020}, where the increase in angular resolution by a factor of 24 with respect to HI4PI reveals a significant preference for H{\sc i} filaments parallel to the Galactic plane at $v_{\rm lsr}$\,$<$\,0\,km\,s$^{-1}$ in the range 17\deg\,$<$\,$l$\,$<$\,67\deg\ and $|b|$\,$<$1\pdeg25.

\subsubsection{Effect of radial velocity resolution}

We also performed the Hessian analysis in the GALFA-HI observations at their native 0.184 km/s resolution using an 5\arcmin\ FWHM derivative kernel.
The results, also presented in Fig.~\ref{fig:HI4PIandGALFAprs}, show no global change in the results of our analysis when considering the higher radial velocity resolution in the GALFA-HI data.
This is not unexpected given that emission from the narrowest H{\sc i} components, corresponding to the CNM, is at least a few kilometers per second wide and the emission structures are likely to be coherent across several narrow velocity channels.

\begin{figure*}[ht!]
\centerline{
\includegraphics[width=0.33\textwidth,angle=0,origin=c]{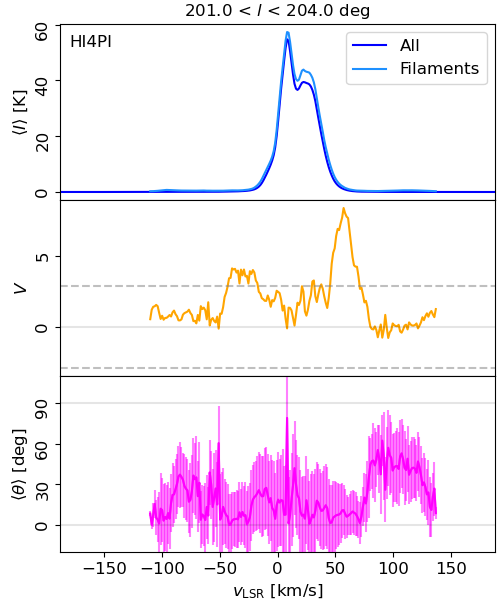}
\includegraphics[width=0.33\textwidth,angle=0,origin=c]{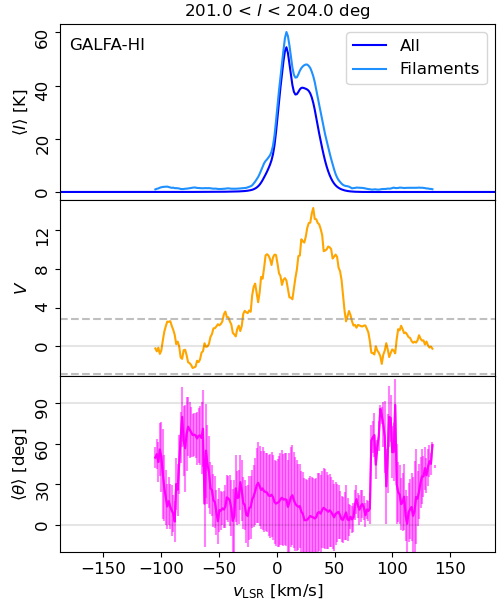}
\includegraphics[width=0.33\textwidth,angle=0,origin=c]{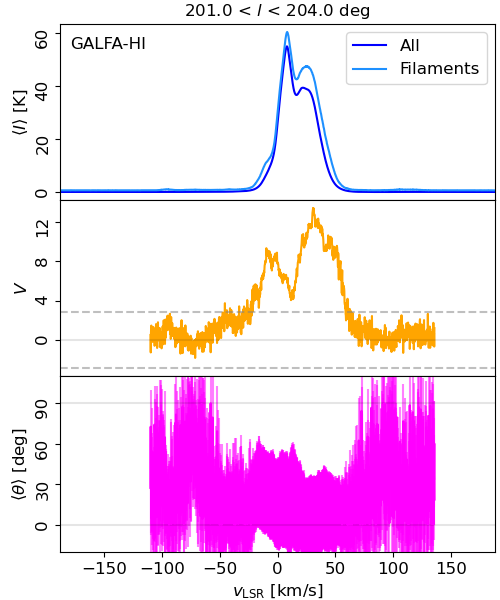}
}
\caption{
Mean intensity ($\left<I\right>$, {\it top}), projected Rayleigh statistic ($V$, {\it middle}), and mean orientation angle of the filaments ($\left<\theta\right>$, {\it bottom}) for the tiles centered on $l$\,$=$\,202\pdeg5 in the HI4PI and GALFA-HI.
\juan{The left panel corresponds to the HI4PI observations analysis with a 18\arcmin\ FWHM kernel for a 1.29-km\,s$^{-1}$ velocity resolution.}
\juan{The center and right panels correspond to the GALFA-HI observations analysis with a 5\arcmin\ FWHM kernel for 1.29-km\,s$^{-1}$ and 0.184-km\,s$^{-1}$ velocity resolutions, respectively.}
}
\label{fig:exampleTile}
\end{figure*}

We illustrate the effect of the angular and radial velocity resolution for one example tile in Fig.~\ref{fig:exampleTile}.
At the same 1.29\,km\,s$^{-1}$ resolution and S/N threshold, the results of the HI4PI analysis at 18\arcmin\ FWHM and GALFA-HI at 5\arcmin\ FWHM indicate a slight change in the orientation trends. 
This change is not unexpected, as the orientation at the large scale does not uniquely determine the orientation of the structures at the small scale.
Most of the differences are found in the velocity channels where the bulk of the emission is located, $v_{\rm lsr}$\,$\approx$\,0\,km\,s$^{-1}$.
At the largest $v_{\rm lsr}$, which also correspond to the largest $d$ and $R_{\rm gal}$, the increase in angular resolution does not introduce any significant changes in the relative orientation trends, thus supporting that the large $V$ values found there are not a product of the distance to the emission sources.
For illustration, we also present the mean orientation angles $\left<\theta\right>$, which further confirms the dominant trend of filaments aligned with the Galactic plane at both angular resolutions.

The results of the GALFA-HI analysis at 0.184\,km\,s$^{-1}$ resolution are very similar to those at 1.29\,km\,s$^{-1}$ resolution.
The orientation trends are coherent across a few kilometers per second in both cases.
The most prominent exception is the loss of signal around at $v_{\rm LSR}$\,$\approx$\,80\,km\,s$^{-1}$ and $v_{\rm LSR}$\,$>$\,150\,km\,s$^{-1}$, which is the result of the S/N decrease in the narrow velocity channels.
In general, we do not find significant differences in the analysis of the narrow velocity channels that would critically change the filament orientation trends obtained at 1.29\,km\,s$^{-1}$ resolution.

\section{Discussion}\label{section:discussion}

\subsection{Atomic filament orientations as a tracer of Galactic dynamics}

The trends reported in Fig.~\ref{fig:lvdiagrams} indicate that most of the filaments in the H{\sc i} emission are parallel to the Galactic plane.
This preferential orientation appears naturally from the combined effect of the gravitational potential and Galactic rotation, as confirmed by numerical simulations of Galactic disks \citep[see, for example,][]{smith2020,tress2020}.
The gravitational potential constrains the gas in the disk.
The Galactic rotation stretches the parcels of gas by shearing them apart.
However, these effects cannot explain by themselves the slight prevalence of vertical H{\sc i} filaments toward the inner Galaxy and the clear predominance of horizontal ones toward the outer Galaxy.

One observational effect that may influence the observed H{\sc i} filament orientation change with Galactocentric radius is the pileup of emission from a broad range of distances into the same channel in velocity space.
The portion of line of sight covered by a fixed-width velocity channel increases with Galactocentric radius.
For example, at $l$\,$=$\,90\deg\ and under the assumption of circular motions, a 1.29-km\,s$^{-1}$ velocity channel corresponds to $\Delta R_{\rm gal}$\,$\approx$\,200\,pc and $\Delta R_{\rm gal}$\,$\approx$\,500\,pc for $v_{\rm LSR}$\,$=$\,100 and 150\,km\,s$^{-1}$, respectively.
The velocity pile up can be a plausible explanation for the decrease in $V$ at $R_{\rm gal}$\,$\gtrsim$\,20\,kpc reported in Fig.~\ref{fig:PRSvsRgal}, by mixing portions of the line of sight where H{\sc i} structures are mostly parallel to the Galactic plane and randomly oriented into the same velocity channel.
However, it does not offer a straight forward explanation for the increase of $V$ with $R_{\rm gal}$ for 10\,$\lesssim$\,$R_{\rm gal}$\,$\lesssim$\,20\,kpc or the high-$V$ bands in Fig.~\ref{fig:deprojections}.

The analysis of the GALFA-HI observations indicates that the regions dominated by filaments parallel to the Galactic plane, $V$\,$>$\,2.87, do not resolve into structures with random orientations.
Instead, the higher angular and velocity resolution observations revealed that this preferential orientation is even more significant, as illustrated in Fig.~\ref{fig:GALFAandHI4PIlvdiagrams}.
The increase in angular resolution does not result in a similar increase in the orientation significance for tiles slightly dominated by filaments perpendicular to the Galactic plane, $V$\,$<$\,0.
We interpret this as indication that vertical H{\sc i} filaments are less frequent than horizontal ones.

The experiments with derivative kernels fixed to a physical length, such as the one presented in Fig.~\ref{fig:MultiKernelTest}, indicate that the progressive transition from $V$\,$<$\,0 to $V$\,$>$\,0 with increasing $R_{\rm gal}$ is not due to a mixture of scales introduced by the fixed angular size of the derivative kernel.
Rather than producing larger $V$, the introduction of derivative kernels with larger angular sizes results in the reduction of $V$, as further illustrated in App.~\ref{appendix:kernelsize}.
Hence, we consider that the progressive transition from $V$\,$<$\,0 to $V$\,$>$\,0 with increasing $R_{\rm gal}$ reported in Fig~\ref{fig:PRSvsRgal} has an origin in the physical conditions of the Galactic disk.

\subsubsection{Vertical H{\sc i} filaments}

\begin{figure}[ht!]
\centerline{\includegraphics[width=0.495\textwidth,angle=0,origin=c]{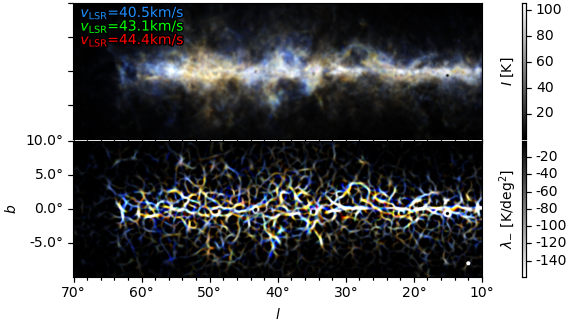}}
\caption{
H{\sc i} emission ({\it top}) and filamentary structures ({\it bottom}) toward a region dominated by vertical structures in the indicated radial velocity range.
}
\label{fig:exampleRegionI}
\end{figure}

Figure~\ref{fig:exampleRegionI} shows an example of the emission and the filaments toward one of the inner Galaxy regions with prevalence of vertical structures, or $V$\,$<$\,$0$.
The majority of the structures that would be identified as HISA are constrained to relatively low Galactic longitudes, where the background emission is brightest, and are unlikely to be the dominant object producing the global filament orientation.
This is also true for other regions in the inner Galaxy, as expected from the distribution and angular size in HISA catalogs \cite[see, for example,][]{gibson2005}.
However, it does not dismiss the importance of a dedicated \juan{study of the HISA structures orientation}.

Most of the structures contributing to $V$\,$<$\,$0$ are worms and arcs above and below the bright filamentary structure that corresponds to the midplane.
These structures extend into high Galactic latitudes, some even beyond the studied Galactic latitude range, $|b|$\,$<$\,10\deg.
\juan{However, the higher-resolution but lower $b$-coverage H{\sc i} observations in the THOR survey indicate that predominantly vertical filamentary structures are also found in the $|b|$\,$<$\,1\pdeg0 range \citep{soler2020}.}

One physical explanation for the observed vertical H{\sc i} filaments is the Galactic fountain, in which hot gas is produced by the cumulative effect of supernovae in the disk; thermal expansion lifts it above the Galactic plane, where it becomes unstable; cool clouds condense and fall back ballistically \citep{shapiroANDfield1976,fraternali2006}.
However, direct SN feedback events can also cause vertical structures.
Strong clustered SN feedback causes large scale bubbles, which are consecutively elongated along the vertical direction due to bouyancy and Parker instability \citep{parker1966}. 
Cosmic rays (CRs) add an additional source of slowly cooling energy, which builds up a vertical pressure gradient that helps lifting the gas out of the midplane \citep{breitschwerdt1993,girichidis2016,ruszkowski2017}. 

Numerical simulations indicate that CR-lifted gas is primarily warm rather than hot, which will increase the amount of H{\sc i} compared to hot only SN-heated gas \citep[see, for example,][]{girichidis2018}.
CRs mainly move along the magnetic field lines, which further amplifies the effects of Parker loops and the related instabilities \citep[][]{hanasz2003}. 
Consequently, the gas along the buoyantly rising magnetic field lines also shows a preferentially vertical orientation.
Overall, strong SN feedback in the presence of a vertical stratification of gas and a dynamically relevant magnetic field will favor the formation of vertical structures. 
Additionally, spiral shock waves may lift and stretch the atomic gas in the vertical direction \citep{suchkov1974,kim2010} and Parker instability that can create buoyant loops of magnetic field at a kiloparsec scale \citep{rodrigues2016,heintz2020}.

The progressive change from mostly vertical H{\sc i} filaments to mostly horizontal with increasing $R_{\rm gal}$, reported in Fig.~\ref{fig:PRSvsRgal}, indicates that the dominant process behind the H{\sc i} filament orientation also has a radial dependence.
Spiral shocks trace the position of the spiral arm structure, thus they should carry an azimuthal dependence and produce differences between arm and inter-arm regions that are not noticeable in our results.
The axisymmetric variations in filament orientation strongly suggest the prevailing role of SN feedback.

Supernovae are very effective in lifting the gas off from the Galactic plane, as seen in numerous numerical experiments \citep[see, for example,][]{joung2006,hennebelle2014,gatto2015,kim2015}.
The radial variations in the clustering and frequency of SNe with Galactocentric radius are a plausible explanation for the lack of vertical filaments in the outer Galaxy.
The number of SN remnants in the Milky Way decreases significantly for $R_{\rm gal}$\,$>$\,5\,kpc \citep{case1998,green2015}.
The distribution of stars in the Galaxy decreases exponentially with a characteristic radius $R_{0}$ that depends on the age of the population and is around 3.5\,kpc for young stars \citep{drimmel2001,frankel2018}.
Both of these observations indicate that for $R_{\rm gal}$\,$>$\,10\,kpc, the number of SNe is significantly lower than in the inner Galaxy.
At large $R_{\rm gal}$ not only the number of SNe is lower, but the area of the disk also is more extensive, so the SN rate per unit area falls off even faster with radius than the number of SNe.
Consequently, due to the lack of events in the outer Galaxy, SNe's energy and momentum input may not be the dominant process shaping the atomic gas.

The SNe do not only contribute to the formation of vertical filaments by lifting the H{\sc i} gas, but also through the infall of the previously ejected gas back on the plane.
\cite{marascoANDfraternali2011} studied the extraplanar H{\sc i} gas of the Milky Way and found that it mainly consists of gas that is falling back to the MW after being ejected by SNe.
The extraplanar H{\sc i} is more conspicuous while falling because it cools while settling down. 
When it is lifted, it is hotter and so less prominent in the H{\sc i} emission, as can be seen in the numerical simulations presented in \cite{kim2015b} and \cite{girichidis2016}.
It is difficult to determine the portion of the vertical filaments that correspond to outgoing or incoming H{\sc i} gas. 
Still, the filament orientations provide a roadmap to study these effects in the Galactic plane and in Galactic-scale simulations that include the impact of stellar feedback. 


\subsubsection{Filaments and holes in other galaxies}

Most the face-on nearby galaxies (2\,$\lesssim$\,$d$\,$\lesssim$\,15\,Mpc) in The HI Nearby Galaxy Survey \cite[THINGS,][]{walter2008} present holes in their inner regions, where the gas density and the star formation rate are higher, and elongated structures at larger radii.
A detailed study of one of them, NGC\,6946, a well studied spiral galaxy of about one-third of the Milky Way's size located at 7.72\,Mpc, revealed 121 H{\sc i} holes in its inner regions and high-velocity gas associated with some of them, thus suggesting the presence of a fountain mechanism \citep{boomsma2008}.
It is not straightforward to demonstrate that the associations of vertical filaments in the inner Milky Way correspond to the holes seen in face-on nearby galaxies, particularly due to the difficulty in disentangling emission components and assigning reliable distances in our own Galaxy.
However, there is significant evidence that toward some portions of the Galactic plane we are seeing the gas swept by the expansion of SN remnants, as it is the case of the H{\sc i} shells identified in the range 10\,kpc\,$<$\,$R_{\rm gal}$\,15\,kpc toward QIV \citep{mcclure-griffiths2002}.

A recent study of the H{\sc i} gas structure in a sample of 33 disk galaxies in THINGS revealed changes in the delta-variance spectrum in the inner and outer regions of the disks 
\citep{dib2021}.
The authors interpret these variations as the dominant effect of SN feedback in the inner parts of the galaxies and the prevalence of other large-scale dynamics in the outer parts.
Certainly, SNe are not independent actors in the shaping the ISM.
Recent numerical experiments indicate that feedback exclusively produced by SNe does not reproduce produce observed star cluster masses and warm gas volume-filling factors \citep{kim2018,rathjen2021}.
Detailed modeling of the Galactic winds propelled by SNe indicate that cool and hot outflows can lift significant amounts of mass, energy, and metals \citep{kim2020}.
Our results indicate that the study of the H{\sc i} filaments can shed light in the study of these Galactic winds and their distribution in the Milky Way.

\subsection{Atomic filaments as a tracers of Galactic structure}\label{subsection:warpANDflare}

\subsubsection{Galactic warp}

In addition to making the change in orientation with $R_{\rm gal}$ noticeable, the identification of filaments in the H{\sc i} emission highlights the disk warp and other variations in the midplane height of the Milky Way.
Figure~\ref{fig:scaleheight} shows that the $\left<z\right>$ features at $R_{\rm gal}$\,$>$\,20\,kpc known as scalloping are highlighted by the filaments.
This may be an indication that these features are not the product of global disk perturbations, as considered in \cite{levine2006}, but can be the product of events such as gas inflow or the interaction with Milky Way satellites.

\begin{figure}[ht!]
\centerline{\includegraphics[width=0.495\textwidth,angle=0,origin=c]{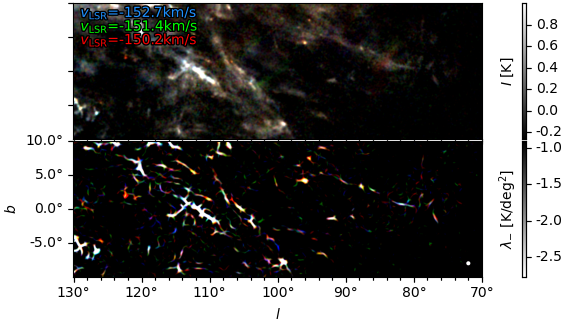}}
\caption{
H{\sc i} emission ({\it top}) and filamentary structures ({\it bottom}) for the region around the $\left<z\right>$ dip toward $\phi$\,$\approx$\,100\deg\ and 20\,$<$\,$R_{\rm gal}$\,$<$\,25\,kpc in Fig.~\ref{fig:scaleheight}.
}
\label{fig:exampleRegionA}
\end{figure}

The most prominent variation in $\left<z\right>$, aside from the warp, is the dip toward $\phi$\,$\approx$\,100\deg\ and 20\,$<$\,$R_{\rm gal}$\,$<$\,25\,kpc.
This feature, shown in Fig.~\ref{fig:exampleRegionA}, corresponds to a horizontal cloud at $b$\,$\approx$\,$-$8\deg\ that contrasts with the location of most of the H{\sc i} emission, which is concentrated at 0\deg\,$<$\,$b$\,$<$\,10\deg.
It is located in one of the few regions of the $lv$-diagram that does not show a preferential filament orientation at 0\deg\ or 90\deg, as it is evidenced by the elongated structures oriented at 45\deg, among which is the giant atomic filament Cattail \citep{li2021}.
The filaments in this region seem to be pointing in the direction of the Magellanic stream and their location both in radial velocity and Galactic longitude seem to reinforce the potential association with the 
interwoven tail of filaments trailing the Magellanic Clouds in their orbit around the Milky Way \citep[see][and references therein]{donghiaANDfox2016}.
Although determining this association is beyond the scope of this work, it exemplifies the use of the H{\sc i} filament properties as a tracer of the Galactic dynamics.

\subsubsection{Filament scale height}

\begin{figure}[ht!]
\centerline{\includegraphics[width=0.495\textwidth,angle=0,origin=c]{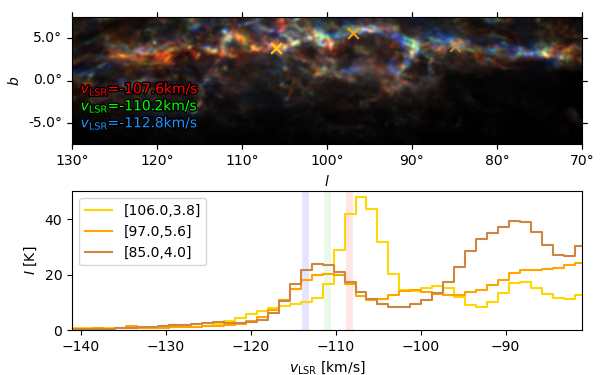}}
\caption{
H{\sc i} emission ({\it top}) and selected spectra toward filamentary structures ({\it bottom}).
The spectra correspond to the positions indicated by the crosses in the upper panel.
The colored vertical bars in the bottom panel correspond to the three velocity channels presented in the corresponding colors in the top panel.
}
\label{fig:spectra}
\end{figure}

Figure~\ref{fig:profilesSigmaZ} indicates a difference of roughly a factor of two in the scale height traced by the bulk of the H{\sc i} emission and just by the H{\sc i} filaments at $R_{\rm gal}$\,$>$\,15\,kpc.
To explain this difference, we need to consider the physical conditions that may lead to the formation of these structures.
Most likely the filaments are produced by large-scale compressive motions and thermal instability that produce overdensities in which the H{\sc i} can cool rapidly \citep{wolfire2003}.
Thus, the filaments may correspond to a mixture of CNM and WNM surrounded by WNM. 
This scenario is supported by the relatively narrow emission line widths toward the filamentary structures, as shown in the example presented in Fig.~\ref{fig:spectra}.

\cite{kalberlaANDkerp1998} estimated a scale height at the location of the Sun of about 150\,pc for the for the CNM and 400\,pc for the WNM, assuming hydrostatic equilibrium between the components.
The authors inferred velocity dispersions of 4 and 24\,km\,s$^{-1}$ for the CNM and WNM, which are roughly consistent with the combination of the turbulent and the thermal sound speeds of the gas in each phase.
That result suggest that each component is separately in vertical equilibrium, so that the total turbulent and thermal midplane pressures balances the vertical weight of the ISM \citep[see, for example,][]{ostriker2011}.
However, both the measured scale heights and the measured velocity dispersions are subject to relatively large uncertainties: for example, \cite{heilesANDtroland2003} reports that the difference in scale heights of the two phases is closer to a factor of two, as result of their finding of large velocity dispersion for the CNM and smaller for the WNM.

Toward the inner Galaxy the scale height of the CNM has also been observed to be smaller than that of the WNM \citep{dickeyANDlockman1990,mcclure-griffiths2007}.
If we assume that the same conditions can be extended to the outer galaxy, the difference in scale heights reported in Fig.~\ref{fig:profilesSigmaZ} is consistent with the expected difference for the two H{\sc i} components.
However, that conclusion is in tension with the results of the observations of H{\sc i} absorption against strong continuum sources, which indicate that the scale height of the \juan{CNM} is similar to that of the \juan{WNM}, and both increase with Galactic radius in the outer disk \citep{dickey2009}.

\cite{liszt1983} shows that one portion of the \juan{WNM} must be associated with the \juan{CNM}, while the other portion is more widely spread in the neutral medium between the \juan{CNM} clouds.
\cite{dickey2009} argues that the scale height derived from the H{\sc i} observed in absorption, which samples mostly the \juan{CNM clouds}, and that from the H{\sc i} emission, which samples both the \juan{CNM and WNM}, are similar in the outer Galaxy, thus suggesting that the scale heights of \juan{the two} H{\sc i} components are not very different.
One reason for the discrepancy with our results are the azimuthal variations that are evident in the radial profiles presented in Fig.~\ref{fig:NHfracVsRgal}.

Figure~7 of \cite{dickey2009} shows that the best agreement between the scale heights sampled by absorption and emission is for the average over the azimuth range 0\deg$<$\,$\phi$\,$<$180.
In that range our analysis results in large dispersions, mostly due to the lower emission toward that portion of the Galaxy at $R_{\rm gal}$\,$>$\,20\,kpc that results in fewer filaments.
In the range 180\deg$<$\,$\phi$\,$<$360, the agreement between the scale heights at $R_{\rm gal}$\,$>$\,17\,kpc in \cite{dickey2009} is less evident.
It is precisely toward that azimuthal range that we find the clearest difference between the $\sigma_{z}$ sampled by the filaments and by the bulk of the H{\sc i} emission.

Our results are not conclusive toward 0\deg$<$\,$\phi$\,$<$180, where we do not find enough filaments to match the results of \cite{dickey2009}.
However, we find a strong indication of different scale heights in the range 180\deg$<$\,$\phi$\,$<$\,360, where only 126 of 650 H{\sc i} opacity observations available at the time of \cite{dickey2009} are located.
Whether or not the difference in the scale heights derived using the H{\sc i} filaments are confirmed by the emission and opacity studies will depend on the analysis of future high-resolution surveys that increase the number of observations against continuum sources.

\subsubsection{Column density in atomic filaments}

We found that the ratio $N^{\rm fil}_{\rm H}/N_{\rm H}$ is about $80\%$ up to $R_{\rm gal}$\,$\approx$\,15\,kpc, as shown in Fig.~\ref{fig:NHfracVsRgal}.
The reason for the decrease in $N^{\rm fil}_{\rm H}/N_{\rm H}$ at $R_{\rm gal}$\,$>$\,15\,kpc toward some lines of sight can be attributed to the lack of filamentary structure in the emission, particularly toward QI and QII. 
The explanation for the roughly constant values of $N^{\rm fil}_{\rm H}/N_{\rm H}$ for $R_{\rm gal}$\,$<$\,15\,kpc is potentially related to the almost constant CNM to WNM mass ratio throughout the Galaxy.
The CNM fraction is almost constant with $R_{\rm gal}$ and corresponds to between 15 and 20\% of the atomic gas in the outer disk \citep{dickey2009}.
The larger values of $N^{\rm fil}_{\rm H}/N_{\rm H}$ can result from the filaments tracing not exclusively CNM but also some WNM associated with the cool gas.
This, however, implies a very coarse average because on scales up to at least hundreds of parsecs the pressure and density of the ISM may be highly variable and far from equilibrium

The decrease on the values of $N^{\rm fil}_{\rm H}/N_{\rm H}$ happens at roughly the same $R_{\rm gal}$ at which we see the end of the regular trend in filament orientation illustrated in Fig.~\ref{fig:PRSvsRgal}.
This two observations can be interpreted as an indication of the change in the physical conditions of the atomic gas in the outer galaxy.
These changes are not axisymmetric and can be caused by a variety of conditions that include the passage of the Milky satellites and variations in the ISM pressure and the stellar potential.

In principle, this observation implies that the formation of the structures that we observe as H{\sc i} filaments is not affected by variations in star formation and stellar feedback with Galactocentric radius.
Potentially converging flows compensate for the decrease in the SN rate with increasing distance from the Galactic center in maintaining the density enhancement that drives the warm gas into the cool structures that we observe as filaments.
However, further interpretation calls for the detailed study of particular regions and the analysis of numerical simulations of Galactic disks.

\subsection{Properties of atomic filaments}

We have employed the word filaments to designate the elongated structures found in H{\sc i} emission.
We used their orientation to describe the structure of the emission and its anisotropy without assigning their existence to a particular kind of object in three-dimensional space.
The fact that the preferential orientation is persistent across 1.29-km\,s$^{-1}$ velocity channels is a strong indication that this trend is produced by three-dimensional density structures in the ISM and not by fluctuations imprinted by the turbulent velocity field \citep[see][and references therein]{clark2019}.

The deep examination of the nature of the filamentary structures in three dimensions, for example, to distinguish between filaments and sheets \citep{heiles1997}, is beyond the scope of this work.
However, the characterization of the emission in terms of filaments highlights a large number of elongated structures that are reminiscent of the plumes in the H{\sc i} emission toward the outskirts of nearby spiral galaxies \citep{walter2008}.
Recent studies have identified prominent elongated structures in the H{\sc i} emission, for example, the 1.1-kpc-long Cattail filament with roughly 6.5\,$\times$\,$10^4$\,M$_{\odot}$ \citep{li2021} and the 1.2-kpc-long Magdalena filament with approximately 7.2\,$\times$\,\juan{10$^{5}$}\,M$_{\odot}$ \citep{syed2022}.
Our results indicate that these elongated structures are the norm toward the outer Galaxy.

The Magdalena filament is singular given its orientation, parallel to the Galactic plane, and its distance from the H{\sc i} emission midplane, roughly 500\,pc.
The kind of perturbations that can produce an object of these characteristics are currently unknown.
However, they may be related to the horizontal orientation of the VHC complexes AC and H and as such the orientation of this structures can be used as an indicator of the dynamics of the Galactic disk and its interaction with the halo.

There are few detections of molecular gas toward the bands dominated by horizontal H{\sc i} filaments in Fig.~\ref{fig:lvdiagrams}.
Carbon monoxide (CO) observations indicate that there is barely any significant emission at these large radii \citep{dame2001,miville-deschenes2017}.
Toward the OSC, \cite{dameANDthaddeus2011} report ten detections in 220 pointings toward high-H{\sc i} density regions.
\cite{sun2015} reported the detection of 72 molecular clouds in the CO(1\,$\rightarrow$\,0) emission toward the extension of the OSC into the second quadrant. 
Similar observations for the outer regions toward the outer third and forth quadrants have not been made.
Thus, it is hard at the moment to estimate if the H{\sc i} filaments inherit their elongation and orientation into molecular clouds that they may help form within them, in contrast with the general trend for the clouds traced by CO in the inner Galaxy \citep{soler2021}.
Yet, the outer Galaxy reveals as an attractive laboratory to test our theories of molecular cloud formation and the coupling between the ISM phases.

\section{Conclusions}\label{section:conclusions}

We studied the orientation of the filamentary structure in the H{\sc i} emission toward the Galactic plane.
We have found that the orientation of the H{\sc i} filaments with respect to the Galactic plane changes progressively with Galactocentric distance, from mostly perpendicular or having no preferred orientation in the inner Galaxy to mostly parallel up to $R_{\rm gal}$\,$\approx$\,15\,kpc and beyond.
The radial dependence in the H{\sc i} filament orientation strongly suggests that SNe's energy and moment input is responsible for lifting the atomic gas in the inner Galaxy.
Toward the outer Galaxy, the horizontal H{\sc i} filaments are most likely the result of Galactic rotation and shear.
Together these observations indicate that the atomic gas in the Milky Way follows a similar distribution as in nearby spiral galaxies, where holes potentially produced by SNe-propelled winds are found toward the inner portions, and arms and plumes dominate the outer ones.

The characterization of the H{\sc i} structure in terms of filaments highlights the features in midplane height and dispersion around the midplane height that might be indicative of the local effects perturbing the Galactic disk, such as variations in the potential or the inflow of gas.
We found that the scale height of the H{\sc i} filaments is lower than that traced by the bulk of the H{\sc i} emission by a factor of two.
If the filaments are tracing cold gas clouds and the portion of warm gas associated with them, our results imply that the scale heights of the cold and warm gas are not close, as previously inferred from H{\sc i} absorption observations.

We estimated that the amount of H{\sc i} in filamentary structures is roughly constant with Galactocentric radius up to $R_{\rm gal}$\,$\approx$\,15\,kpc.
This observation is potentially related to the mass ratio between the cold and warm H{\sc i} phases.
However, conclusively proving that hypothesis calls for further characterization of the H{\sc i} phases, for example, through Gaussian decomposition and dedicated studies of the filament properties.

In this paper, we have presented a study of the characteristics of the H{\sc i} emission.
We did so without assuming that these characteristics define an idealized physical object in three-dimensional space, such as a spiral arm or a thin strand of gas.
Through this generality, we identified a general anisotropy related to a physical quantity: the Galactocentric radius.
Some of the emission features that we classified as filaments may correspond to thread-like gas clouds, but that was not the starting assumption of this study.
Our conclusions were driven by a dominant trend found through the data analysis rather than adapting the data to a preconceived notion about the gas structure.
Given the diversity of processes shaping the ISM, this data-driven approach offers a promising path to \juan{study the cycles of mass and energy that lead to the formation and destruction of star-forming molecular clouds}.
\juan{Even if the observed lack of correlation between star formation and neutral atomic hydrogen at low surface densities has traditionally excluded the H{\sc i} observations from the global study of star-formation rates in galaxies \cite[see, for example,][]{kennicutt1989,leroy2008,schruba2011}, the H{\sc i} still carries crucial information on the dynamics that drive the diffuse gas into the condensations that become new stars.}

\begin{acknowledgements}
JDS, RKS, and SCOG are funded by the European Research Council via the ERC Synergy Grant ``ECOGAL -- Understanding our Galactic ecosystem: From the disk of the Milky Way to the formation sites of stars and planets'' (project ID 855130).
RSK, HB and SCOG acknowledge funding from the Deutsche Forschungsgemeinschaft (DFG) via the Collaborative Research Center (SFB 881, Project-ID 138713538) ``The Milky Way System'' (subprojects A1, B1, B2 and B8) and from the Heidelberg cluster of excellence (EXC 2181 - 390900948) ``STRUCTURES: A unifying approach to emergent phenomena in the physical world, mathematics, and complex data'', funded by the German Excellence Strategy.
RJS acknowledges funding from an STFC ERF (grant ST/N00485X/1) and HPC from the DiRAC facility (ST/P002293/1).
\juan{The team at ZAH acknowledges computing resources and data storage facilities provided by the State of Baden-W\"{u}rttemberg through bwHPC and the German Research Foundation (DFG) through grant INST 35/1134-1 FUGG and INST 35/1503-1 FUGG, and they thank for computing time from the Leibniz Computing Center (LRZ) in project pr74nu.}

\juan{We thank the anonymous referee for the thorough review and appreciate the suggestions, which helped us improve the manuscript.}
JDS thanks the following people who helped with their encouragement and conversation: Jonathan Henshaw, Jonas Syed, Eleonora Zari, Milena Benedettini, Robert Benjamin, and Thomas Henning.
Part of the crucial discussions that lead to this work took part under the program Milky-Way-Gaia of the PSI2 project funded by the IDEX Paris-Saclay, ANR-11-IDEX-0003-02. 
The computations for this work were made at the Max-Planck Institute for Astronomy (MPIA) {\tt astro-node} servers.

This work has been written during a moment of strain for the world and its inhabitants. 
It would not have been possible without the effort of thousands of workers facing the COVID-19 emergency around the globe.
Our deepest gratitude to all of them.

{\it Software}: {\tt astropy} \citep{astropy2018}, {\tt SciPy} \citep{2020SciPy-NMeth}, {\tt magnetar} \citep{magnetar2020}.

\end{acknowledgements}

\bibliographystyle{aa}
\bibliography{AA202243334.bbl}

\clearpage
\appendix

\section{The Hessian analysis method}\label{appendix:method}

\subsection{Studying angle distributions using circular statistics}\label{appendix:circstats}

Throughout this paper we have characterized the distribution of orientations using circular statistics, the branch of statistics that deals with observations that are directions \citep[see, for example,][]{batschelet1981,mardia2009directional}.
The probability distributions against which we are comparing the data are angles.
The most natural characteristic to describe the angle distributions is the mean, which is defined as
\begin{equation}\label{eq:meanangle}
\left<\theta\right> = \arctan\left(\frac{\sum w_{i}\sin\theta_{i}}{\sum w_{i}\cos\theta_{i}}\right),
\end{equation}
where the index $i$ runs over the elements in a group of angle observations and $w_{i}$ is the statistical weight corresponding to each of them.

The mean angle, $\left<\theta\right>$, is insufficient to fully describe the distribution of angles. 
For example, the angle distribution is ill-defined if the distribution is uniform or not unimodal.
To test whether the angle distribution is uniform and unimodal one can use the Rayleigh statistic, which is defined as
\begin{equation}\label{eq:rs}
Z = \left[\left(\sum w_{i}\cos\theta_{i}\right)^{2}+\left(\sum w_{i}\sin\theta_{i}\right)^{2}\right]^{1/2}.
\end{equation}
This value can be understood as the net displacement of a random walk in the complex plane with step sizes $w_{i}$.
For a uniform distribution of angles, $Z$\,$\approx$\,0.

In our application there is one direction that produces an expected anisotropy in the filament orientation: the Galactic plane.
That is why we used the projected Rayleigh statistic ($V$, Eq.~\ref{eq:myprs}) to test the orientation angle distribution against 0\deg\ and 90\deg, or parallel and perpendicular to the Galactic plane, respectively.
The observations also reveal that these two orientations are predominant in the H{\sc i} filaments.

\begin{figure}[h]%
\centering{
\includegraphics[width=0.495\textwidth]{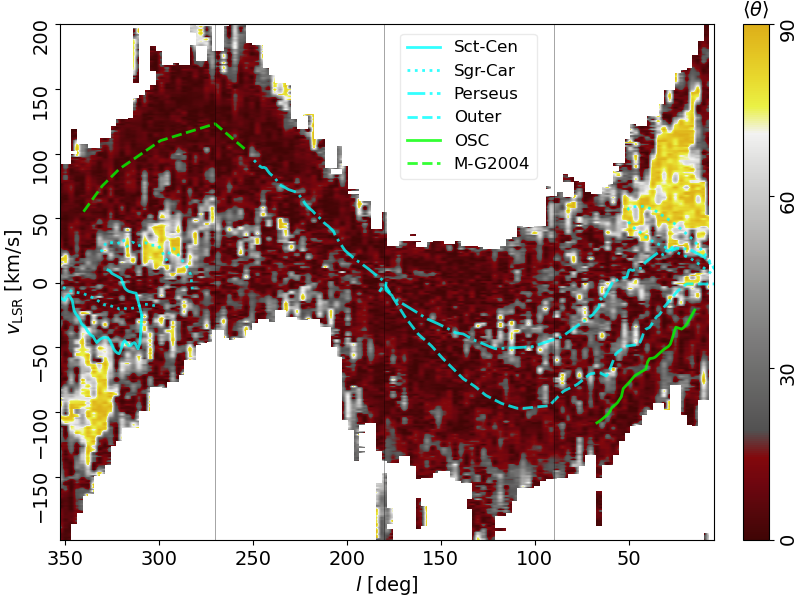}
}
\centering{
\includegraphics[width=0.495\textwidth]{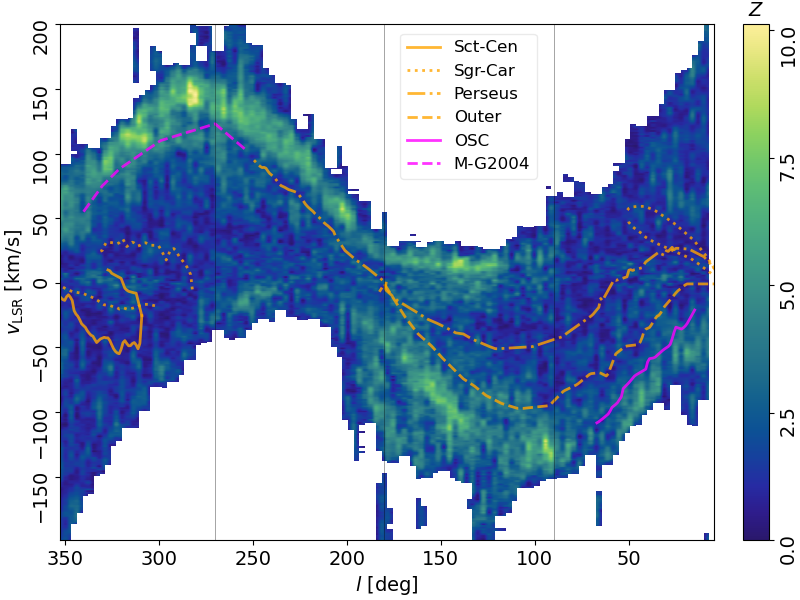}
}
\caption{Same as Fig.~\ref{fig:lvdiagrams}, but for the filament orientation quantified by the mean orientation angle ($\left<\theta\right>$, Eq.~\ref{eq:meanangle}, top) and the Rayleigh statistic ($Z$, Eq.~\ref{eq:rs}, bottom).
}\label{fig:MeanThetaRSlvdiagram}
\end{figure}

Figure~\ref{fig:MeanThetaRSlvdiagram} shows the results of our filament orientation analysis in terms of $\left<\theta\right>$ and $Z$.
It is evident that the majority of the tiles have filamentary structures with mean orientations $\left<\theta\right>$\,$\approx$\,0\deg.
However, this observation alone is not enough to conclude the significance of this trend.
The values of $Z$ indicate that there is a significant tendency for the angle distributions to be unimodal.
The highest significance is found in two bands located at the minimum and maximum $v_{\rm LSR}$ toward the first two and the last two Galactic quadrants, respectively.
These bands correspond to the high-$V$ bands shown in Fig.~\ref{fig:lvdiagrams}.

\begin{figure}[h]%
\centering{
\includegraphics[width=0.495\textwidth]{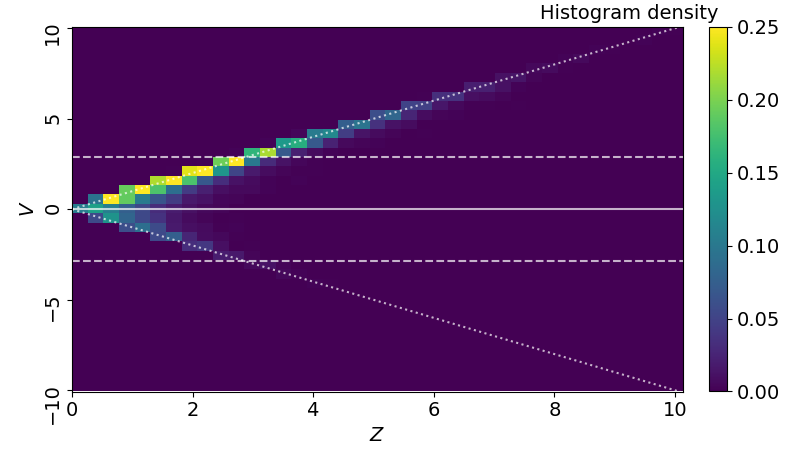}
}
\caption{Two-dimensional histogram of the Rayleigh statistic ($Z$) and projected Rayleigh statistic ($V$) calculated for the filamentary structures in the HI4PI observations.}\label{fig:scatterVvsZ}
\end{figure}

A summary of the trends obtained using the circular statistic is shown in Fig.~\ref{fig:scatterVvsZ}.
We found that the large majority of the tiles with significantly unimodal distributions, $Z$\,$\gtrsim$\,3, correspond to large values of $|V|$.
This confirms that 0\deg\ and 90\deg\ correspond to the most represented orientations in the H{\sc i} filaments and justified the use of these two reference angles in the application of the calculation of $V$.

\subsection{Selection of parameters in the Hessian analysis method}\label{appendix:hessian}

In the main body of this paper we have reported the results of a particular set of parameters in the Hessian analysis and the study of the filament orientations.
In this section we present the parameter study that led to that selection.
In what follows we detail the considerations related to the selection of the tile sizes, the derivative kernel size employed for the calculation of the Hessian matrix, and the thresholds in intensity and curvature used to select the filamentary structures.
We considered the variations introduced by the changes in one of these parameters and leaving the other fixed to the values presented in Table.~\ref{table:HessianParametersGALFAHI}.

\subsubsection{Tile sizes}\label{appendix:tilesize}

\begin{figure}[h]%
\centering{
\includegraphics[width=0.49\textwidth]{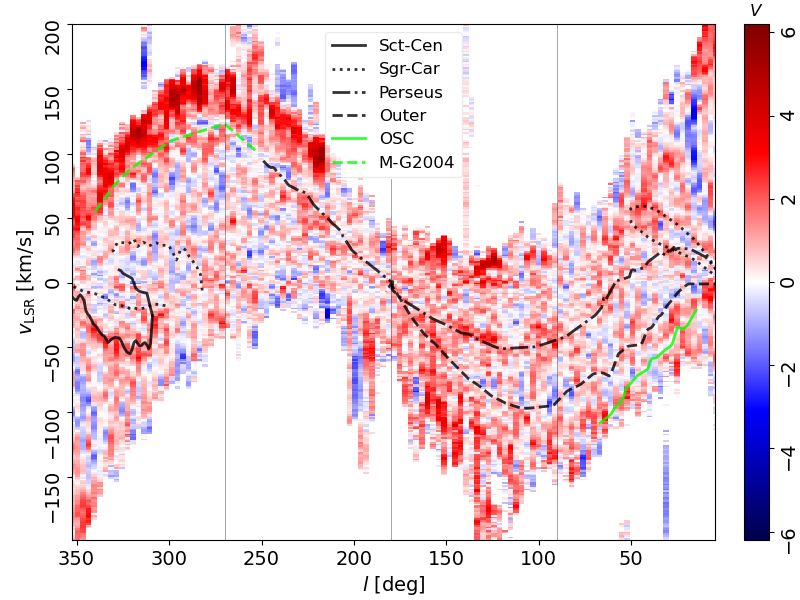}
\includegraphics[width=0.49\textwidth]{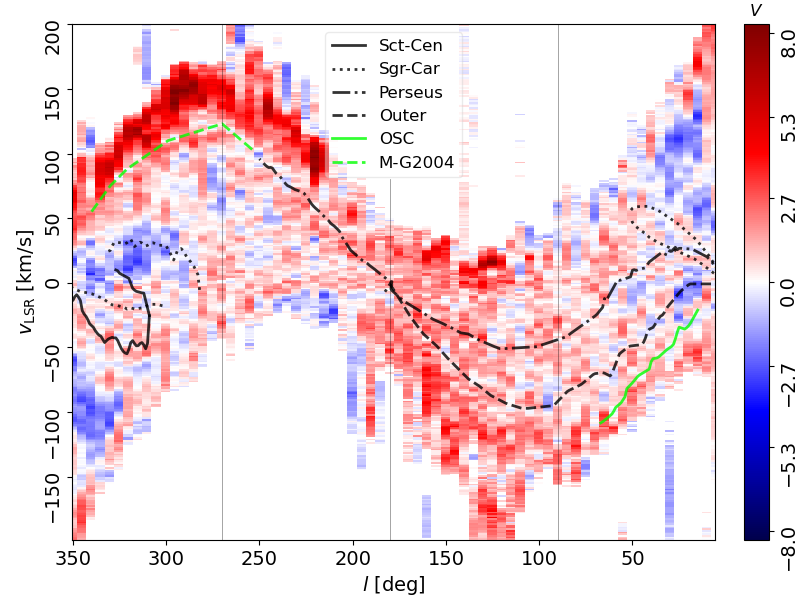}
\includegraphics[width=0.49\textwidth]{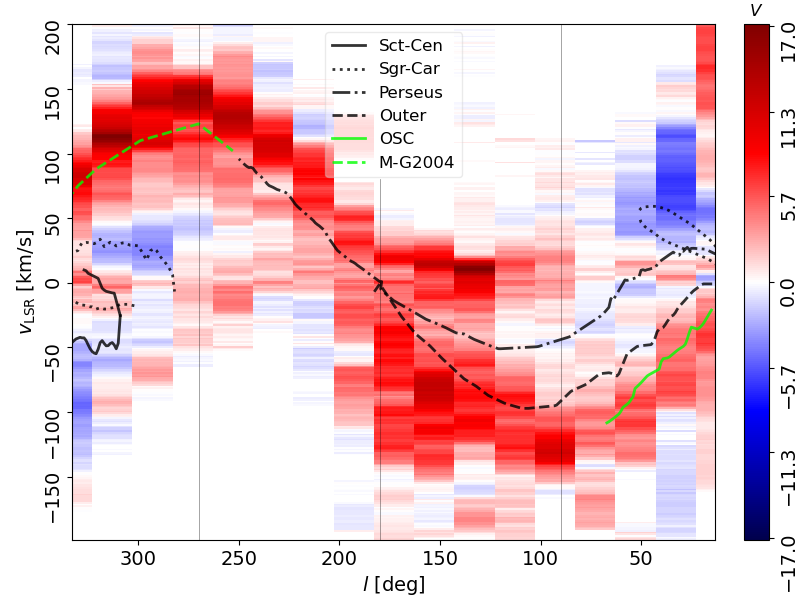}
}
\caption{Longitude-velocity ($lv$) diagrams of the H{\sc i} filament orientation quantified by the projected Rayleigh statistic ($V$, right).
Each of the pixel elements in the diagrams corresponds to a 3\deg\,$\times$\,3\deg\,$\times$\,1.29\,km\,s$^{-1}$ ({\it top}), 5\deg\,$\times$\,5\deg\,$\times$\,1.29\,km\,s$^{-1}$ ({\it middle}), and 20\deg\,$\times$\,20\deg\,$\times$\,1.29\,km\,s$^{-1}$ ({\it bottom}).
These results correspond to the Hessian analysis performed using a 18\arcmin\ FHWM derivative kernel.
The overlaid curves correspond to the spiral arm features discussed in the main body of the paper.
}\label{fig:PRSlvdiagramSquareTilesTest}
\end{figure}

We selected the $|b|$\,$\leq$\,10\deg\ range for the analysis aiming to resolve the warp and the flaring of the Galactic disk.
However, we examined the dependence of the H{\sc i} filament orientation trends on the tile size.
Figure~\ref{fig:PRSlvdiagramSquareTilesTest} shows the $V$ $lv$-diagrams corresponding the square tiles centered on $b$\,$=$\,0\deg\ and with side lengths of 3\deg, 5\deg, and 20\deg.
 
The 3\deg\,$\times$\,3\deg\ tiles cover the range included in the interferometric Galactic plane surveys THOR, $|b|$\,$<$\,1\pdeg25, and SGPS, $|b|$\,$<$\,1\pdeg4.
The corresponding $V$ $lv$-diagram, shown in the top panel of Fig.~\ref{fig:PRSlvdiagramSquareTilesTest}, indicates that the high-$V$ band reported in the main body of the paper is not found in that range toward the first quadrant.
This is due to the Galactic warp, which reaches midplane heights around 3.0\,kpc for $R_{\rm gal}$\,$\gtrsim$\,15\,kpc toward the first quadrant, as discussed in Sec.~\ref{subsection:warpANDflare}.

The 5\deg\,$\times$\,5\deg\ tiles cover the range included in the 1\arcmin-resolution interferometric Canadian Galactic Plane Survey \citep[CGPS][]{taylor2003}.
The corresponding $V$ $lv$-diagram, shown in the middle panel of Fig.~\ref{fig:PRSlvdiagramSquareTilesTest}, indicates that the high-$V$ band in the first quadrant is present in this range, but its significance is relatively low.
The high-$V$ band toward the last two quadrants is clearly distinguishable in both the 3\deg\,$\times$\,3\deg\ and 5\deg\,$\times$\,5\deg\ tiles analysis, because the Galactic warp in that direction is lower.
Finally, the bottom panel of Fig.~\ref{fig:PRSlvdiagramSquareTilesTest} shows the results corresponding to the 20\deg\,$\times$\,20\deg\ tiles.

\juan{The decrease in the tile size leads to a reduction of the maximum $V$ values, as it is evident from the comparison of the panels in Figure~\ref{fig:PRSlvdiagramSquareTilesTest}, which is the result of the lower number of pixels or, equivalently, independent filament orientations.
However, the positive $V$ in radial velocities corresponding to the outer galaxy and negative $V$ toward the inner galaxy, which are highlighted as the main results of this paper, are persistent for the considered tile size selections.
The selection of broader tiles washes away some of the azimuthal variations. 
Thus, we compromise between $b$-coverage and resolution in $l$ by using rectangular tiles.
}

\begin{figure}[h]%
\centering{
\includegraphics[width=0.49\textwidth]{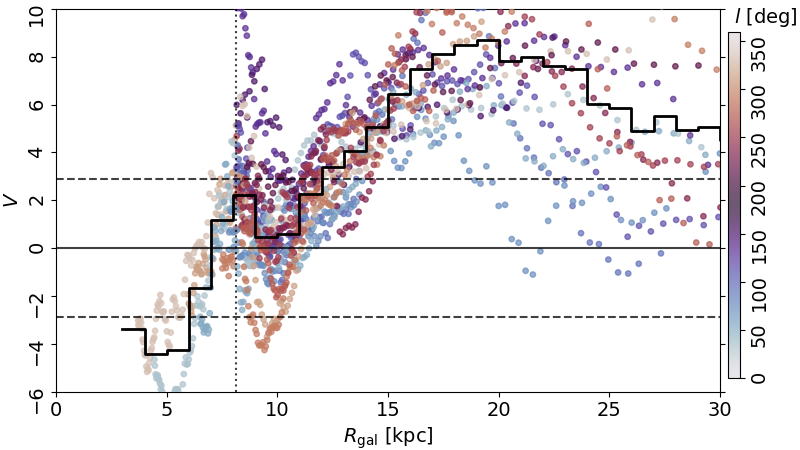}
}
\caption{Same as Fig.~\ref{fig:PRSvsRgal}, but for tile sizes 20\deg\,$\times$\,20\deg\,$\times$\,1.29\,km\,s$^{-1}$.
}\label{fig:PRSvsRgalSquareTiles}
\end{figure}

Figure~\ref{fig:PRSvsRgalSquareTiles} illustrates that the selection of square tiles increases the significance in the signal produced by mostly vertical structures.
However, this increase comes at the cost of reducing the number of tiles in Galactic longitude and significantly increasing the signal coming from the local emission.
All things considered, we chose rectangular tiles to highlight the significance of the filamentary structure in the outer Galaxy and produce a more detailed face-on reconstruction of the filament orientation trends.

\subsubsection{Tile aspect ratio}

\begin{figure}[h]%
\centering{
\includegraphics[width=0.49\textwidth]{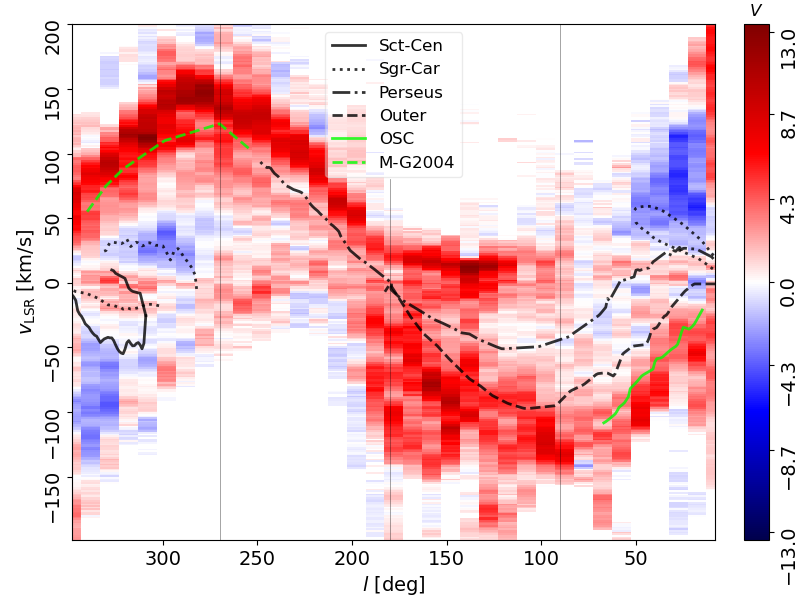}
\includegraphics[width=0.49\textwidth]{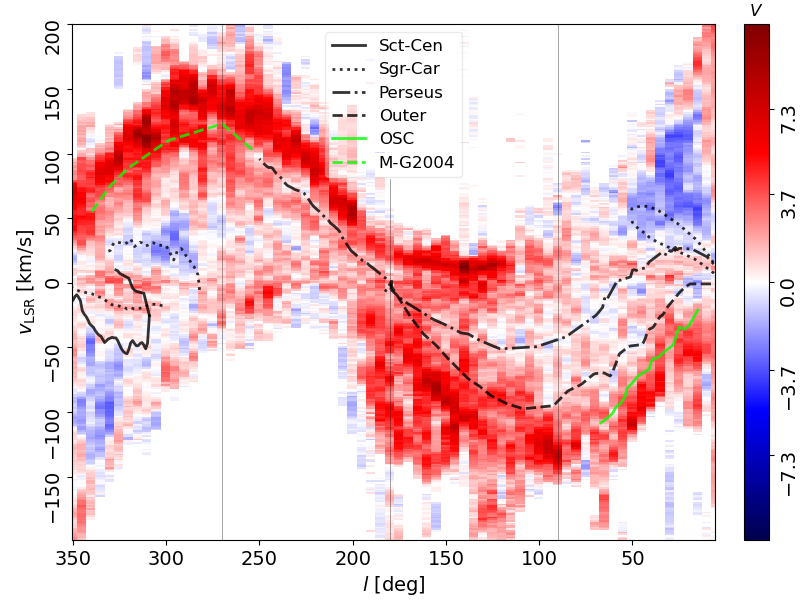}
}
\caption{Same as Fig.~\ref{fig:PRSlvdiagramSquareTilesTest}, but for tile sizes 10\deg\,$\times$\,20\deg\,$\times$\,1.29\,km\,s$^{-1}$ and 5\deg\,$\times$\,20\deg\,$\times$\,1.29\,km\,s$^{-1}$, shown in the top and bottom panels, respectively.
}\label{fig:PRSlvdiagramRectangularTilesTest}
\end{figure}

Square tiles are a natural selection to avoid any bias toward one of the preferential orientations studied in this analysis, that is, 0\deg\ or 90\deg, as discussed in \cite{soler2020}.
Yet the use of rectangular tiles improves the resolution in $l$ and the study of azimuthal effects.
The 3\deg\,$\times$\,20\deg\,$\times$\,1.29\,km\,s$^{-1}$ tiles selected for the analysis in the main body of the paper result from the compromise between azimuthal resolution and the $V$ significance .
To reach that selection, we empirically studied the effect of the tile aspect ratio on the values of $V$.

Figure~\ref{fig:PRSlvdiagramRectangularTilesTest} presents the results obtained using 10\deg\,$\times$\,20\deg\,$\times$\,1.29\,km\,s$^{-1}$ and 5\deg\,$\times$\,20\deg\,$\times$\,1.29\,km\,s$^{-1}$ tiles.
We found that the main features in the $V$ $lv$-diagram persist with the change in the tile aspect ratio.
The increase in the $l$ resolution resolves the negative-$V$ patches in the bottom panel of Fig.~\ref{fig:PRSlvdiagramSquareTilesTest} into regions surrounded by $V$\,$=$\,$0$ ranges.
The rectangular tiles also constrain the $l$ range to which the dominant features in the map are assigned.
For example, the increase in the $l$ resolution resolved the separation between the high-$V$ band in the second quadrant and the HVC complex located around 155\deg\,$<$\,$l$\,$<$\,190\deg\ and $-340$\,$<$\,$v_{\rm LSR}$\,$<$\,$-135$\,km\,s$^{-1}$.

We found that for tiles with 3\deg\ coverage in $l$, the values of $V$ are reduced below the level of significance in most of the $l$ diagram.
This effect is particularly acute for the $V$\,$<$\,0, where the narrow tiles reduce the coverage of close-to-vertical structures and introduce a bias for $V$\,$>$\,0.
The high $V$-bands, which are mostly produced by a series of horizontal filaments distributed over a relatively broad range in $b$, are less affected by the tile aspect ratio.

\begin{figure}[h]%
\centering{
\includegraphics[width=0.49\textwidth]{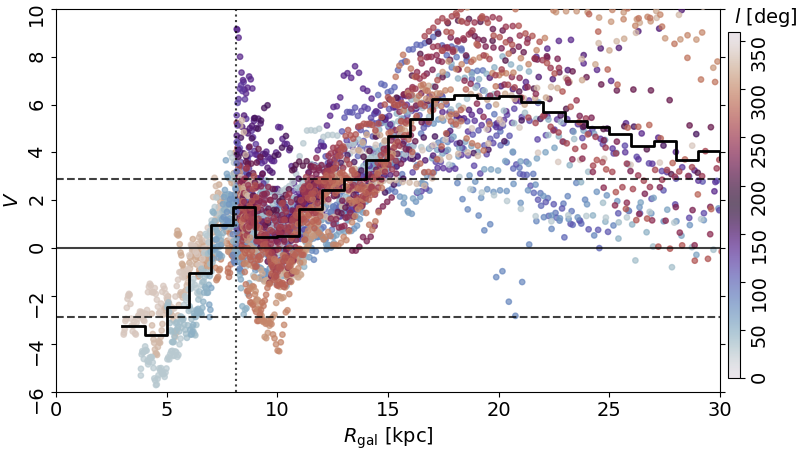}
}
\caption{Same as Fig.~\ref{fig:PRSvsRgal}, but for tile sizes 10\deg\,$\times$\,20\deg\,$\times$\,1.29\,km\,s$^{-1}$.
}\label{fig:PRSvsRgalRectangularTiles}
\end{figure}

Figure~\ref{fig:PRSvsRgalRectangularTiles} illustrates the effect of a broader tile selection in the trends reported in Fig.~\ref{fig:PRSvsRgalRectangularTiles}.
The 10\deg\,$\times$\,20\deg\,$\times$\,1.29\,km\,s$^{-1}$ tiles maintain a high significance in the regions where the filaments are preferentially oriented perpendicular to the Galactic plane.
This tile aspect ratio emphasizes the change in the orientation of the filaments, from mostly perpendicular to mostly parallel to the Galactic plane with increasing Galactocentric radius.
However, the 3\deg\,$\times$\,20\deg\,$\times$\,1.29\,km\,s$^{-1}$ tiles used to report the results in the mainbody of the paper increase the sampling in Galactic longitude, thus producing a more robust identification of the increasing values of $V$ with $R_{\rm gal}$.

\subsubsection{Kernel sizes}\label{appendix:kernelsize}

We have chosen a 18\arcmin\ FWHM derivative kernel to profit from the angular resolutions of the observations.
The use of larger kernels implies necessarily a lowering of the significance of the filament orientations, given the reduction in the number of independent filament orientations.
However, we tested the results of the analysis of 3\deg\,$\times$\,20\deg\,$\times$\,1.29\,km\,s$^{-1}$ tiles using larger derivative kernels.

\begin{figure}[h]%
\centering
\includegraphics[width=0.49\textwidth]{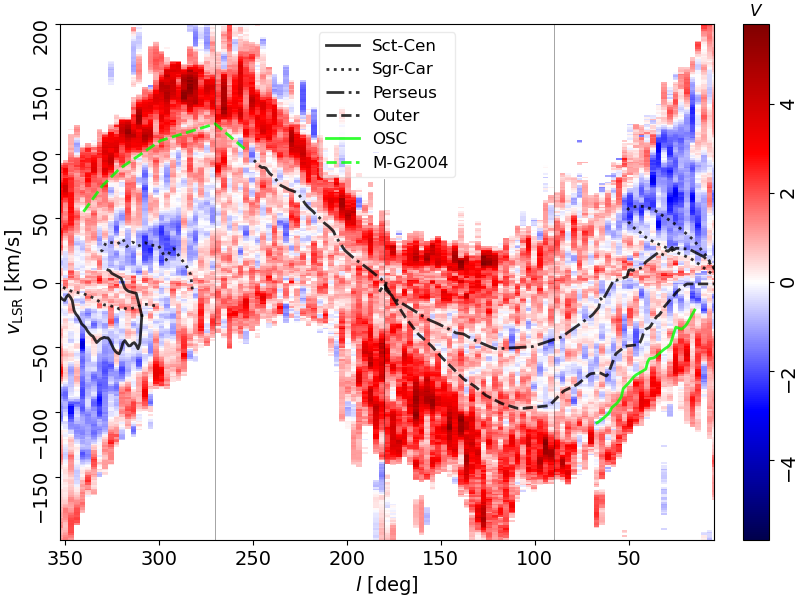}
\includegraphics[width=0.49\textwidth]{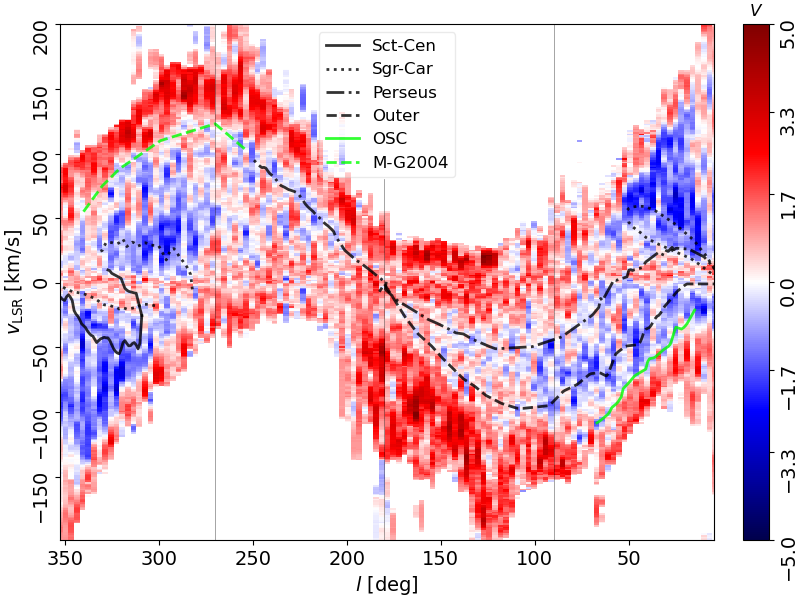}
\caption{Same as Fig.~\ref{fig:PRSlvdiagramSquareTilesTest}, but for 3\deg\,$\times$\,20\deg\,$\times$\,1.29\,km\,s$^{-1}$ tiles analyzed using 32\parcm4 and 48\parcm6 FWHM derivative kernels, shown in the top and the bottom panels, respectively.}\label{fig:PRSlvdiagramTestKernelSize}
\end{figure}

Figure~\ref{fig:PRSlvdiagramTestKernelSize} show the $V$ $lv$-diagrams obtained using derivative kernels two and three times larger than the beam size, 32\parcm4 and 48\parcm6 FWHM, respectively.
Besides the expected reduction in the $V$ values, the larger derivative kernels displace the high-$V$ bands to larger radial velocities.
This is due to the loss of significance produce by the reduction in the number of independent filament orientation measurements.
Rather than producing more horizontal structures, which would increase the values of $V$, the smoothing with a larger derivative kernel reduces the preferential orientation in the filamentary structures.

\begin{figure}[h]%
\centering
\includegraphics[width=0.49\textwidth]{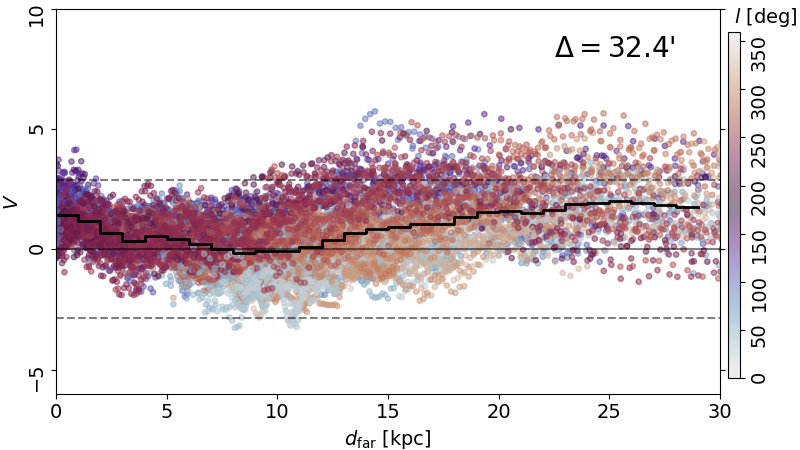}
\includegraphics[width=0.49\textwidth]{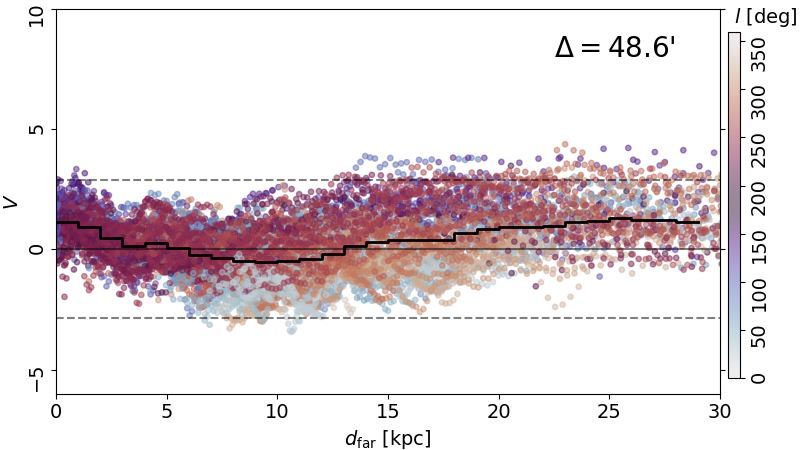}
\caption{Same as Fig.~\ref{fig:PRSvsDist}, but for 32\parcm4 and 48\parcm6 FWHM derivative kernels, shown in the top and the bottom panels, respectively.}\label{fig:PRSvsDistTestKernelSize}
\end{figure}

Figure~\ref{fig:PRSvsDistTestKernelSize} illustrates that the larger derivative kernels systematically reduce the values of $V$ independently of the kinematic distance corresponding to each tile.
Rather than artificially introducing high-$V$ values at closer distances, the smoothing results in a loss of the significance in the relative orientation trends.
This reinforces the argument that the high-$V$ bands are not a product of the angular resolution of the HI4PI survey, which we proved using the 4\arcmin\ resolution observations in GALFA-HI.
For the sake of simplicity, we present only the results corresponding to the far distance, but the same is inferred for the nearby solution to the near kinematic distance.

\begin{figure}[h]%
\centering
\includegraphics[width=0.49\textwidth]{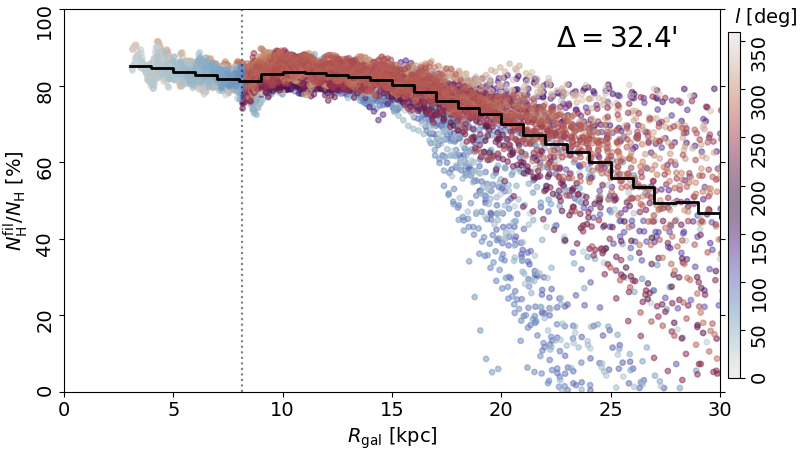}
\includegraphics[width=0.49\textwidth]{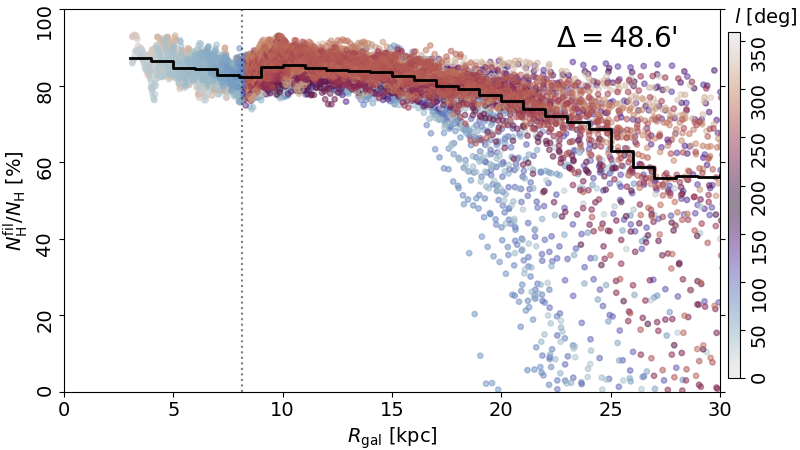}
\caption{Same as Fig.~\ref{fig:NHfracVsRgal}, but for 32\parcm4 and 48\parcm6 FWHM derivative kernels, shown in the top and the bottom panels, respectively.}\label{fig:NHfracVsRgalTestKernelSize}
\end{figure}

Figure~\ref{fig:NHfracVsRgalTestKernelSize} shows the ratio between the column density in filaments and in the bulk of the H{\sc i} emission ($N_{\rm H}^{\rm fil}/N_{\rm H}$) for the two derivative kernels.
We found that the increase in the derivative kernel size does not substantially change the fraction of column density in filaments or the $R_{\rm gal}$ above which we found a drop in $N_{\rm H}^{\rm fil}/N_{\rm H}$.
This result suggests that the drop in $N_{\rm H}^{\rm fil}/N_{\rm H}$ is not an effect introduced by the distance.
If distance was producing the drop, we would see a systematic shift in the $R_{\rm gal}$ values where $N_{\rm H}^{\rm fil}/N_{\rm H}$ starts to decrease.
The presumed shift in $R_{\rm gal}$ should be lower for the larger derivative kernel but this is not what is found in the data, where we found the drop at the same $R_{\rm gal}$ for the three considered kernel sizes.

Since we fixed the $I$ S/N threshold to test the effect of the derivative kernel size, we also conclude that the drop in $N_{\rm H}^{\rm fil}/N_{\rm H}$ is not simply caused by a drop in the emission.
The derivative kernels effectively smooth the velocity channel maps, thus bringing a larger portion of the emission above the $I$ S/N threshold.
However, changing this smoothing by up to a factor of three above the beam size does not substantially change the trends in $N_{\rm H}^{\rm fil}/N_{\rm H}$.  
Therefore, we conclude that the drop in $N_{\rm H}^{\rm fil}/N_{\rm H}$ is related to a transition toward less-structured H{\sc i} emission rather than a change introduced by the decrease in H{\sc i} emission with increasing $R_{\rm gal}$.

\subsubsection{Intensity threshold}

\begin{figure}[h]%
\centering
\includegraphics[width=0.45\textwidth]{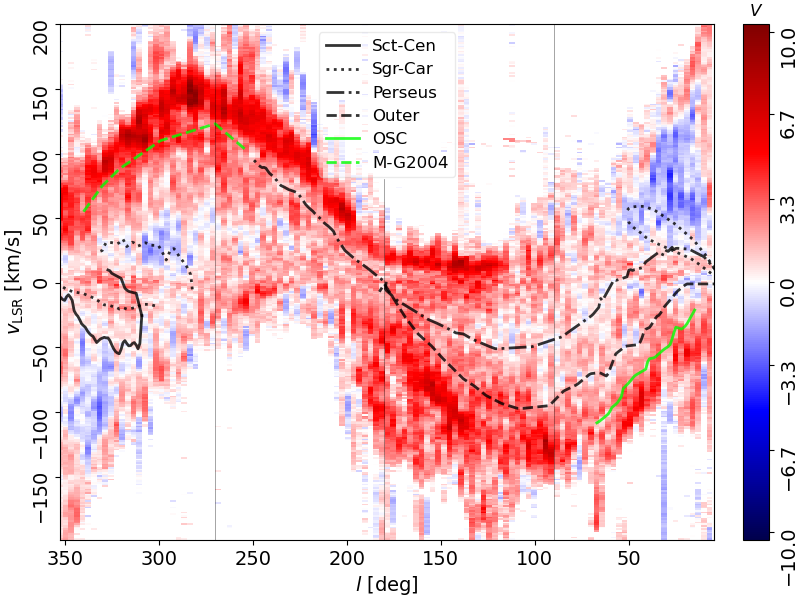}
\includegraphics[width=0.45\textwidth]{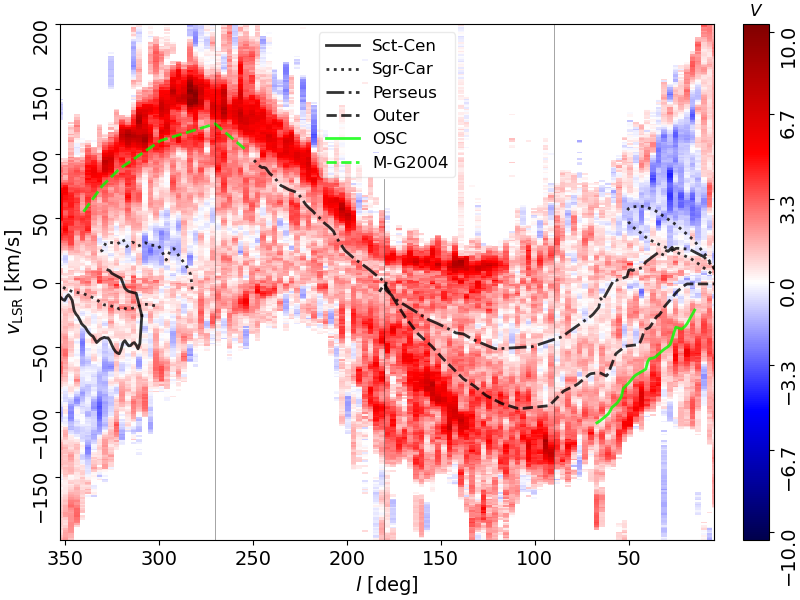}
\includegraphics[width=0.45\textwidth]{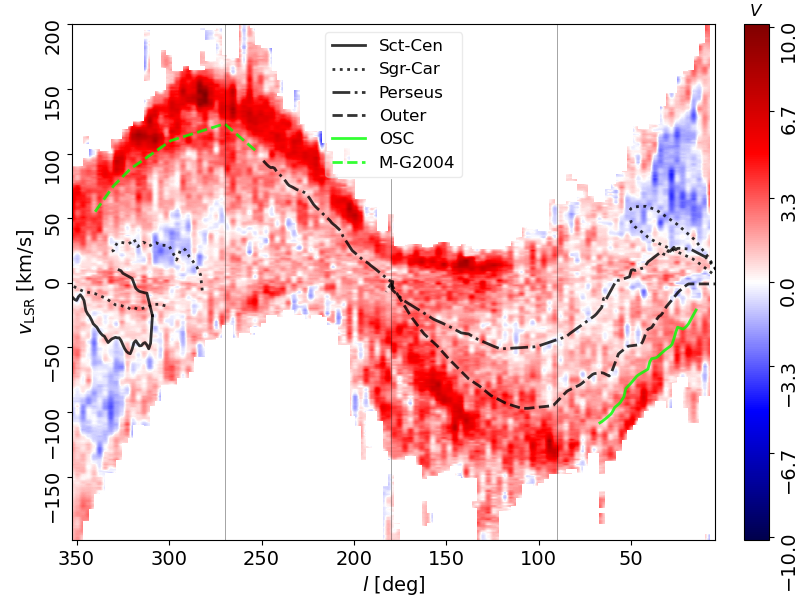}
\caption{Same as Fig.~\ref{fig:PRSlvdiagramTestKernelSize}, but for 3\deg\,$\times$\,20\deg\,$\times$\,1.29\,km\,s$^{-1}$ tiles and intensity thresholds of 3$\sigma$ (\juan{top}), 5$\sigma$ (\juan{middle}), and 9$\sigma$ (\juan{bottom}).}\label{fig:PRSlvdiagramTestSNR}
\end{figure}

We empirically evaluated the effect of the intensity threshold in the filament orientation analysis.
Figure~\ref{fig:PRSlvdiagramTestSNR} shows two examples of this selection above and below the threshold used for the results presented in the main body of this paper.
The results corresponding to the 3$\sigma_{I}$ intensity threshold show a vertical stripe around $l$\,$\approx$\,140\deg.
These features are typical for low-intensity thresholds and are produced by features in low-intensity velocity channels.
Most of these features disappear with a 5$\sigma_{I}$ threshold, as shown in the right-hand side panel of Fig.~\ref{fig:lvdiagrams}.

We do not find significant differences in the results produced by a 9$\sigma_{I}$ threshold.
The increase in the $I$ threshold tends to eliminate low-intensity velocity channels at the extremes of the velocity range rather than modifying the general trends in the $V$ $lv$ diagram.
This relatively low dependence on the intensity threshold is partly due to the stringent criterion in curvature, $\lambda_{-}$, which is dominant in determining the structures selected for the filament orientations analysis.

\subsubsection{Curvature threshold}

The lowest eigenvalue of the Hessian matrix, $\lambda_{-}$, plays a crucial role in the identification of the filamentary structure \citep[see, for example,][]{planck2014-XXXII,schisano2014}.
The filamentary structures correspond to the pixels where $\lambda_{+}\lambda_{-}$\,$<$\,0 and $\lambda_{-}$\,$<$\,$0$.
In the observations, however, there is a potential contribution of the noise to the curvature that we account for by choosing a lower curvature $\lambda^{\rm thres}_{-}$.

We present an example of the results obtained when using no curvature threshold, just $\lambda_{-}$\,$<$\,$0$, in Fig.~\ref{fig:PRSlvdiagramTestLambdaMinusThreshold}.
The example tile shows a collection of low-$I$ and relatively low curvature, $\lambda_{-}$\,$\approx$\,$0$, filamentary structures.
These faint filaments objects appear systematically over the whole map and do not show an evident preferred orientation.

The top panel of Fig.~\ref{fig:PRSlvdiagramTestLambdaMinusThreshold} shows the net effect of the low-curvature structures.
Globally, the $V$ $lv$-diagram presents the same high-$V$ bands and the low $V$ regions reported in Fig.~\ref{fig:lvdiagrams}.
However, the $|V|$ values are lower, which corresponds to a lower significance in the filament orientation trends.
We also found a series of vertical stripes in the $V$ $lv$-diagram which are particularly prominent in the region $v_{\rm LSR}$\,$<$\,0 toward the first two quadrants and $v_{\rm LSR}$\,$>$\,0 toward the last two.
These stripes are mitigated by applying a curvature threshold.

The bottom panel of Fig.~\ref{fig:PRSlvdiagramTestLambdaMinusThreshold} shows the results obtained by using a curvature threshold based on the $\lambda_{-}$ values found in the lowest intensity-S/N channels toward the position of each tile.
This selection increases the significance of the relative orientation trends, quantified by $V$, 
We selected the median of these per-tile $\lambda_{-}$ threshold as the value used for the results reported in the main body of this paper.


\begin{figure}[ht]
\centerline{
\includegraphics[width=0.45\textwidth,angle=0,origin=c]{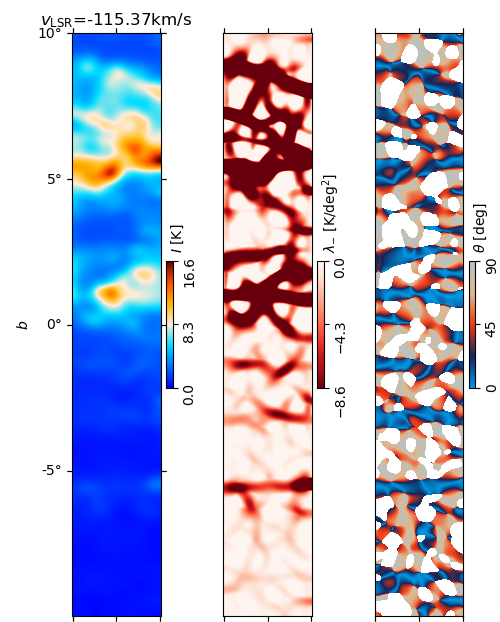}
}
\centerline{
\includegraphics[width=0.45\textwidth,angle=0,origin=c]{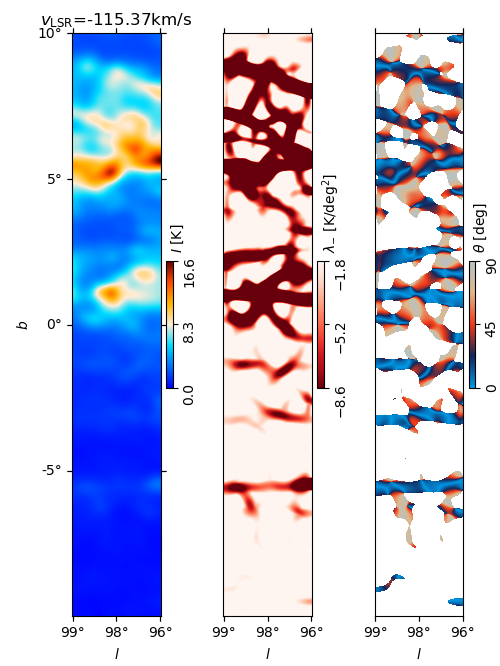}
}
\caption{Example of the filamentary structures identified using the Hessian method with curvature thresholds $\lambda_{-}$\,$<$\,0 ({\it top}) and $\lambda_{-}$\,$<$\,$-1.8$\,K/deg$^{2}$ ({\it bottom}).
The three maps represent: 
{\it Left.} H{\sc i} intensity after the smoothing by a Gaussian kernel with the size of the derivative kernel.
{\it Center.} Second eigenvalue of the Hessian matrix, $\lambda_{-}$, calculated using Eq.~\ref{eq:lambda}.
{\it Left.} Filament orientations, $\theta$, calculated using Eq.~\ref{eq:theta}. 
The radial velocity corresponds to a kinematic distance $d$\,$=$\,$12.7$\,kpc and Galactocentric radius $R_{\rm gal}$\,$=$\,16.1\,kpc, as derived using the assumptions in App.~\ref{app:deprojection}
}\label{fig:ExampleTile}
\end{figure}

\begin{figure}[h]%
\centering{
\includegraphics[width=0.49\textwidth]{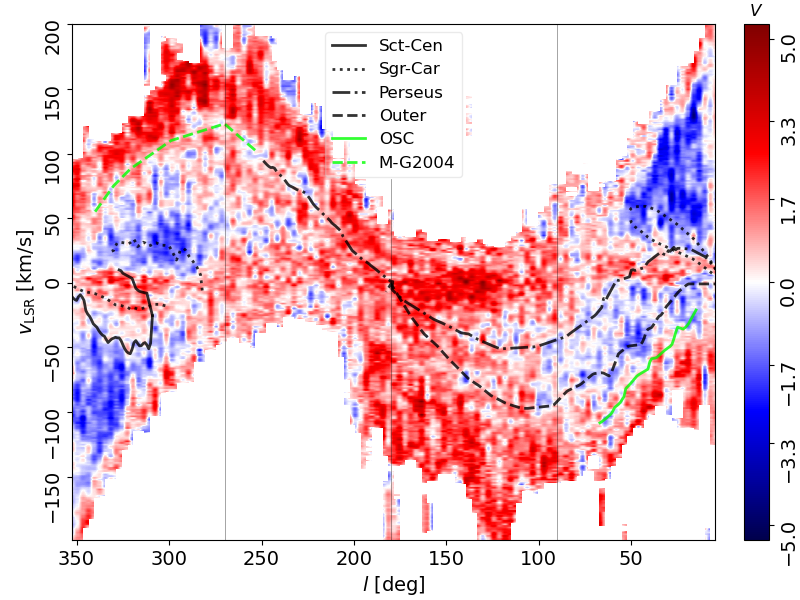}
\includegraphics[width=0.49\textwidth]{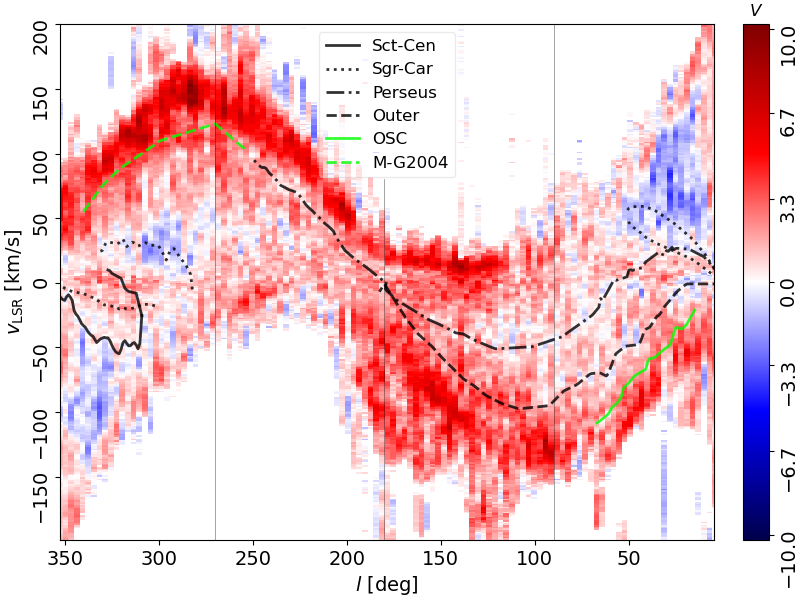}
}
\caption{Same as Fig.~\ref{fig:PRSlvdiagramTestSNR}, but for $\lambda_{-}$\,$<$\,$0$ (top) and a tile-dependent $\lambda_{-}$-threshold (bottom).}\label{fig:PRSlvdiagramTestLambdaMinusThreshold}
\end{figure}

\section{De-projection of the spectral observations}\label{app:deprojection}

\subsection{Construction of the face-on view}

We de-projected the results our analysis results in the $l$, $b$, and $v_{\rm LSR}$ grid into Galactocentric cylindrical coordinates $R_{\rm gal}$, $\phi$, and $z$.
We calculated a Galactocentric radius, $R_{\rm gal}$ for each tile's central position and radial velocity using the Kinematic Distance Utilities: {\tt KDUtils}\footnote{\url{https://github.com/tvwenger/kd}} \citep{wenger2018}.
In practice, these routines solve
\begin{equation}\label{eq:rgal}
R_{\rm gal} = R_{0}\sin(l)\frac{V(R_{\rm gal})}{v_p+V_{0}\sin(l)},
\end{equation}
where $R_0$ and $V_{0}$ are the Galactocentric radius and tangent velocity of the solar orbit around the Galactic center, $V(r)$ is the rotation curve, and $v_p$ is the radial velocity, which at the center of the tile is equal to $v_{\rm LSR}$.
In this study we used the rotation curve derived in \cite{reid2019}. 

\begin{figure}[h]%
\centering
\includegraphics[width=0.49\textwidth]{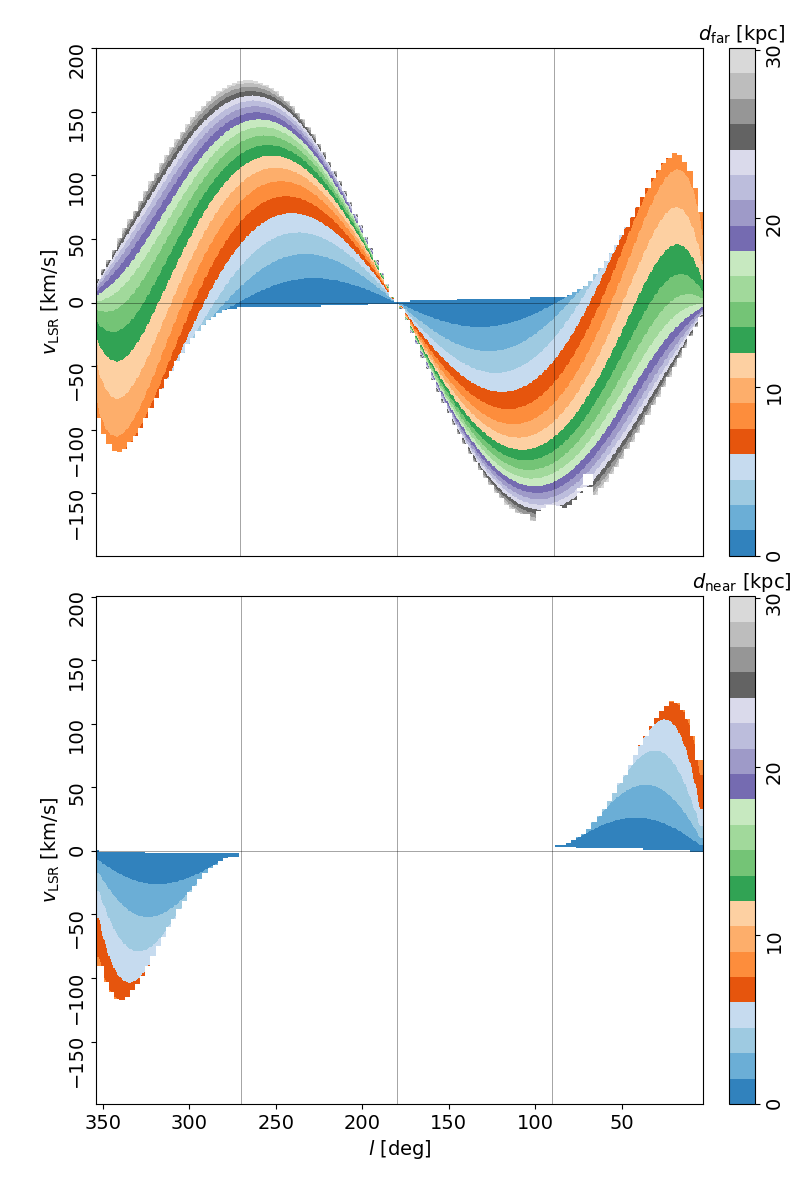}
\caption{Near and far kinematic distances assigned to the 3\deg\,$\times$\,20\deg\,$\times$\,1.29\,km\,s$^{-1}$ tiles used to construct the $lv$-diagrams in Fig.~\ref{fig:lvdiagrams}.
These results correspond to the solutions of Eq.~\ref{eq:distance} using the rotation curve presented in \cite{reid2019}.
}\label{fig:KinematicDistances}
\end{figure}

We also evaluated the Galactocentric azimuth angle
\begin{equation}\label{eq:azimuth}
\phi  = \arccos\left(\frac{R_{0}-d\cos l}{R_{\rm gal}}\right),
\end{equation}
where $d$ is the distance given by
\begin{equation}\label{eq:distance}
d = R_{0}\cos(l) \pm \sqrt{R_{\rm gal}^2 - R^{2}_{0}\sin^2(l)}.
\end{equation}
In the inner Galaxy ($R_{\rm gal}$\,$<$\,$R_{0}$) this expression yields to two distances along the line of sight, the near and the far
kinematic distances, located on either side of the tangent point.
Figure~\ref{fig:KinematicDistances} illustrates the near- and far-distances obtained for the positions of the 3\deg\,$\times$\,20\deg\,$\times$\,1.29\,km\,s$^{-1}$ tiles in the $lv$-diagram.

The ambiguity introduced by the two solutions to \juan{Eq}.~\ref{eq:distance} makes the determination of kinematic distances very challenging in the inner Galaxy. 
Thus, we restricted the results of our de-projection to the range $R_{\rm gal}$\,$>$\,10\,kpc to avoid the artifacts produced by the motions around the solar orbit.
We also exclude the longitude ranges $l$\,$<$\,15\deg, $l$\,$>$\,345\deg, and 165\deg\,$<$\,$l$\,$<$\,195\deg, where the radial velocities are not dominated by the circular motions and it is difficult to establish reliable distances \citep[][]{levine2006warp}.

\begin{figure}[ht!]
\centerline{\includegraphics[width=0.495\textwidth,angle=0,origin=c]{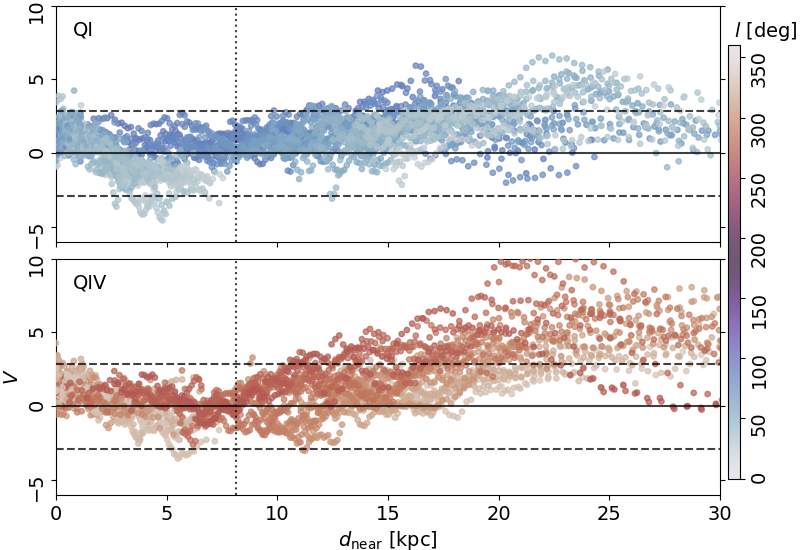}}
\caption{\juan{Same as Fig.~\ref{fig:PRSvsDistNEAR}, but for the near heliocentric kinematic distances.}}
\label{fig:PRSvsDistNEAR}
\end{figure}

\begin{figure}[h]%
\centering{
\includegraphics[width=0.49\textwidth]{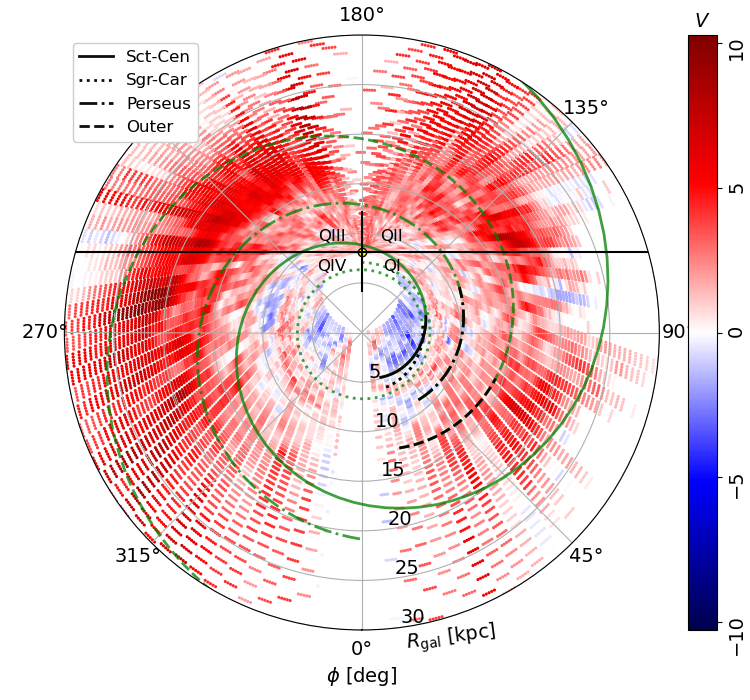}
}
\caption{Same as Fig.~\ref{fig:scaleheight}, but without masking $R_{\rm gal}$\,$<$\,10\,kpc and assigning all kinematic heliocentric distances to the largest solution to Eq.~\ref{eq:distance} .
}\label{fig:prsUnmasked}
\end{figure}

\juan{To test the potential effect of distance in the filament orientation trends, we presented the $V$ distribution for the ``far'' heliocentric kinematic distances, $d_{\rm far}$ in Fig.~\ref{fig:PRSvsDist}.
For the sake of completeness, we present the same diagram considering the ``near'' heliocentric distances in Fig.~\ref{fig:PRSvsDistNEAR}, which differ from $d_{\rm far}$ only toward QI and QIV.
We also present the unmasked face-on reconstruction of H{\sc i} filament orientations obtained by using the far heliocentric distances in Fig.~\ref{fig:prsUnmasked}.}

\subsection{Galactic warp and flaring}

For each pixel in a tile, we calculate the height $z$ with respect to $b$\,$=$\,0\deg\ by computing
\begin{equation}\label{eq:height}
z = d \arctan(b).
\end{equation}
This approach assumes a single representative distance for each tile, that is, it does not account for the variation between $b$\,$=$\,0\deg\ and $b$\,$=$\,$\pm10$\deg.
\juan{The maximum differences in $R_{\rm gal}$ and $d$ between the tile top and center are around 0.074\,kpc and 0.076\,kpc, respectively.
These are found in tiles at large $R_{\rm gal}$ and $d$, where the variations in the two quantities due to the tile height are negligible compared to the uncertainties introduced by the resolution in radial velocity.}

Our de-projection approach is different from the PPV grid interpolation and mapping into a fixed PPP grid followed in, for example, \cite{levine2006warp}, \cite{kalberla2008}, and \cite{koo2017}.
However, it does not yield significant differences in the results obtained for the midplane displacement (warp) and dispersion around the midplane (flaring), as shown in Sec.~\ref{subsection:results:warpANDflare}, and it is computationally more efficient.
Our analysis does not require the continuous sampling of the PPP grid necessary for the Fourier mode analysis in \cite{levine2006warp} or the three-dimensional mass reconstruction presented in \citep{nakanishi2016}, but rather aims to illustrate the variation in filament orientation with increasing $R_{\rm gal}$.
Given the one-to-one mapping from $v_{\rm LSR}$ into $R_{\rm gal}$ in the outer galaxy under the assumption of circular motions, our approach is close to a discrete sampling of the PPP space.

\subsection{Warp and flaring}\label{ssec:warpANDflaring}

For each one of the tiles de-projected into the Galactocentric cylindrical coordinates, we calculated the mid-plane displacement with respect to the $b$\,$=$0\deg\ position, or warp, by calculating the first-moment of the height $z$ weight by the intensity,
\begin{equation}\label{eq:warp}
\left<z\right> = \frac{\sum I_{ij}z_{ij}}{\sum I_{ij}},
\end{equation}
where the indices $i$ and $j$ run over the tile pixels.

We also calculated the dispersion around the mid-plane height, or flaring, by computing the second-moment of the height $z$ weight by the intensity,
\begin{equation}\label{eq:flaring}
\sigma_z = \left[\frac{\sum I_{ij}(z_{ij}-\left<z\right>)^2}{\sum I_{ij}}\right]^{1/2}
\end{equation}
where $\left<z\right>$ is calculated using Eq.~\ref{eq:warp}.

\section{Comparison between HI4PI and GALFA-HI}\label{appendix:HI4PIandGALFAHI}

The only survey with enough coverage of the Galactic plane to expand the results of the HI4PI analysis to higher resolutions is GALFA-HI.
With 4\arcmin\ FWHM resolution, GALFA-HI provides a factor of four improvement with respect to HI4PI in the ranges 36\pdeg5\,$<$\,$l$\,$<$\,70\pdeg5 and 175\pdeg3\,$<$\,$l$\,$<$\,209\pdeg4
We applied the Hessian matrix analysis to the GALFA-HI data using the parameters presented in Table.~\ref{table:HessianParametersGALFAHI}.
The analysis was applied to the data in its native velocity resolution, 0.184\,km\,s$^{-1}$, and to the data smoothed in the spectral axis to match the HI4PI velocity resolution. 
As with the HI4PI data, the curvature threshold ($\lambda^{C}_{-}$) was estimated empirically using noise-dominated velocity channels.

\begin{figure*}[ht]
\centerline{
\includegraphics[width=0.45\textwidth,angle=0,origin=c]{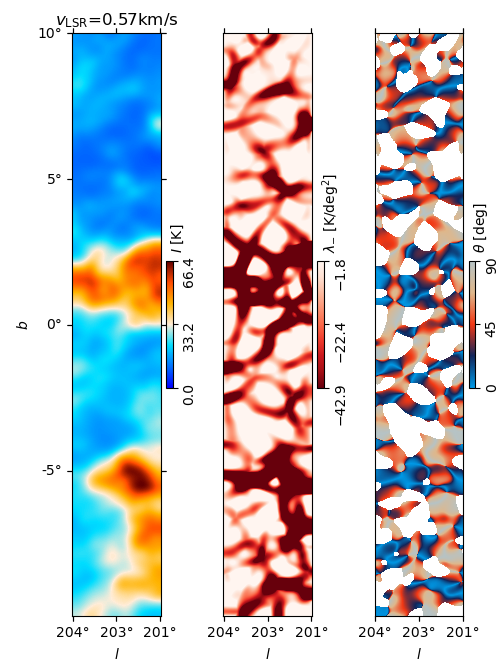}
\includegraphics[width=0.45\textwidth,angle=0,origin=c]{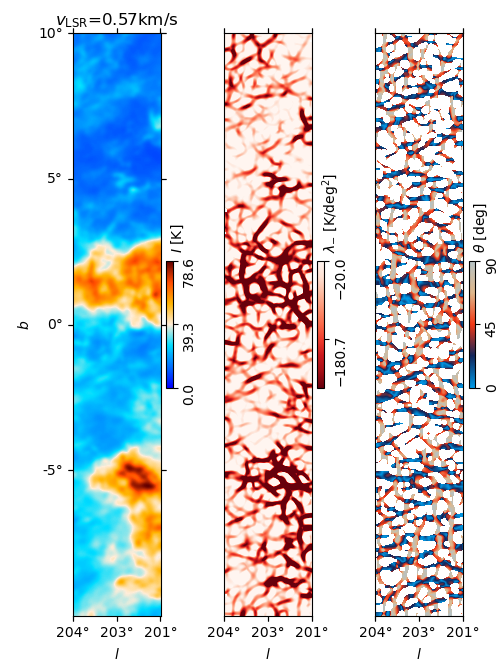}
}
\centerline{
\includegraphics[width=0.45\textwidth,angle=0,origin=c]{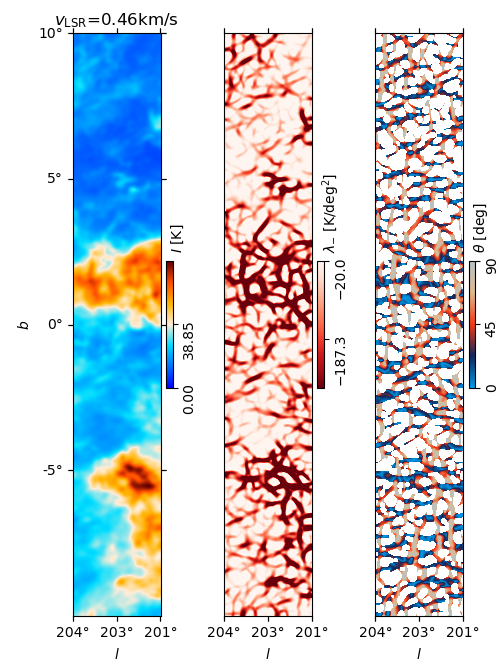}
\includegraphics[width=0.45\textwidth,angle=0,origin=c]{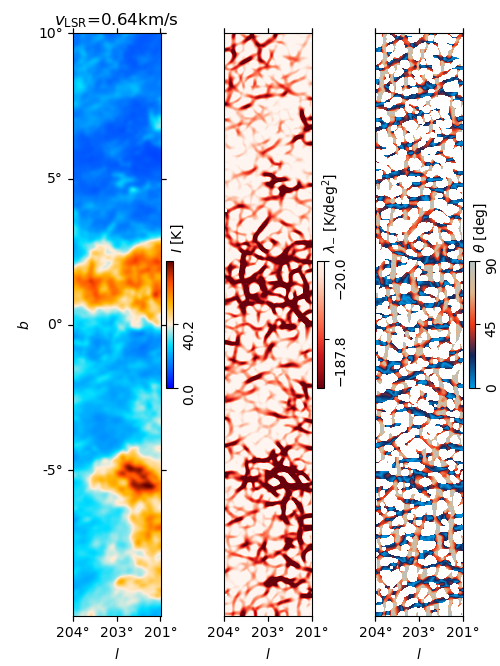}
}
\caption{Hessian matrix analysis applied to the HI4PI and the GALFA-HI observations.
{\it Top.} Results obtained on a 3\deg\,$\times$\,20\deg\,$\times$\,1.29\,km\,s$^{-1}$ tile in the HI4PI and the GALFA-HI observations, shown in the left and right panels, respectively.
{\it Bottom.} Results obtained on two 3\deg\,$\times$\,20\deg\,$\times$\,0.184\,km\,s$^{-1}$ tiles in the GALFA-HI observations.
In each of the panels, the left-hand side map shows the H{\sc i} emission convolved with a 2D Gaussian with the same diameter of the derivative kernel.
The map in the center shows the second eigenvalues of the Hessian matrix, $\lambda_{-}$, which highlights the filamentary structure.
The right-hand side map shows the orientation angle of the filamentary structure.
}
\label{fig:GALFAandHI4PIexamples}
\end{figure*}

We present an example of the Hessian analysis applied to one 3\deg\,$\times$\,20\deg\,$\times$\,1.29\,km\,s$^{-1}$ tile in the HI4PI and GALFA-HI observations in the top panels of Fig.~\ref{fig:GALFAandHI4PIexamples}.
This example tile illustrates the connection in the emission structures traced by the two surveys at different angular resolutions and projected into the same radial velocity resolution.

We also present an example of the Hessian analysis applied to the GALFA-HI observations in its native 0.184\,km\,s$^{-1}$ velocity resolution in the bottom panels of Fig.~\ref{fig:GALFAandHI4PIexamples}.
These tiles, which correspond to the same line of sight of those in the top panels, illustrate the similarity of the intensity structures in contiguous narrow velocity channels.
This similarity, caused by the linewidth of the H{\sc i} emission, is evident for the whole velocity range in Fig.~\ref{fig:HI4PIandGALFAprs}.

\begin{figure*}[ht]
\centerline{
\includegraphics[width=0.48\textwidth,angle=0,origin=c]{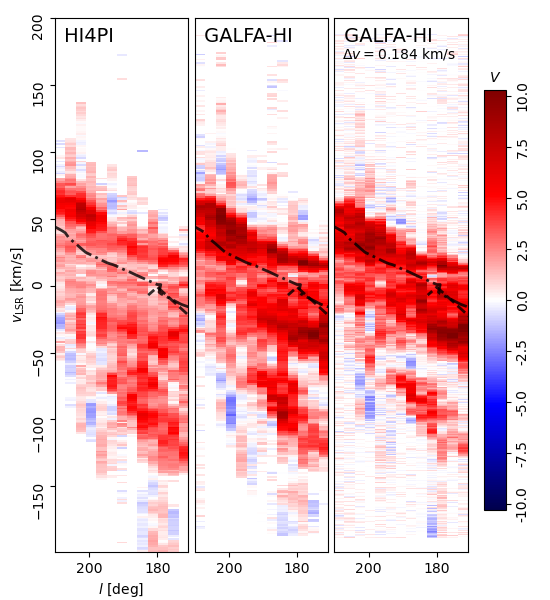}
\includegraphics[width=0.48\textwidth,angle=0,origin=c]{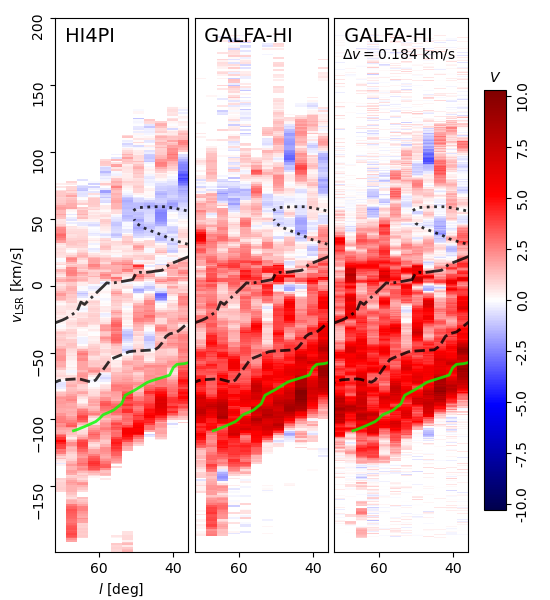}
}
\caption{Same as Fig.~\ref{fig:GALFAandHI4PIlvdiagrams}, but for an 8\arcmin\ FWHM kernel.
}
\label{fig:GALFAandHI4PIlvdiagrams8arcmin}
\end{figure*}

\begin{figure}[ht!]
\centerline{
\includegraphics[width=0.45\textwidth,angle=0,origin=c]{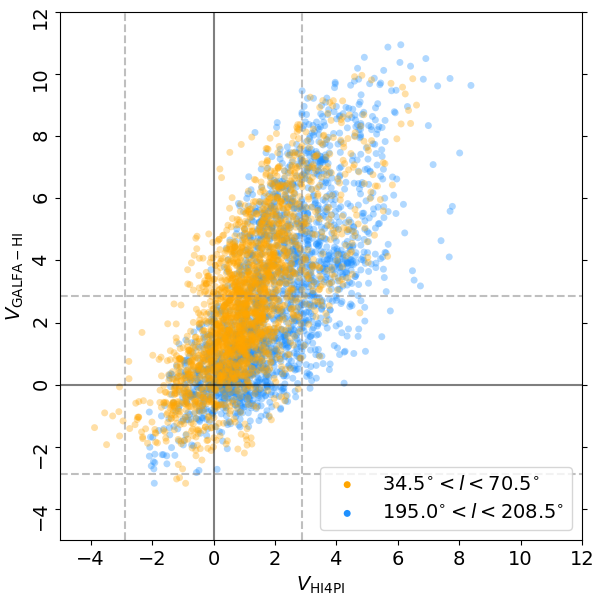}
}
\caption{Same as Fig.~\ref{fig:HI4PIandGALFAprs}, but for an 8\arcmin\ FWHM kernel.}
\label{fig:HI4PIandGALFAprs8arcmin}
\end{figure}

\juan{We also tested the results of a larger derivative kernel and a corresponding $\lambda^{C}_{-}$, as detailed in Table~\ref{table:HessianParametersGALFAHI8arcmin}.
We present the equivalent of Fig.~\ref{fig:GALFAandHI4PIlvdiagrams} for the 8\arcmin\ FWHM kernel in Fig.~\ref{fig:GALFAandHI4PIlvdiagrams8arcmin}.
As expected from the lower number of gradients derived with a large kernel, the values of $V$ are reduced.
The equivalent of Fig.~\ref{fig:HI4PIandGALFAprs} for the 8\arcmin\ FWHM kernel, shown in Fig.~\ref{fig:HI4PIandGALFAprs8arcmin}, indicates a significant increase in the values of $V$ with respect to those obtain for the HI4PI data.} 

\begin{table}
\caption{Hessian analysis parameters for the GALFA-HI data.}              
\label{table:HessianParametersGALFAHI8arcmin}      
\centering                                      
\begin{tabular}{l l l}          
\hline\hline                        
Parameter & & Value \\    
\hline                                   
Tile size & & 3\deg\,$\times$\,20\deg\,$\times$\,0.184\,km\,s$^{-1}$ \\ 
              & & 3\deg\,$\times$\,20\deg\,$\times$\,1.290\,km\,s$^{-1}$ \\      
Kernel size & & 8\arcmin\ FWHM \\
Intensity threshold & & 0.3\,K \\
Curvature threshold ($\lambda^{C}_{-}$) & & $-20.0$\,K/deg$^{2}$ \\
\hline                                             
\end{tabular}
\end{table}

\clearpage
\section{Filamentary structures in high velocity clouds}\label{appendix:HVCs}

\begin{figure}[h]%
\centering
\includegraphics[width=0.49\textwidth]{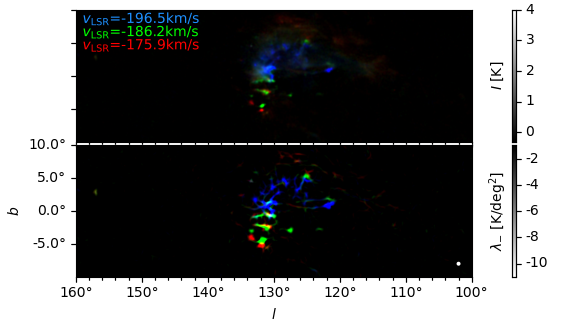}
\caption{Atomic hydrogen (H{\sc i}) emission (top) and filamentary structures identified using the Hessian matrix method (bottom) toward the HVC complex H.
}\label{fig:InterestingObject1}
\end{figure}

\begin{figure}[h]%
\centering
\includegraphics[width=0.49\textwidth]{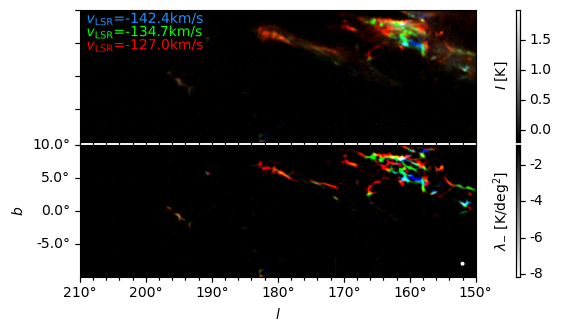}
\caption{Same as Fig.~\ref{fig:InterestingObject1}, but for the anticenter HVC complex.
}\label{fig:InterestingObject2}
\end{figure}

We present the result of the Hessian matrix analysis toward two high velocity cloud (HVC) complexes in Fig.~\ref{fig:InterestingObject1} and Fig.~\ref{fig:InterestingObject2}.

\raggedright

\end{document}